\documentclass[aps,superscriptaddress,amsmath,amssymb]{revtex4-2}

\setcitestyle{super}

\usepackage{bm}

\usepackage{array}
\newcolumntype{.}{D{.}{.}{8}}
\usepackage{dcolumn}

\usepackage[utf8]{inputenc}
\usepackage{physics}
\usepackage{tabularx,booktabs,array,dcolumn}
\usepackage{setspace}
\usepackage{vmargin}
\usepackage{xcolor}
\usepackage{graphicx,psfrag,subfigure}

\usepackage{silence}
\usepackage{float}
\usepackage[all]{xy}
\usepackage{multirow}

\usepackage{makecell}
\usepackage{amsmath}
\usepackage{amstext}
\usepackage{amsfonts}
\usepackage{amssymb}
\usepackage{color}
\usepackage{longtable}
\usepackage{rotating}
\usepackage{multirow}
\usepackage{tikz}
\usepackage[version=4]{mhchem}
\usepackage{listings}
\usepackage{float}
\usepackage{threeparttable}
\usepackage{url}
\usepackage{here}
\usepackage[absolute,overlay]{textpos}
\usepackage{adjustbox}  

\usepackage[unicode]{hyperref}
\hypersetup{
   unicode=true,          
   plainpages=false,
   colorlinks=true,       
   linkcolor=blue,          
   linkcolor=blue,          
   citecolor=blue,        
   filecolor=blue,      
   urlcolor=blue           
}
\usepackage{url}
\usepackage{siunitx}  

\newcommand{\be}{\begin{equation}}
\newcommand{\ee}{\end{equation}}
\newcommand{\bea}{\begin{eqnarray}}
\newcommand{\eea}{\end{eqnarray}}

\newcommand{\eqa}{\begin{equation}}
\newcommand{\eqz}{\end{equation}}
\newcommand{\eqma}{\begin{eqnarray}}
\newcommand{\eqmz}{\end{eqnarray}}

\newcommand{\bos}[1]{\boldsymbol{#1}}
\newcommand{\cm}{$\text{cm}^{-1}$}

\newcommand{\Cs}{$C_\mathrm{s}$}
\newcommand{\Ctv}{$C_{\mathrm{3v}}\mathrm{(M)}$}
\newcommand{\mr}[1]{\mathrm{#1}}

\newcommand{\ROBO}{{\scshape Robosurfer}}
\newcommand{\nut}{$\tilde{\nu}$}

\newcommand{\ild}{\raisebox{0.5ex}{\texttildelow}}

\usepackage{xcolor}

\begin{document}

\title{%
  Exact quantum dynamics of methanol: full-dimensional \emph{ab initio} potential energy surface of spectroscopic quality and variational vibrational states
}

\author{Ayaki Sunaga}
\affiliation{%
ELTE, E\"otv\"os Lor\'and University, Institute of Chemistry, P\'azm\'any P\'eter s\'et\'any 1/A 1117 Budapest, Hungary
}
\author{Tibor Győri}
\email{tibor.gyori@chem.u-szeged.hu}
\affiliation{%
MTA-SZTE Lendület ``Momentum'' Computational Reaction Dynamics Research Group, Interdisciplinary Excellence Centre and Department of Physical Chemistry and Materials Science, Institute of Chemistry, University of Szeged, Rerrich Béla tér 1, Szeged H-6720, Hungary
}
\author{G\'abor Czakó}
\email{gczako@chem.u-szeged.hu}
\affiliation{%
MTA-SZTE Lendület ``Momentum'' Computational Reaction Dynamics Research Group, Interdisciplinary Excellence Centre and Department of Physical Chemistry and Materials Science, Institute of Chemistry, University of Szeged, Rerrich Béla tér 1, Szeged H-6720, Hungary
}
\author{Edit M\'atyus}%
\email{edit.matyus@ttk.elte.hu}
\affiliation{%
ELTE, E\"otv\"os Lor\'and University, Institute of Chemistry, P\'azm\'any P\'eter s\'et\'any 1/A 1117 Budapest, Hungary
}%

\date{\today}
\begin{abstract}
\noindent %
The methanol molecule is a sensitive probe of astrochemistry, astrophysics, and fundamental physics. The first-principles elucidation and prediction of its rotation-torsional-vibrational motions are enabled in this work by the computation of a full-dimensional, \emph{ab initio} potential energy surface (PES) and numerically exact quantum dynamics.
An active-learning approach is used to sample explicitly correlated coupled-cluster electronic energies, and the datapoints are fitted with permutationally invariant polynomials to obtain a spectroscopic-quality PES representation.
Variational vibrational energies and corresponding tunnelling splittings are computed up to the first overtone of the C-O stretching mode by direct numerical solution of the vibrational Schrödinger equation with optimal internal coordinates and efficient basis and grid truncation techniques. 
As a result, the computed vibrational band origins finally agree with experiment within $5\ \text{cm}^{-1}$, allowing for the exploration of the large-amplitude quantum mechanical motion and tunnelling splittings coupled with the small-amplitude vibrational dynamics. These developments open the route towards simulating rovibrational spectra used to probe methanol in outer space and in precision science laboratories, as well as for probing interactions with external magnetic fields.
\end{abstract}

\maketitle 


%
\section{Introduction}
CH$_3$OH is the smallest alcohol and a prototype of quantum mechanical large-amplitude motions and corresponding tunnelling splittings. 
It is also one of the most abundant organic molecules in the Universe, which explains its importance in astrophysics. 
Its particular rovibrational level structure, dominated by the large-amplitude internal rotation, 
has been proposed as a sensitive probe for the proton-to-electron mass ratio variation at astronomical scales~\cite{Jansen2011PRL_ch3oh,Levshakov2011APJ_methanol,Bagdonaite2013Science,Vorotyntseva2023MNRAS_ch3oh}.
Most recently, the internal rotation of substituted methanol compounds was considered for enhanced parity-violating effects~\cite{Sunaga2025JCP}.
The hyperfine-Zeeman splittings of methanol, based on a reduced dimensionality model, were used to measure (weak) magnetic fields on distant astronomical objects~\cite{Lankhaar2018NA_hyperfine}.
The methanol molecule also plays a central role in astrochemistry; the ratio of its isotopologues has been studied in the context of the formation of organic compounds in outer space~\cite{Kleiner2019ACS,Jorgensen2016AA_methanol,Oyama2023AJ,Riedel2023AA_methanol,Jeong2025AJSS_methanol,Tamanai2025AJ_methanol}. 

Methanol has been a prototype for fundamental spectroscopy and quantum nuclear motion theory, mostly for the non-trivial couplings of its small-amplitude degrees of freedom, the large-amplitude internal rotation, and the overall molecular rotation.
Its fundamental vibrations were detected with infrared~\cite{Borden1938JCP_methanol} and Raman~\cite{Halford1937JCP_methanol} spectroscopy already in 1937-38 and were assigned based on the $C_\text{s}$ point group~\cite{Tanaka1957SA_Cs,Falk1961JCP_Cs,Timidei1970ZNA_Cs,Shimanouchi1972_Tables_of_vib,Serrallach1974JMS}.
The torsion-rotation coupling was later analysed in a series of work~\cite{Herbst1984JMS_torrot_exp,Xu1996JMS_torrot_exp,Lees2002PRA,Lees2004JMS,Temsamani2003JMS,Xu2008JMS}, recently followed by high-precision spectroscopy using an ultra-narrow quantum cascade laser~\cite{Santagata2019Optica_methanol}. 
The molecular symmetry group~\cite{Bunker2006_molsym} can be used for a more complete understanding of the degenerate and tunnelling splitting levels, but 
arguments were put forward for using the complete permutation-inversion group~\cite{Quack1977MP_parity,Quack2011_Handbook}, 
so as not to lose important parity information of the eigenstates.

We also note that combination bands of methanol have been studied with matrix isolation spectroscopy~\cite{Perchard2007CP,Perchard2008CP_2400,Dinu2024ACSAu}.
Regarding (rectilinear) normal coordinate computations, which are limited to describing small-amplitude vibrations, we mention vibrational self-consistent field (VSCF) and related vibrational configuration interaction-type (VCI) computations~\cite{Bowman1978JCP,Yagi2004JCP_methanol,Urena2005JMS,Scribano2010JCP,Schroder2022_VCI}. The computed VCI fundamental-mode energies, not considering the torsional motion and the tunneling splittings, agree well with the matrix isolation data~\cite{Dinu2020TCA_4cm}. Further computational methodologies were developed using a tree-tensor-network-state ansatz based on the state-shifted Hamiltonian~\cite{Larsson2025JPCL_TTN}.

There has been recent progress in a numerically exact solution of the rovibrational Schrödinger equation on an \emph{ab initio} (or perhaps spectroscopically refined) potential energy surface (PES), allowing also for large-amplitude motions~(LAMs)~\cite{Csaszar2012PCCP_fourth,Tennyson2016JCP_review,Carrington2017JCP_review,Chen2021arXiv_ElVibRot,Matyus2023CC}.
In relation to scaling up the vibrational dimensionality, we mention recent developments in collocation~\cite{Simmons2023JCP_new_collocation,wodraszka2024JCP_collocation} and contraction~\cite{Kallullathil2023JCP_contraction, Simko2024JCP,Simko2025JCP_H2O_tri}.
The full-dimensional (12D) variational computation of methanol was first reported by Lauvergnat and Nauts~\cite{Lauvergnat2014SA_methanol,Nauts2018MP_methanol}.
In our group, 12D variational computations were also performed using the GENIUSH-Smolyak approach for methane-atom(ion) complexes~\cite{Avila2019JCP_methodology,Avila2019JCP_CH4F-,Avila2020PCCP,Papp2023MP}, by including the atom(ion)-methane large-amplitude relative motions fully coupled with the methane's small-amplitude internal vibrations. These developments were followed by applications for the formic acid molecule~\cite{Daria2022JMS_HCOOH,Avila2023PCCP_HCOOH} and methanol~\cite{Sunaga2024JCTC} using path-following curvilinear normal coordinates. The efficient basis and grid truncation made it possible to converge the vibrational energies of methanol (12D) within 1-2~\cm\ with respect to the vibrational basis and grid up to the combination bands at ca. 2200~\cm\ beyond the zero-point vibrational energy, ZPVE. In this range, the large, 10-20~\cm\ deviations from experiment indicated that the underlying PES~\cite{Qu2013MP_CH3+OH} (henceforth labeled as PES13) required improvement \cite{Sunaga2024JCTC}.

Regarding the PES development, the challenging aspects of treating the torsional motion are
(i) accounting for the three equivalent minima and (ii) covering a wider geometry region beyond small-amplitude motions (SAMs) about local minima.
A solution for (i) has been well established in the community: the permutation-invariant-polynomial (PIP) method, which was initially reported by Joel Bowman’s group~\cite{Huang2005JCP,Braams2009IRPC_pip,Chen2018ARPC_PIP} and is continuing to see recent improvements~\cite{Qu2018JCTC_PIP,Moberg2021JCTC_PIP,Houston2023JCP_PESPIP,Drehwald2025JCP_MOLPIP,Bowman2025ChemRxiv_PIP}.
Solving (ii) involves expanding the region of interest in the 12D hypervolume, which necessitates the use of either increasingly many \emph{ab initio} energies to adequately sample the hypervolume or some sparse sampling technique that takes advantage of the fitting method's ability to interpolate between the points of the fitting set. Since high-accuracy \emph{ab initio} energies are computationally expensive, the latter approach is preferable; however, the practical implementation details remain challenging and will be addressed in this work.
\par
One of the strategies for efficiently choosing geometry points for the fitting is the active-learning-like approach of the {\scshape Robosurfer} program system~\cite{Gyori2020JCTC}. The accuracy of PESs developed with the PIP fitting has been demonstrated~\cite{Avila2020PCCP,Papp2023MP} from the equilibrium structures up to the methane-atom(ion) dissociation limits of the methane complexes {CH$_4\cdot$Ar} and {CH$_4\cdot$F$^{-}$}, of which the latter applications used \ROBO. In this study, we use and improve upon this combination of methods to obtain a new PES for methanol, which is reported in the first part of the paper.

%
%
\section{Construction of the potential energy surface}\label{sec:PES}
\subsection{Background and methodology}
\par
The concept of a PES arises from the Born-Oppenheimer approximation. For every molecular geometry, there is a potential energy value that can (in principle) be obtained by solving the Schrödinger equation of the electrons. While no analytical solution is known for many-electron systems and numerically exact solutions are unfeasible in most cases, modern quantum chemistry offers a number of efficient and accurate approximations. These typically involve the simultaneous use of: (I) an approximate many-electron wavefunction where the correlated motion of electrons stemming from electron-electron repulsion is only partially considered and relativistic effects are either ignored or approximated, and (II) a finite set of carefully optimized basis functions which are linearly combined to construct the molecular orbitals.

In this study, we combine a well-established coupled-cluster approximation of electronic motion, CCSD(T)-F12b~\cite{Adler2007JCP_CCSDT_F12}, with the cc-pVTZ-F12~\cite{Peterson2008JCP_cc-pVnZ-F12} basis set to compute potential energy values at a reasonable computational cost. 
\par
In principle, a variational vibrational computation could be carried out by querying the \emph{ab initio} energy computation method described above at every quadrature point of the numerical integration grid used for computing the matrix elements of the vibrational Hamiltonian. But in practice, this would be prohibitively slow, as the potential energy is required at (hundreds of) millions of geometries, and each \emph{ab initio} energy may take a couple of minutes of CPU time. Therefore, to make the \emph{ab initio} computation of (ro)vibrational energy levels possible, another layer of approximation needs to be introduced: a surrogate model PES, fitted to a limited number of \emph{ab initio} energies.
\par
In this study, we use the Braams--Bowman implementation of permutationally invariant polynomials (PIP)~\cite{Braams2009IRPC_pip} for fitting. Briefly, the polynomials are expanded in terms of Morse variables
\begin{equation}\label{eq:PES_morse}
y_{ij}=\exp(-r_{ij}/a),
\end{equation}
where $r_{ij}$ is the distance between atoms $i$ and $j$ and $a$ is a distance transformation parameter. The number of coefficients (and as a result, the flexibility of the fitting function) is primarily controlled by limiting the degrees of the polynomials, as per standard practice. The energy expression of the PES is
\begin{equation}\label{eq:PES_energy}
  E=\sum_{k=1}^{Z}c_kT_kM_k,
\end{equation}
where $Z$ is the number of permutationally invariant monomials, while $c_k$ and $T_k$ are the coefficient and monomial prefactor of the $k$-th monomial $M_k$, respectively. The PIP method ensures that the values of the energy returned by the PES function are invariant to any permutation of like atoms, in our case, all four H-atoms. A special feature of the PIP implementation used in this study is the ability to increase the fitting function flexibility by adding extra polynomials that are restricted in the number of atomic positions they are permitted to depend on.
We refer to these as `extra $q$-body polynomials,' and we utilise extra 2-, 3-, and 4-body polynomials in the functional form of our new potential energy surface (PES). These extra polynomials and the main polynomial can each use different values for $a$, Eq.~\eqref{eq:PES_morse}, to achieve optimal fitting of the \emph{ab initio} energy-geometry pairs in the fitting set.
The polynomial coefficients are obtained through weighted linear least squares (LLS) fitting; the weighting function is provided in Eq.~S3 of the Supplementary Material.

\par
For constructing the fitting set, we used the \ROBO\ program system\cite{Gyori2020JCTC}, and for computing the required \emph{ab initio} energies, we used the {\scshape Molpro}~\cite{Lindh1991JCP_seward,MOLPRO2012,MOLPRO_Werner2020JChemPhys,MOLPRO2023} quantum chemistry package. Without going into the details of its operation, \ROBO\ automates the development of PESs through an iterative, active-learning-like process, which gradually expands the fitting set with additional samples (geometry-energy pairs) until the quality of the fitted PES is judged to be satisfactory.

\par
For the PES development, we have taken the approach of starting with a small fitting set, low degree polynomials, a high threshold (in terms of weighted error) for adding new geometries to the fitting set ($E_\text{targ}$), and gradually increasing both the flexibility of the fitting function and the accuracy we demand from the PES, as the fitting set grew. In addition to readjusting the parameters of the automated PES development process multiple times, we have also, on a handful of occasions, manually injected geometries into the development process, which were sampled around outliers in the fit. The PES development was started from a fitting set of 530 geometries, $E_\text{targ}=1$~kcal/mol (349.8~\cm), a 4\textsuperscript{th} degree polynomial fit and no extra monomials, resulting in 317 polynomial coefficients.
By the end of the PES development, the fitting set grew to 39~401 geometries, $E_\text{targ}$ was lowered to 0.05~kcal/mol (17.5~\cm), and the complexity of the fitting function was increased to the state shown in Table~\ref{tbl:a_fit_order}. Further details of the fitting and PES development methods, as well as a highly detailed account of the progress of PES development and fitting error minimisation efforts, are provided in Section~S1 of the Supplementary Material.

\subsection{Properties and validation of the final PES}
Our new PES, named PES25, is the result of fitting the final geometry set of 39~401 points with the PIP method and the parameters shown in Table~\ref{tbl:a_fit_order}, yielding 9652 polynomial coefficients. For numerical stability reasons discussed in Section~S2 of the Supplementary Material, we obtained the final coefficient values of PES25 using the DGELSY linear least squares solver, which is a LAPACK subroutine~\cite{laug}. This results in slightly higher fitting errors, but produces a PES with smoother derivatives.

%
\begin{table}[htbp!]
\caption{Employed distance transformation parameters and polynomial degrees in the final fitting function}\label{tbl:a_fit_order}
\centering
\begin{tabular}{@{}c cc@{}}
\hline\hline\\[-0.4cm]
Polynomial & degree & $a$ [bohr]\tabularnewline
\hline\\[-0.4cm]
main (6-body) & 7 & 1.9 \tabularnewline
extra 2-body & 2 & 2.5 \tabularnewline
extra 3-body & 3 & 1.7 \tabularnewline
extra 4-body & 4 & 3.0 \tabularnewline
\hline\hline
\end{tabular}
\end{table}
Our fidelity assessment of the fit is summarised in Table~\ref{tbl:RMSWE}. The RMS fitting error of PES25 over its fitting set, 27.3 \cm, is noticeably higher than those of our previous spectroscopic PIP-PESs; e.g., the RMSEs of Refs~\citenum{Avila2020PCCP} and~\citenum{Papp2023MP} are ca. 1 \cm\ and 3 \cm, respectively. While the RMS weighted fitting error of PES25 is considerably better at 17.3 \cm, that is still an order of magnitude larger than the errors of the aforementioned PESs. This is attributable to a multitude of factors.
%
%
\begin{table}[!htbp]
\caption{%
  Assessment of PES25: number of geometries, root-mean-square error (RMSE) of the fitting set, for energy ranges relative to the \emph{ab initio} equilibrium energy. 
  The total RMSE and the RMS weighted error (RMSWE) over all geometry points are also 
  listed.
  }\label{tbl:RMSWE}
\centering
\begin{tabular}{@{}r rc@{}}
\hline\hline\\[-0.4cm]
Energy range [\cm] & \# of geometries & RMSE [\cm] \tabularnewline
\hline\\[-0.4cm]
0--13011 & 29029 & 13.8\tabularnewline
13011--21947 & 7587 & 36.9\tabularnewline
21947--43894 & 2785 & 69.8\tabularnewline
Total points & 39401 & \tabularnewline
\hline\\[-0.4cm]
Total RMSE &  & 27.3\tabularnewline
Total RMSWE &  & 17.3\tabularnewline
\hline\hline
\end{tabular}
\end{table}
\par
First, it is generally perilous to judge the quality of a statistical model (such as a fitted PES) only from its error over its fitting set, as the accuracy of the model can be vastly different at different geometries. 
This is illustrated by the fact that over the course of developing this PES we have accumulated over 174~000 geometries which were not included in the fit due to their weighted fitting error being less than 17.5~\cm. While this is still not a statistically independent test set, it does imply that the PES works well even for geometries not included in the fitting set.
\par
Second, there are matters of geometry generation. The {CH$_4\cdot$Ar} PES in Ref.~\citenum{Avila2020PCCP} was developed before the development of \ROBO, and its fitting set was generated through the random distortion of methane's equilibrium geometry, while the C--Ar distance and CH$_4$ orientation were randomly sampled. Since this was done without any steps to filter out geometries that would have already been well-described by the PES without them, it is likely that the RMS fitting error of the fitting set of Ref.~\citenum{Avila2020PCCP} benefited from the sampling method. 
While the PES reported in Ref.~\citenum{Papp2023MP} was developed with \ROBO, PES25 is the result of far more sampling effort, which almost always raises the RMS errors of the fitting set, as it discovers and adds points of high error into the fit.
%
%
\begin{table*}[!htbp]
\caption{
Geometries at one of the three permutationally equivalent equilibrium structures and one saddle point obtained with PES13\cite{Qu2013MP_CH3+OH} and PES25 (this work) and \textit{ab initio} computations. Bond lengths are in \AA, bond angles are in degrees. The internal coordinates $r,\theta$, and $\tau$ are defined in Fig.~\ref{fig:methanol}.
}\label{tbl:structure}
\centering
\scalebox{0.95}{
\begin{tabular}{@{}l SSSS |SSSS@{}}
\hline\hline\\[-0.4cm]
 & \multicolumn{4}{c|}{Equilibrium structure} & \multicolumn{4}{c}{Saddle point}\tabularnewline
 & \multicolumn{2}{c}{this work} & \multicolumn{2}{c|}{Ref.~\citenum{Qu2013MP_CH3+OH}} & \multicolumn{2}{c}{this work} & \multicolumn{2}{c}{Ref.~\citenum{Qu2013MP_CH3+OH}}\tabularnewline
 & {PES25} & {\textit{ab initio}$^a$} & {PES13} & {\textit{ab initio}$^b$} & {PES25} & {\textit{ab initio}$^a$} & {PES13} & {\textit{ab initio}$^b$}\tabularnewline
 \hline\\[-0.4cm]
$r_1$(CO)               & 1.4196 & 1.4199 & 1.419 & 1.422 & 1.4236 & 1.4235 & 1.423 & 1.423\tabularnewline
$r_2$(OH)               & 0.9583 & 0.9582 & 0.959 & 0.959 & 0.9558 & 0.9559 & 0.957 & 0.956\tabularnewline
$r_3$(CH$_1$)           & 1.0876 & 1.0875 & 1.089 & 1.088 & 1.0908 & 1.0909 & 1.094 & 1.093\tabularnewline
$r_{4,5}$(CH$_n,n=2,3$) & 1.0931 & 1.0931 & 1.095 & 1.095 & 1.0909 & 1.0907 & 1.093 & 1.093\tabularnewline
$\theta_1$(HOC)               & 108.19 & 108.20 & 108.3 & 108.6 & 108.68 & 108.65 & 109.1 & 108.7\tabularnewline
$\theta_2$(OCH$_1$)           & 106.65 & 106.74 & 106.7 & 106.8 & 112.05 & 112.04 & 112.7 & 112.1\tabularnewline
$\theta_{3,4}$(OCH$_n,n=2,3$) & 111.94 & 111.90 & 111.9 & 112.0 & 109.50 & 109.49 & 109.4 & 109.5\tabularnewline
$\tau_1$(H$_1$COH) & 180.00 & 180.00 & 180.0 & 180.0 & 0.00 & 0.00 &  & \tabularnewline
$\tau_2$(H$_2$COH) & 61.39 & 61.39 &  &  & -120.37  & -120.39  &  & \tabularnewline
$\tau_3$(H$_3$COH) & 298.61 & 298.61 &  &  & 120.37 & 120.39 &  & \tabularnewline
\hline\hline
\end{tabular}
}
\begin{flushleft}
$^a$~CCSD(T)-F12b/cc-pVTZ-F12
$^b$~CCSD(T)-F12b/aug-cc-pVDZ  
\end{flushleft}
\end{table*}

\par
Third, the weighting function (Eq.~S3 in the Supplementary Material) used for PES25 to fit the surface and select the geometries places much more importance on moderately high-energy geometries than the two previous PESs. 
As an example, while the development of PES25 considered all geometries up to 13 011 \cm\ equally important (a reasonable guess for a PES intended to describe vibrations up to ZPVE + 2200 \cm), the weighting used in Ref.~\citenum{Papp2023MP} would have only given a geometry at 13 011 \cm\ a weight of \ild0.4. 
As a result of this increased region of interest (in terms of energy), achieving a low RMS fitting error becomes much more challenging and demands more flexibility from the fitting function. And finally, of course, different molecular systems can be more or less challenging to create a PES for. It is possible that methanol is inherently a more difficult system to fit than {CH$_4\cdot$Ar} or {CH$_4\cdot$F$^{-}$}.

The quality and improvement of PES25 over PES13 can also be assessed by comparing the \textit{ab initio} computations and the literature values of Ref.~\citenum{Qu2013MP_CH3+OH}. The molecular structure parameters are listed in Table~\ref{tbl:structure}. The PES25 values agree well with the \textit{ab initio} values with a maximum deviation of $3\times10^{-4}$ \AA\ for bond lengths and $0.09^\circ$ for bond angles. 
The largest changes are due to the different basis set, $2\times10^{-3}$ \AA\ for bond lengths and $0.4^\circ$ for bond angles (Table~\ref{tbl:structure}, \emph{ab initio} data). The accuracy of the PES fitting is sufficient to reflect the improvement in terms of the basis set size.
%
%
\begin{figure}[!htbp]
    \centering
    \includegraphics[width=0.7\linewidth]{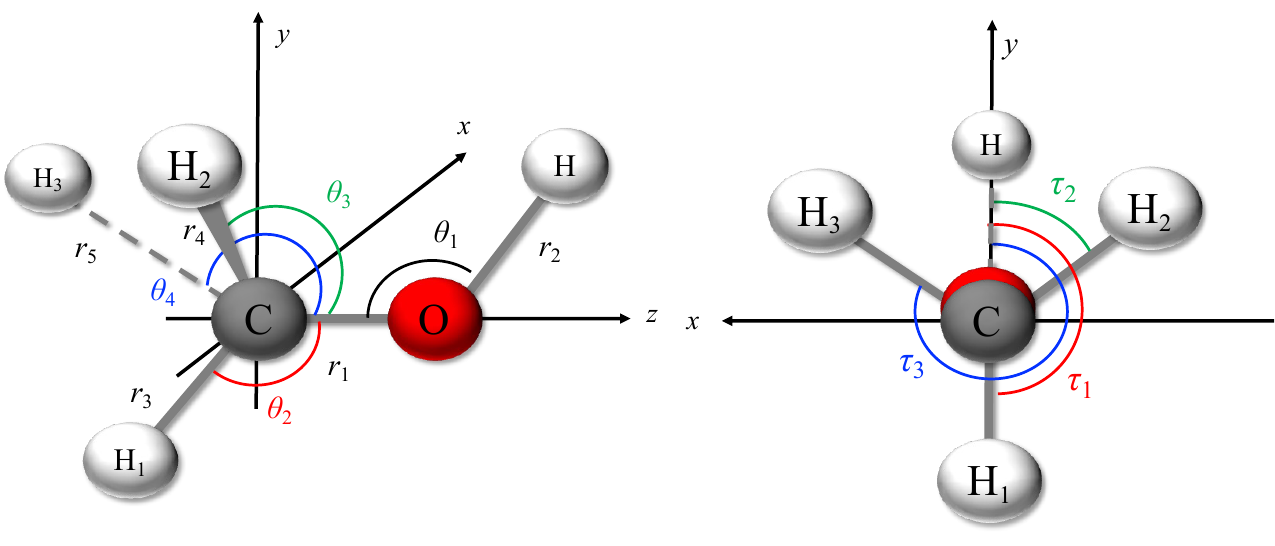}
    \caption{Definition of the primitive internal coordinates of the methanol molecule and the body-fixed Cartesian frame used in this work.}
    \label{fig:methanol}
\end{figure}

\par
%
%
\begin{table*}[!htbp]
\caption{
Harmonic frequencies and harmonic zero-point vibrational energy (ZPVE) in \cm\ at one of the three permutationally equivalent equilibrium structures and one saddle point obtained with the PES25 and \textit{ab initio} computations. 
The harmonic frequencies of Ref.~\citenum{Qu2013MP_CH3+OH} are listed in decreasing energetic order. Mean absolute errors (MAE) and maximum (MAX) errors of the harmonic frequencies computed with the PESs and obtained from the \textit{ab initio} computations are also shown.
}\label{tbl:harm_freq}
\centering
\begin{tabular}{@{}lll rrrr |rrrr@{}}
\hline\hline\\[-0.4cm]
& & & \multicolumn{4}{c|}{Equilibrium structure} & \multicolumn{4}{c}{Saddle point}\tabularnewline
& &  & \multicolumn{2}{c}{this work} & \multicolumn{2}{c|}{Ref.~\citenum{Qu2013MP_CH3+OH}} & \multicolumn{2}{c}{this work} & \multicolumn{2}{c}{Ref.~\citenum{Qu2013MP_CH3+OH}}\tabularnewline
description & coord & sym  & {PES25} & {\textit{ab initio}$^a$} & {PES13} & {\textit{ab initio}$^b$} & {PES25} & {\textit{ab initio}$^a$} & {PES13} & {\textit{ab initio}$^b$}\tabularnewline
 \hline\\[-0.4cm]
 $\nu$(\ce{OH}) & $r_\mathrm{OH}$ & $a'$  & 3863.1 & 3866.4 & 3869.5 & 3864.9 &    3898.5 & 3900.5 & 3901.8 & 3901.5\tabularnewline
 $\nu$(\ce{CH3})$_{\mathrm{asym}}$  & $r_\mathrm{CH}$ &$a'$   & 3134.4 & 3136.7 &     3150.3 & 3133.0 & 3104.9 & 3107.9 & 3120.8 & 3106.1\tabularnewline
 $\nu$(\ce{CH3})$_{\mathrm{asym}}$ & $r_\mathrm{CH}$&$a''$  & 3073.1 & 3075.7 & 3068.8 & 3071.9 &     3093.8 & 3101.9 & 3090.2 & 3100.5\tabularnewline
 $\nu$(\ce{CH3})$_{\mathrm{sym}}$& $r_\mathrm{CH}$ &  $a'$  & 3014.6 & 3015.9 & 3016.2 & 3013.4 &     3026.0 & 3029.1 & 3018.4 & 3029.6\tabularnewline
 $\delta$(\ce{CH3})$_{\mathrm{asym}}$ & $\varphi_2$ &  $a'$ & 1525.0 & 1521.6 & 1527.1 & 1514.8 &     1528.2 & 1529.2 & 1518.4 & 1526.8\tabularnewline
$\delta$(\ce{CH3})$_{\mathrm{asym}}$ & $\varphi_1$ &$a''$  & 1512.4 & 1511.1 & 1499.0 & 1504.6 &     1505.7 & 1500.8 & 1508.6 & 1499.2\tabularnewline
 $\delta$(\ce{CH3})$_{\mathrm{sym}}$ & $\theta_\mathrm{HCO}$ &  $a'$ & 1485.4 & 1485.4 & 1476.0 & 1481.3 &     1487.9 & 1489.5 & 1498.2 & 1487.7\tabularnewline
 $\delta$(\ce{COH}) & $\theta_\mathrm{COH}$ &  $a'$  & 1378.8 & 1383.7 & 1376.8 & 1380.5 &     1355.9 & 1366.1 & 1369.0 & 1366.0\tabularnewline
$\rho$(\ce{CH3})$_{\mathrm{}}$& $\theta_\mathrm{HCO}$ &  $a''$  & 1177.7 & 1181.7 & 1196.2 & 1178.4 &     1186.6 & 1191.5 & 1191.4 & 1192.3\tabularnewline
$\rho$(\ce{CH3})$_{\mathrm{}}$ &$\theta_\mathrm{HCO}$ & $a'$  & 1091.9 & 1090.1 & 1067.0 & 1089.5 &     1097.5 & 1096.7 & 1080.3 & 1100.3\tabularnewline
 $\nu$(\ce{CO})$_{\mathrm{}}$&  $r_\mathrm{CO}$ & $a'$  & 1063.3 & 1062.2 & 1065.2 & 1064.3 &     1062.2 & 1063.1 & 1058.5 & 1069.1\tabularnewline
 tor(OH) &  $\tau$ & $a''$  & 298.7 & 293.1 & 287.0 & 289.2 &     300.2$i$ & 296.8$i$ & 301.9$i$ & 284.8$i$\tabularnewline
\multicolumn{3}{c}{ZPVE}  & 11309.5 & 11311.8 & 11299.6 & 11292.9 & 11173.8 & 11188.1 & 11177.8 & 11189.6\tabularnewline
 \hline
\multicolumn{3}{c}{MAE}  & 2.6 &  & 8.2 &  & 3.6 &  & 9.7 & \tabularnewline
\multicolumn{3}{c}{MAX} & 5.7 &  & 22.5 &  & 10.2 &  & 20.0 & \tabularnewline
\hline\hline
\end{tabular}
\begin{flushleft}
$^a$~CCSD(T)-F12b/cc-pVTZ-F12
$^b$~CCSD(T)-F12b/aug-cc-pVDZ  
\end{flushleft}
\end{table*}
The harmonic frequencies and deviations from the \textit{ab initio} computations are summarised in Table \ref{tbl:harm_freq}. Both MAE and MAX errors of PES25 are less than those of Ref.~\citenum{Qu2013MP_CH3+OH}. Most of the harmonic frequencies obtained with the two \textit{ab initio} computations agree within a few \cm, and the largest deviation of more than 6.0~\cm\ is found for the modes related to the H-O-C-H$_n(n=1,2,3)$ dihedral angles. 
Interestingly, the torsional harmonic frequency changes by more than 10~\cm\ by changing from PES13 to PES25. However, most of the variational vibrational energies corresponding to the torsional motion are less sensitive to this change (Table~S7 of the Supplementary Material). The largest deviation, 8.3~\cm, is observed in the 1800 \cm\ region beyond the ZPVE (Table~S8 of the Supplementary Material). 
\par
As a final test of the quality of PES25, we have prepared one-dimensional slices of the PES along selected internal coordinates, which extend into regions moderately far from the equilibrium structure. As shown in Figs.~S11--S13 of the Supplementary Material, the fitting errors along these slices match the RMS weighted fitting error of the fitting set.

%
%
\section{Variational vibrational computations}

\subsection{Methodological details}
We solve the vibrational Schrödinger equation, 
\be
  [\hat{T}^{\mathrm{v}} + V] \Psi_n = E_n\Psi_n
  \label{eq:vibSch}
\ee
using the $V$ potential energy surface developed in the first part of this work. The vibrational kinetic energy operator is
\be
  \hat{T}^{\mathrm{v}}
  =
  -\frac{\hbar^2}{2} 
  \sum_{k=1}^{D} \sum_{l=1}^{D} 
    G_{k l} \frac{\partial}{\partial \rho_k} \frac{\partial}{\partial \rho_l}
  -\frac{\hbar^2}{2} 
  \sum_{k=1}^{D} 
    B_k \frac{\partial}{\partial \rho_k}
  +U
\label{eq:keo}
\ee
with
\be
 B_k=\sum_{l=1}^{D} \frac{\partial}{\partial \rho_l} G_{l k}
\ee
and
\be
 U=\frac{\hbar^2}{32} \sum_{k=1}^{D}\sum_{ l=1}^{D}\left[\frac{G_{k l}}{\tilde{g}^2} \frac{\partial \tilde{g}}{\partial \rho_k} \frac{\partial \tilde{g}}{\partial \rho_l}+4 \frac{\partial}{\partial \rho_k}\left(\frac{G_{k l}}{\tilde{g}} \frac{\partial \tilde{g}}{\partial \rho_l}\right)\right] \; ,
\ee
where
$D$ is the vibrational dimensionality (here $D=12$), 
$\tilde{g}=\det\bos{g}$ is the determinant and $G_{kl}=(\bos{g}^{-1})_{kl}$ labels elements of the inverse of the mass-weighted metric tensor corresponding to the coordinate transformation from laboratory-frame Cartesian coordinates to rotational and internal, vibrational coordinates (the overall molecular translation is exactly separated~\cite{Matyus2023CC}).

The $\rho_k$ internal coordinate definition is summarised as follows.
To have a large-amplitude torsional coordinate that carries the permutation symmetry of the three hydrogens of the methyl unit, we define the following linear combinations of the $\tau_1,\tau_2$ and $\tau_3$ primitive dihedral angles (Fig.~\ref{fig:methanol})~\cite{Meyer1969JCP_tau,Bell1994JMS_tau,Lauvergnat2014SA_methanol},
\begin{align}
  \tau &=\frac{1}{3}\left(\tau_1+\tau_2+\tau_3\right) \label{eq:tau}, \\ 
  \varphi_1 &=\frac{1}{\sqrt{2}}\left(\tau_2-\tau_3\right) \label{eq:vphi1}, \\ 
  \varphi_2 &=\frac{1}{\sqrt{6}}\left(2 \tau_1-\tau_2-\tau_3\right) \; . \label{eq:vphi2}
\end{align}
The dynamically relevant interval for $\tau$ is $[0,2\pi)$; $\varphi_1$ and $\varphi_2$ are small-amplitude coordinates.
To describe the small-amplitude vibrations, we start with the bond distances and angles of the valence coordinates (Fig.~\ref{fig:methanol}) and add $\varphi_1$ and $\varphi_2$ of Eqs.~\eqref{eq:vphi1}--\eqref{eq:vphi2},
\be
\bos{\xi}=\left(r_1, r_2, r_3, r_4, r_5, \theta_1, \theta_2, \theta_3, \theta_4, \varphi_1, \varphi_2 \right) \; .
\ee
Then, we define the $q_k$ curvilinear normal coordinates by the linear combination of the $D^\text{s}=11$ small-amplitude coordinates as
\be
  \Delta \xi_i=\sum_{k=1}^{D^\text{s}} L_{i k} q_k 
  \quad\text{with}\quad 
  \Delta \xi_i=\xi_i- \xi_i^{\text {ref }}(\tau) \; ,
  \quad i=1,\ldots,D^\text{s}\; .
  \label{eq:defqk}
\ee 
$\bos{\xi}^{\text{ref}}(\tau)$ is the minimum-energy path on the PES for $\tau$. $\bos{L}$ diagonalizes the $\bar{\bos{G}}^\text{s} \bar{\bos{F}}^\text{s}$ matrix, where $\bar{\bos{G}}^\text{s}$ and $\bar{\bos{F}}^\text{s}$ are obtained by averaging the elements of the $D^\text{s}\times D^\text{s}$\  $\bos{G}^\text{s}(\tau_i)$ matrix and the $\bos{F}^\text{s}(\tau_i)$, the second derivative matrix of the potential energy function, over the three minimum structures, $\tau_i = \pi/3, \pi, 5\pi/3$.
Eq.~\ref{eq:defqk} defines the small-amplitude $q_k$ coordinates, and so, the vibrational coordinates used in the vibrational kinetic energy operator, Eq.~\eqref{eq:keo}, are $\bos{\rho}=(q_1,\ldots,q_{D^\text{s}};\tau)$. As a result of this construction, the coupling of the small-amplitude coordinates is small (at the harmonic level), and thus, we can use an efficient basis and grid pruning.
Further details (and possible coordinate choices) are described in Ref.~\citenum{Sunaga2024JCTC}. We note that the $\bos{\xi}^\text{ref}(\tau)$ minimum-energy path and the $\bos{L}$ coefficients were (re-)computed with the PES25 developed in the first part of this work, the associated harmonic frequencies are provided in the Supplementary Material (Table~S6).

%
%
\begin{table*}[!htbp]
\caption{%
  Deviation of the vibrational band origins, \nut\ in \cm, of CH$_3$OH computed with the GENIUSH-Smolyak program with PES13~\cite{Qu2013MP_CH3+OH} and PES25 (this work) from available experimental data.
}\label{tbl:error_vib}
\centering
\begin{tabular}{@{}l r@{--}l SSS S SSS@{}}
\hline
\hline\\[-0.4cm]
%
 \multicolumn{1}{c}{\raisebox{-0.3cm}{Description}} & 
 \multicolumn{2}{c}{Energy range} & 
 \multicolumn{3}{c}{ {\nut\ (PES13)\cite{Sunaga2024JCTC}} } && 
 \multicolumn{3}{c}{{\nut\ (PES25)}}\tabularnewline\\[-0.60cm]
\cline{4-6} \cline{8-10}\\[-0.4cm]
 & \multicolumn{2}{c}{[\cm]} & {MAE} & {RMSE} & {MAX} && {MAE} & {RMSE} & {MAX}\tabularnewline
\hline\\[-0.4cm]
(a) torsional excitations: &     0 & 1050  & 1.6 & 1.9 & 3.2 && 0.9 & 1.1 & 2.1\tabularnewline
(b) SAM-torsion couplings:  &  1050 & 2000  & 6.9 & 8.4 & 16.1 && 1.5 & 2.0 & 4.0\tabularnewline
(c) combination bands:  &  2000 & 2200  & 9.3 & 11.9 & 21.3 && 1.5 & 1.8 & 3.2\tabularnewline
\hline
\hline
\end{tabular}
\end{table*}

We solve Eq.~\eqref{eq:vibSch} in a variational procedure; harmonic oscillator basis functions are used for the $q_k$ small-amplitude degrees of freedom, and Fourier basis functions for the $\tau$ large-amplitude degree of freedom. A direct product basis representation would be too large for this 12D system, so we exploit that the coupling of the small-amplitude vibrations is small \cite{Daria2022JMS_HCOOH,Avila2023PCCP_HCOOH,Matyus2023CC,Sunaga2024JCTC} and we can truncate the direct-product basis. For the computation of the Hamiltonian matrix, we use the Smolyak non-product grid to numerically compute integrals for the small-amplitude coordinates  \cite{Avila2009JCP,Avila2011JCP,Avila2019JCP_methodology,Avila2019JCP_CH4F-,Avila2020PCCP,Avila2023PCCP_HCOOH}.
The $\bos{g}$ matrix and all further kinetic energy coefficients are computed over the multi-dimensional grid of the vibrational coordinates ($q_k$ and $\tau$) and the $t$-vector formalism is used as described in Refs.~\citenum{Matyus2009JCP,Matyus2023CC} and \citenum{Avila2019JCP_methodology}.
Regarding the number of basis functions and grid points, extensive convergence tests on PES13 have been reported in Ref.~\citenum{Sunaga2024JCTC}.
In this work, the following basis and grid parameters are used in the GENIUSH-Smolyak computation (the definition of these computational parameters can be found in Refs.~\citenum{Avila2019JCP_methodology,Avila2019JCP_CH4F-,Daria2022JMS_HCOOH,Avila2023PCCP_HCOOH,Sunaga2024JCTC}): $n_\tau=32$ and $M_\tau=54$ for the LAM basis and grid size; and $b=7$ and $H=21$ for the SAM basis truncation and grid pruning which converge the vibrational energies within 0.5~\cm\ up to 2000~\cm\ from the ZPVE and within 1.5~\cm\ up to 2200~\cm.

%
%
\subsection{Vibrational energies}
First of all, we compare the vibrational energies computed with the newly developed PES25, and with the previously available PES13~\cite{Qu2013MP_CH3+OH}.
For the comparison, we match the vibrational states based on their assignment, and this assignment also helps us compare with data from (gas-phase) experiment. The vibrational energies and assignments with PES13 are taken from Ref.~\citenum{Sunaga2024JCTC}, and the new, detailed assignments with PES25 are collected in Tables~S7-S9 of the Supplementary Material.
The mean absolute error (MAE), root mean square error (RMSE), and maximum error (MAX) resulting from this analysis are shown in Table~\ref{tbl:error_vib}. 

The comparison, in Table~\ref{tbl:error_vib}, is shown over three energy ranges defined according to the vibrational excitation types:
(a) torsional states, 0--1050~\cm;
(b) vibration-torsion coupling states, 1050--1600~\cm; and 
(c) combination bands, 2000--2200~\cm. 
To prepare Table~\ref{tbl:error_vib}, only gas-phase experimental data were used. There are further, matrix-isolation vibrational band origins in neon and nitrogen matrices available in the literature. They are included in Table~S9 of the Supplementary Material, and the agreement is mostly good with our variational results. However, direct, quantitative comparison (as for gas-phase data) is hindered by unpredictable matrix effects, which may be larger than the inaccuracy of our computations.

We also report that while the computed and experimental vibrational band origins (VBOs) are within a few \cm\ with PES13 or PES25 for the lowest-energy (a) range; the RMSE (MAX) of the VBOs increases to 12~(21)~\cm\ for PES13 for the combination band range (c). At the same time, for the newly developed PES25, the RMSE (MAX) of the VBOs remains small, and even for the high-energy (c) range, it is only 1.8~(3.2)~\cm, far lower than the RMS fitting errors of the fitting set of PES25.

The computed VBOs using PES25 and the deviation from the available experiments are listed in Table~\ref{tbl:vib1_modified}.

\begin{figure}[!htbp]
    \centering
    \includegraphics[width=0.60\linewidth]{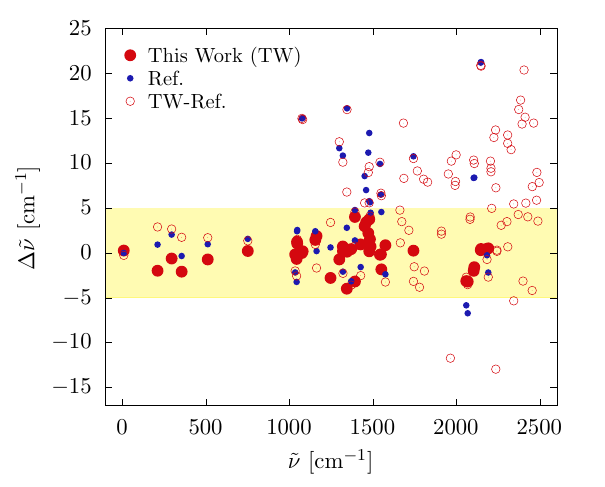}
    \caption{%
    Variational vibrational band origins computed in this work, with the newly developed PES25, at the CCSD(T)-F12b/cc-pVTZ-F12 level of electronic structure theory and the GENIUSH-Smolyak program, reproduce experimental data with $\Delta\tilde\nu$ difference less than $\pm 5$~\cm. They significantly improve upon earlier results (Ref.) computed with PES13\cite{Qu2013MP_CH3+OH} and reported in Ref.~\citenum{Sunaga2024JCTC}.
    The vibrational band origins are referenced to the ZPVE of each computation. The ZPVEs with the PES13 and the PES25 are 11108.0~\cm\cite{Sunaga2024JCTC} and 11119.6~\cm\ (this work).
    }
    \label{fig:13-25}
\end{figure}

There are several VBOs, for which there is currently no (gas-phase) experimental value available. For an overall assessment, we compare the VBOs of the PES13 and PES25 in Figure~\ref{fig:13-25}, which reveals large, $>10$~\cm, deviations beyond ca.~1500~\cm.
The figure also shows the deviations from the gas-phase experimental data summarised in Table~\ref{tbl:error_vib}. 

These observations suggest that PES25 represents a major improvement over earlier PES representations and provides a (near) spectroscopic quality description not only for the torsional excitation range but also for the vibration-torsion coupling and combination bands.

At this point, it is appropriate to consider the major sources of remaining errors due to approximations and missing physical corrections. This short analysis may guide future improvements and set expected uncertainties for numerical applications.
First of all, the current PES relies on frozen-core electronic energies computed at the CCSD(T) level, and it has been known that the correlation effects of core electrons largely cancel the high-order correlation effects beyond CCSD(T) \cite{Cortez2007JCTC_AE_HO_cancel,Tew2007JPCA_AE_HO_cancel}. Any future improvement will require improvement of both aspects. 
At the same time, the core-electron correlation effects have been reported to be important for accurate rotational constants (affecting them at the level of hundreds of MHz \cite{Puzzarini2008JCP_AE,Das2024IJQC_AE}). Hence, corrections will be considered for an accurate description of the rotational-vibrational spectrum.

Furthermore, the present work uses a triple-zeta electronic basis set. It was noted in Ref.~\citenum{Rauhut2009JCP_QZ}  that the harmonic oscillator wave numbers' (finite basis set) error can be reduced by ca. 50~\% upon enlarging the electronic basis set to quadruple-zeta quality [from 1.7~\cm~(cc-pVTZ-F12) to 0.8~\cm~(cc-pVQZ-F12) root mean square error in that particular study]. A comparable improvement is anticipated for a fully-coupled anharmonic treatment.

An additional source of error in the vibrational energies originates from the fitting error of the PES, which was already discussed in detail in Sec.~\ref{sec:PES} (and in the Supplementary Material). 

Further small errors are due to missing finite-nuclear mass corrections, \emph{i.e.,} the diagonal Born-Oppenheimer correction and the (less known) effective rotational-vibrational masses (in the present work, we used atomic masses). Relativistic and quantum-electrodynamical effects will also be relevant when the electronic structure representation is substantially improved. 
These small physical effects are often neglected in a standard electronic structure computation, but have been of interest in the spectroscopy community~\cite{Pyykko2001PRA_H2O_qed,Polyansky2003Science_water,Temelso2004JPCA_REL_DBOC}.
Rigorous information on their contributions is available from recent studies of few-electron diatomic molecules~\cite{PiJe09,SiPaKoPu23,PaKo24,FeKoMa20,MaRaJeMa25}.

%
%

%
%
\begin{table*}
\caption{
 Variational vibrational energies, $\tilde{\nu}$ in \cm, referenced to the zero-point vibrational energy (ZPVE, 11119.6~\cm) of CH$_3$OH in 12D computed with the GENIUSH-Smolyak program ($b=7$ basis set) and PES25 developed in the present work. 
}\label{tbl:vib1_modified}
\centering
\scalebox{0.85}{%
\begin{tabular}{@{}lllcr@{}r| lllcr@{}r| lllcr@{}}
\hline\hline\\[-0.4cm]
\# & $\nu_{\tau}$$^\text{a}$ & SAMs$^\text{b}$ & $\Gamma$$^\text{c}$ & \multicolumn{1}{c}{$\tilde{\nu}$} & $\delta$$^\text{d}$ & %
\# & $\nu_{\tau}$$^\text{a}$ & SAMs$^\text{b}$ & $\Gamma$$^\text{c}$ & \multicolumn{1}{c}{$\tilde{\nu}$} & $\delta$$^\text{d}$ & %
\# & $\nu_{\tau}$$^\text{a}$ & SAMs$^\text{b}$ & $\Gamma$$^\text{c}$ & \multicolumn{1}{c}{$\tilde{\nu}$} \\
\hline\\[-0.4cm]		     	    
1        & 0   & 0   & $A_1$             & 0         &             &		     51-52 & 2   & $8_1$              & $E$ & 1549.9   & [--1.8] &			         103-104 & 3   & $5_1$                  & $E$ & 2203.0            \\       	    
2-3      & 0   & 0   & $E$               & 8.9             & [0.3]       &		     53-54 & 1   & $6_1$              & $E$ & 1574.5   &  [0.9]      &			         105     & 3   & $4_1,10_1$             & $A_2$ & 2205.8            \\     	    
4-5      & 1   & 0   & $E$               & 210.9             & [--2.0]     &		     55-56 & 1,2 & $5_1,7_1,11_1$     & $E$ & 1661.1   &         &			     	 106     & 3   & $4_1,10_1$             & $A_1$ & 2206.2            \\        	    
6        & 1   & 0   & $A_2$             & 295.1             & [--0.6]     &		     57-58 & 1   & $5_1$              & $E$ & 1663.9   &         &			     	 107-108 & 3,4 & $7_1,11_1$             & $E$ & 2210.4            \\          	    
7        & 2   & 0   & $A_1$             & 355.3             & [--2.1]     &		     59    & 1   & $4_1,10_1$         & $A_2$ & 1672.2   &         &			     	 109-110 & 0   & $7_2,11_2$             & $E$ & 2223.8            \\          
8-9      & 2   & 0   & $E$               & 511.0 & [--0.7]     &   60    & 1   & $4_1,10_1$         & $A_1$ & 1682.5   &         &			                             	 111     & 0,1 & mix & $A_2$ & 2233.7            \\      		           
10-11    & 3   & 0   & $E$               & 750.8             & [0.2]       &		     61-62 & 1   & $4_1,10_1$         & $E$ & 1685.0   &         &			     	 112-113 & 3,4 & $7_1,11_1$             & $E$ & 2234.6            \\          	    
12       & 0   & $8_1$ & $A_1$           & 1034.5            & [--0.2]     &		     63    & 1,2 & $6_1,7_1,11_1$     & $A_2$ & 1714.7   &         &			     	 114-115 & 3   & $4_1,10_1$             & $E$ & 2235.1            \\    
13-14    & 0   & $8_1$ & $E$             & 1043.3            & [--0.7]     &		     64    & 2   & $6_1,7_1,11_1$     & $A_1$ & 1742.2   &         &			     	 116     & 5   & 0                      & $A_2$ & 2241.4            \\        
15       & 3   & 0   & $A_2$             & 1045.4            & [1.2]       &		     65-66 & 2   & $6_1$              & $E$ & 1742.4 &  [0.3]      &                             	 117     & 6   & 0                      & $A_1$ & 2241.5            \\        	    
16       & 4   & 0   & $A_1$             & 1046.5            & [1.1]       &		     67    & 1   & $5_1$              & $A_2$ & 1746.6   &         &			     	 118-119 & 1   & $8_2$                  & $E$ & 2267.5           \\          
17       & 0   & $6_1,7_1,11_1$ & $A_1$   & 1074.7            & [0.0]       &		     68-69 & 1   & $4_1,10_1$         & $E$ & 1765.1 &         &                             	 120     & 0   & mix & $A_1$ & 2301.6            \\      	    	           	    
18-19    & 0   & $6_1,7_1,11_1$ & $E$     & 1079.1            & [0.2]       &		     70-71 & 3   & $8_1$              & $E$ & 1778.5   &         &			     	 121     & 4   & $6_1$                  & $A_1$ & 2306.2            \\        	    
20-21    & 0   & $7_1,11_1$ & $E$         & 1155.0            & [1.5]       &		     72-73 & 5   & 0                  & $E$ & 1803.0   &         &			     	 122     & 3   & $6_1$                  & $A_2$ & 2306.2            \\  
22       & 0   & $7_1,11_1$ & $A_2$       & 1162.1            & [1.9]       &		     74    & 2   & $5_1$              & $A_1$ & 1807.7   &         &			     	 123-124 & 0   & mix         & $E$ & 2306.3            \\	       	           
23-24    & 1   & $8_1$ & $E$             & 1245.8            & [--2.8]          &		     75-76 & 2   & $4_1,10_1$         & $E$ & 1827.1   &         &			     	 125-126 & 0   & mix         & $E$ & 2326.3            \\	       	           	    
25-26    & 0   & $6_1$ & $E$             & 1298.2            &  [--0.7]          &		     77    & 3   & $7_1,11_1$         & $A_1$ & 1909.7   &         &			     	 127     & 1   & $8_2$                  & $A_2$ & 2342.2            \\  
27       & 0,1 & $6_1,7_1,11_1$ & $A_1$   & 1319.9            &  [0.7]          &		     78    & 3   & $7_1,11_1$         & $A_2$ & 1909.8   &         &			     	 128     & 0   & mix         & $A_1$ & 2342.3            \\	     
28       & 1   & $7_1,8_1,11_1$ & $A_2$   & 1320.4            & [0.2]       &		     79-80 & 2,3 & $6_1,7_1,11_1$     & $E$ & 1950.9   &         &			     	 129-130 & 1   & mix         & $E$ & 2368.8            \\	       	           	    
29-30    & 1   & $6_1$ & $E$             & 1343.5            & [--4.0]     &		     81-82 & 2   & $5_1$              & $E$ & 1963.1   &         &			       	 131     & 1   & $8_2$                  & $A_2$ & 2372.0            \\     
31       & 1   & $8_1$ & $A_2$           & 1344.9            & [0.1]       &		     83-84 & 2   & $4_1,10_1$         & $E$ & 1969.1   &         &			     	 132     & 0   & $6_2$                  & $A_1$ & 2383.6            \\  
32       & 0,2 & $6_1,7_1,11_1$ & $A_1$   & 1369.3            & [0.4]       &		     85    & 2   & $4_1,10_1$         & $A_2$ & 1991.2   &         &			     	 133-134 & 0,1 & $6_2,7_2,11_2$         & $E$ & 2392.3            \\          
33       & 2   & $8_1$ & $A_1$           & 1392.1            & [--3.2]     &		     86    & 2   & $4_1,10_1$         & $A_1$ & 1993.8   &         &			     	 135     & 0   & mix & $A_1$ & 2397.9            \\      		           
34-35    & 4   & 0   & $E$               & 1392.2            & [4.0]       &		     87-88 & 3   & $6_1$              & $E$ & 1997.5   &         &			     	 136     & 0   & mix & $A_2$ & 2403.9            \\       	    	           	    
36-37    & 1   & $7_1,11_1$ & $E$         & 1426.2            & [1.0]    &		     89    & 0   & $8_2$              & $A_1$ & 2058.1   & [--3.1] &			     	 137-138 & 0,1 & $6_2,7_2,11_2$         & $E$ & 2410.4            \\    
38       & 0   & $5_1$ & $A_1$             & 1450.3            & [3.0]       &		     90-91 & 0   & $8_2$              & $E$ & 2066.8   & [--3.2] &			     	 139     & 2   & $8_2$                  & $A_1$ & 2414.9            \\  
39-40    & 0   & $5_1$ & $E$               & 1458.7            & [3.4]       &		     92    & 3   & $8_1$              & $A_2$ & 2080.9   &         &			     	 140-141 & 4   & $8_1$                  & $E$ & 2426.6            \\          
41-42    & 0   & $4_1,10_1$ & $E$          & 1471.7            & [2.2]       &		     93    & 4   & $8_1$              & $A_1$ & 2081.9   &         &			     	 142-143 & 1   & mix & $E$ & 2452.6            \\	       	    	           
43-44    & 2   & $7_1,11_1$ & $E$          & 1477.4            &             &		     94    & 0   & mix & $A_1$ & 2102.2   & [--2.0] &	      		                     	 144     & 1   & mix         & $A_1$ & 2453.4            \\       	           
45       & 0   & $4_1,10_1$ & $A_2$        & 1477.8            & [3.7]       &		     95-96 & 0   & mix & $E$ & 2106.4   & [--1.6] &		                             	 145-146 & 0   & mix & $E$ & 2461.5            \\	          	     
46-47    & 0   & $4_1,10_1$ & $E$          & 1481.8            & [1.5]       &		     97    & 0   & mix & $A_1$ & 2144.8   & [0.3]   &		                             	 147     & 0   & $5_1+8_1$             & $A_1$ & 2478.9            \\   
48       & 0   & $4_1,10_1$ & $A_1$        & 1485.4            & [0.7]       &		     98-99 & 0   & mix & $E$ & 2146.9   & [0.4]   &                                          	 148     & 0,1 & mix & $A_2$ & 2481.0            \\      		     
49       & 1   & $6_1$ & $A_2$           & 1541.7            & [--0.2]           &		100-101 & 0   & mix & $E$ & 2182.5   & [0.5] 	&		                     	 149     & 0   & $5_1+8_1$             & $E$ & 2487.2            \\           
50       & 2   & $6_1$ & $A_1$           & 1547.4            &  [--0.1]          &		102     & 0   & mix & $A_2$ & 2190.1   & [0.5] 	&			     	 150     & 0,2 & mix & $E$ & 2494.3            \\                       
\hline\hline								           
\end{tabular}		
} 
\begin{flushleft}
{\footnotesize%
  $^\text{a}$~Assignment of the torsional excitation similarly to Ref.~\citenum{Sunaga2024JCTC}. \\
  $^\text{b}$~Assignment in terms of the (small-amplitude, SAM) curvilinear normal modes $1_n, 2_n, ..., 11_n$ $(n=0,1,\ldots)$. Zero excitation ($n=0$) is noted as `0'. The label `Mix' stands for strongly mixed states. \\
  $^\text{c}$~$\Gamma$: $C_{3\text{v}}(\text{M})$ irreducible representation label. Further details are described in Sec.~S1 of the Supporting Information of Ref.~\citenum{Sunaga2024JCTC}. \\
  $^\text{d}$~%
    Deviation from the experimental vibrational band origin~\cite{Serrallach1974JMS,Moruzzi1995_ch3oh,Lees2002PRA,Fehrensen2003JCP,Temsamani2003JMS,Lees2004JMS}, $\delta=\tilde{\nu}_{\rm{exp}} -\tilde{\nu}$ in \cm.   
    Gas-phase experimental data were not available to us for the states \#103--150. Matrix-isolation data are cited in Table~S9 of the Supplementary Material.
}
\end{flushleft}

\end{table*}

\clearpage

\section{Summary, conclusion, and outlook}
We have reported the development of an \textit{ab initio} potential energy surface and variational vibrational states for the methanol molecule. 

The electronic structure computations were carried out at the CCSD(T)-F12b/cc-pVTZ-F12 level of theory, the dataset for the PES was generated with the {\scshape Robosurfer} code, and the PES was fitted using the permutationally invariant polynomials (PIP) method. Properties of the PES, including molecular structure parameters, harmonic frequencies, and one-dimensional slices, were thoroughly investigated, and all details reproduce the \textit{ab initio} values well.

The vibrational states of methanol were computed using this newly developed PES and the GENIUSH-Smolyak approach up to the coupling between the torsion and the overtone of the 
C-O stretching vibration (ca. 2500~\cm\ beyond the zero-point vibrational energy). 
The resulting vibrational band origins improve upon, by 10-20~\cm, earlier vibrational results (computed on PES13) and are in good agreement ($\pm 5$~\cm) with available gas-phase experimental data. In the variational vibrational computations, we have confirmed that the vibrational energies are converged within 1-2~\cm\ with respect to the vibrational basis for these ranges.

As a result, we describe the quantum dynamics of methanol here using a fully \emph{ab initio} approach, confirmed with experiment from the torsional excitation range, through the vibration-torsion coupling, up to the start of the combination bands range at around 2500~\cm\ beyond the zero-point vibrational energy. This result opens the door for future rovibrational computations, for which we plan to implement a path-following, body-fixed frame approach. Furthermore, the development of electric dipole and polarizability surfaces is in progress. This will allow us a fully \textit{ab initio} computation and line-by-line study of the high-resolution infrared and Raman spectra of methanol.
All in all, the present work is an important step towards simulating the `complete' rovibrational spectrum of methanol (and isotopologues) for further applications in astrochemistry, astrophysics, and fundamental physics.

%
%
\begin{acknowledgments}
AS thanks the European Union’s Horizon 2022 research and innovation programme for funding under the Marie Skłodowska-Curie Grant Agreement No. 101105452 (QDMAP).
AS and EM thank the financial support of the Hungarian National Research, Development, and Innovation Office (FK 142869).
TG and GC thank the National Research, Development and Innovation Office-NKFIH, K-146759; project no. TKP2021-NVA-19, provided by the Ministry of Culture and Innovation of Hungary from the National Research, Development and Innovation Fund, financed under the TKP2021-NVA funding scheme; and the Momentum (Lendület) Program of the Hungarian Academy of Sciences for financial support.
We acknowledge KIFÜ (Governmental Agency for IT Development, Hungary) for awarding us access to the Komondor HPC facility based in Hungary.
We thank Viktor Tajti (Szeged) for the extension of the QCT MD code to compute unimolecular trajectories.
\end{acknowledgments}

%

\clearpage
\begin{center}
{\large
\textbf{Supplemental Material}
}\\[0.25cm]
{\large
\textbf{Exact quantum dynamics of methanol: full-dimensional \emph{ab initio} potential energy surface of spectroscopic quality and variational vibrational states}
} \\[0.5cm]

Ayaki Sunaga,$^1$ Tibor Gy\H{o}ri,$^{2,\ast}$ Gábor Czakó,$^{2,\dag}$ Edit Mátyus$^{1,\dagger}$ \\
\emph{$^1$~ELTE, E\"otv\"os Lor\'and University, Institute of Chemistry, P\'azm\'any P\'eter s\'et\'any 1/A 1117 Budapest, Hungary} \\
\emph{$^2$~MTA-SZTE Lendület ``Momentum'' Computational Reaction Dynamics Research Group, Interdisciplinary Excellence Centre and Department of Physical Chemistry and Materials Science, Institute of Chemistry, University of Szeged, Rerrich Béla tér 1, Szeged H-6720, Hungary} \\

$^\ast$tibor.gyori@chem.u-szeged.hu\quad
$^\dag$gczako@chem.u-szeged.hu\quad
$^\dagger$edit.matyus@ttk.elte.hu 

~\\[0.15cm]
(Dated: 25 July 2025)
\end{center}

\setcounter{section}{0}
\renewcommand{\thesection}{S\arabic{section}}
\setcounter{subsection}{0}
\renewcommand{\thesubsection}{S\arabic{section}.\arabic{subsection}}

\setcounter{equation}{0}
\renewcommand{\theequation}{S\arabic{equation}}

\setcounter{table}{0}
\renewcommand{\thetable}{S\arabic{table}}

\setcounter{figure}{0}
\renewcommand{\thefigure}{S\arabic{figure}}

%
%
\section{Construction of the potential energy surface}
\subsection{\textit{Ab initio} methodology}\label{subsec:molpro}
All \textit{ab initio} electronic structure computations were carried out using the {\scshape Molpro}  program package (version 2023.2) \cite{Lindh1991JCP_seward,MOLPRO2012,MOLPRO_Werner2020JChemPhys,MOLPRO2023}, with the CCSD(T)-F12b method~\cite{Adler2007JCP_CCSDT_F12} and the cc-pVTZ-F12 basis set~\cite{Peterson2008JCP_cc-pVnZ-F12}. All auxiliary basis set options were left at their defaults, resulting in the use of the aug-cc-pVTZ/JKFit~\cite{Weigend2002PCCP_jkfit,aug-cc-pVnZ-JKfit_note}, cc-pVTZ-F12/OptRI~\cite{Yousaf2008JCP_OptRI} and aug-cc-pVTZ/MP2Fit~\cite{Weigend2002JCP_MP2Fit} auxiliary basis sets. Core electrons in the 1s orbitals of C and O were frozen. Optimized \textit{ab initio} geometries were also obtained with {\scshape Molpro}~\cite{Eckert1997JCC_MOLPRO_opt}, at one of the three equivalent minima and torsional saddle points (both have \Cs\ symmetry). {\scshape Molpro}'s default integral handling and convergence thresholds are a compromise between execution speed and accuracy, and not optimal for the present work. For this reason, we employed the tighter thresholds summarized in Table~\ref{tbl:threshold}. An even tighter set of thresholds was employed for the \textit{ab initio} geometry optimizations and harmonic frequency computations, and the convergence thresholds of the geometry optimization were also tightened. The required \textit{ab initio} gradients and Hessians were both obtained via numerical differentiation, and the gradients used a four-point scheme. The {\scshape Molpro} output files of the \textit{ab initio} geometry optimizations and frequency computations can be found in the Supplementary Material (harm\_freq.tar.lz).

%
%
\begin{table}[htbp!]
\caption{%
Global thresholds employed for the \textit{ab initio} computations: PES fitting set (single-point computations) and harmonic frequencies (and the associated geometry optimizations). \textit{oneint} and \textit{twoint} refer to the one-electron and two-electron integration screening parameters, while \textit{compress} refers to a threshold for {\scshape Molpro}'s integral compression algorithm. \textit{energy} is a threshold used by {\scshape Molpro} to set multiple other thresholds, such as convergence thresholds for Hartree-Fock and post-Hartree-Fock iterations.
}\label{tbl:threshold}
\begin{tabular}{l cc}
\hline\hline
 & PES & optimization and frequency\tabularnewline
 \hline
oneint & $1.0\times10^{-14}$ & $1.0\times10^{-20}$  \tabularnewline
twoint & $1.0\times10^{-14}$ & $1.0\times10^{-20}$ \tabularnewline
compress & $1.0\times10^{-14}$ & $1.0\times10^{-20}$ \tabularnewline
energy & $1.0\times10^{-10}$ & $1.0\times10^{-12}$ \tabularnewline
\hline\hline
\end{tabular}
\end{table}

%
%
\subsection{{\normalsize R\scriptsize OBOSURFER} and PES fitting methodology}\label{subsec:robosurfer_background}

\subsubsection{Brief description of the operation of Robosurfer}\label{subsubsec:robosurfer_bg}
The PES development was performed using a development version of the {\scshape Robosurfer} program system~\cite{Gyori2020JCTC}. Without going into the details of the operation of {\scshape Robosurfer}, it automates the development of PESs through an iterative, active-learning-like process, which gradually expands the fitting set with additional samples (geometry-energy pairs) until the quality of the fitted PES is judged to be satisfactory.
\par
We note that besides the fitting set, {\scshape Robosurfer} also maintains a secondary set of geometries we call ``spares". These are geometries for which the \textit{ab initio} energy has been computed, but are yet to be moved into the fitting set. The fitting errors of spare geometries are examined by {\scshape Robosurfer} every iteration, and any geometry with a weighted (Eq.~\ref{eq:weight})  fitting error greater than the configured target accuracy ($E_\text{targ}$) is moved into the fitting set, while geometries already well-described by the PES are left in the spares set.
\par
For this reason, adding geometries to the spares set is the preferred method of expanding the fitting set, as it reduces the risk of accidental imbalanced sampling and the needless inflation of the fitting set. The accumulation of low-error points in the spares set is useful not only for avoiding repeated \textit{ab initio} computations in regions of low fitting error, but some of them can end up moved into the fitting set in subsequent iterations if their weighted fitting error ever exceeds $E_\text{targ}$.  The brief outline of a single {\scshape Robosurfer} iteration is as follows:
\begin{enumerate}
\item The PES is fitted based on the geometries in the fitting set, and large numbers of new geometries are rapidly generated using the PES, e.g. by running quasi-classical trajectories (QCT) and geometry optimizations.
\item A set of heuristics is used to estimate which of the new geometries are the most likely to have large fitting errors on the current PES. The most important heuristics are the geometrical and energetic distances of new geometries from the union of the fitting and spare sets. These heuristics select new geometries that are sufficiently different from points already in the fitting set or spares.
\item \textit{Ab initio} computations are performed at the selected geometries and the results (geometries with energies) are appended to the spares set.
\item The fitting error (deviation between the \textit{ab initio} energy and the value of the fitted PES) is evaluated for every geometry in the spares set. If the geometry with the highest weighted fitting error has an error larger than the configured error target, it is moved to the fitting set, a new PES is fitted, and the errors are reevaluated. These ``microiterations" are continued until the peak weighted fitting error of the spares drops below the targeted fitting error $E_\text{targ}$.
\end{enumerate}

\subsubsection{General {\normalsize R\scriptsize OBOSURFER} configuration and fitting methodology}\label{subsubsec:robosurfer_config}
The PES is fitted using the Braams--Bowman implementation of permutationally invariant polynomials~\cite{Braams2009IRPC_pip}. Briefly, the polynomials are expanded in terms of Morse variables, $y_{ij}=\exp(-r_{ij}/a)$, where $r_{ij}$ is the distance between atoms $i$ and $j$ and $a$ is a distance transformation parameter influencing the diffuseness of the polynomial basis.
The expansions are truncated by limiting the degrees of the polynomials, as per standard practice. Naturally, this main polynomial expansion includes monomials that may depend on the positions of all $N$ atoms; for this reason, we refer to these as the $N$-body monomials. The energy expression of the PES is
\begin{equation}\label{eq:PES_energy}
E=\sum_{k=1}^{Z}c_kT_kM_k,
\end{equation}
where $Z$ is the number of permutationally invariant monomials, while $c_k$ and $T_k$ are the coefficient and monomial prefactor of the $k$-th monomial $M_k$, respectively.
\par
The flexibility of the fitting function can be further enhanced by the addition of what we call `extra monomials' to the fitting basis.
In this work, we use extra monomials obtained with the same polynomial expansion method, but with different distance transformation parameters, maximum degrees, monomial prefactor functions, and a limit on how many atoms they can involve.
We group these extra monomials based on the maximum number of atomic positions their expansion may depend on, resulting in 2-body, 3-body, ..., $q$-body, ...,  $N-1$-body extra monomials.
The prefactors are usually given by
\be\label{eq:extra_t_new}
T_k(\mathbf R_k, \omega_q)=
\begin{cases}
\begin{aligned}
\left(1-\frac{D(\mathbf R_k)}{\omega_q}\right)^{3}\end{aligned}&, \text { if } D(\mathbf R_k) \leq \omega_q 
\\
\begin{aligned}
0\end{aligned}&, \text { if } D(\mathbf R_k) >\omega_q
\end{cases},
\ee
where $\mathbf R_k$ is the subset $r_{ij}$ of interatomic distances the $k$-th monomial depends on, $\omega_q$ is a ``range separation" parameter which can be used to smoothly switch off the extra $q$-body monomials at large $r_{ij}$, and $D(\mathbf R_k)$ is the root mean square (RMS) of the $r_{ij}$ values the $k$-th monomial depends on. There is one possible special case in prefactor calculation: 2-body extra monomials can be configured to have their prefactors multiplied with $1/r_{ij}$, effectively conferring a hyperbolic shape to these basis functions of the fit, which may improve the description of strong electrostatic interactions such as the repulsive wall at short $r_{ij}$.
\par
In our experience, for bimolecular reaction PESs it can be beneficial to optimize $\omega_q$ for each set of $q$-body extra monomials separately, however this may result in small unphysical artifacts on the fitted PES near $\omega_q$. For this reason (and since, unlike PES13, we are not aiming to describe the dissociation of methanol), we use $\omega_q=1000$ bohr in this work for all $q$. The $N$-body monomials are given $T_k$ prefactors numerically equivalent to 1 by using $\omega_N=10^{36}$ bohr for them.
\par
The polynomial coefficients are obtained through weighted linear least squares (LLS) fitting. The weights are computed with a three-segment piecewise function consisting of a constant 1.0 section at low energies, followed by a linear slope intended for reducing the weight of fitting points in somewhat less important regions of the PES, finally ending with a double-hyperbolic section intended to rapidly discount the importance of geometries beyond the maximum energy considered important for the purpose of the PES. The formula for the weight is 
\begin{equation}\label{eq:weight}
\scalebox{0.95}{$
w(E)=
\begin{cases}
\begin{aligned}
w_{\text{end}}\,\frac{E_{\text{dwt},0}}{E+E_{\text{dwt},0}-E_{\text{slopeEnd}}} \frac{E_{\text{dwt},1}}{E+E_{\text{dwt},1}-E_{\text{slopeEnd}}}
\end{aligned}
&,\enspace \text{if } E>E_{\text{slopeEnd}}\\[1ex]
\begin{aligned}
1+\frac{w_{\text{end}}-1}{E_{\text{slopeEnd}}-E_{\text{slopeStart}}}\left(E-E_{\text{slopeStart}}\right)
\end{aligned}
&,\enspace \text{if } E_{\text{slopeStart}}<E\leq E_{\text{slopeEnd}}\\[1ex]
1
&,\enspace \text{if } E\leq E_{\text{slopeStart}}
\end{cases},
$}
\end{equation}
where $E$ is the \textit{ab initio} energy of the geometry, $E_{\text {dwt},0}=E_{\text {dwt},1}=0.1$ hartree (21 947 \cm), $E_{\text {slopeStart }}=37.2$ kcal/mol (13 011 \cm), $E_{\text {slopeEnd }}=45.7$ kcal/mol (15 984 \cm), and $w_{\text {end}}=0.75$. $E$, $E_{\text{slopeStart}}$ and $E_{\text{slopeEnd}}$ are relative to the energy at the equilibrium structure. This weight function was also used for weighting the fitting errors of spare geometries when considering if they should be moved into the fitting set. We set the hard upper limit for the energy of geometries that may be included in the fit to 130 kcal/mol (45 468 \cm) and the semihard upper limit\footnote{
Geometries with an \textit{ab initio} energy higher than the semihard limit are not rejected outright, but they are only considered for moving into the fitting set if they are required for fixing deep unphysical minima (holes) on the fitted PES. This is detected based on the energy of the PES at the geometry in question, if the PES returns an energy lower than $E_{\text{slopeStart}}$ a hole is assumed.
} to 70 kcal/mol (24 483 \cm). For solving the LLS equations we use the DGELS and DGELSY LAPACK subroutines, the RCOND condition number threshold of DGELSY is set to $10^{-11}$. Over the course of development, we have used both the Intel MKL\cite{mkl_url} and OpenBLAS\cite{6413635, 10.1145/2503210.2503219, openblas_url} linear algebra implementations.
\par
The {\scshape Robosurfer} iterations used two geometry sources: (I) QCT initialized both in the vibrational ground state and with any one of the harmonic vibrational modes excited by one quantum, and (II) geometry optimizations ran on the preliminary fitted PESs. The starting points of the geometry optimizations were randomly chosen from the subset of the fitting set that had an \textit{ab initio} energy $<E_{\text{slopeEnd}}$.
We chose not to employ the {\scshape Holebuster} subprogram for the development of this PES, as it has a tendency to strongly favor the global exploration of configuration space and there is a tradeoff between global fitting accuracy and fitting accuracy in our region of interest near the equilibrium structure.

\subsection{PES development with {\normalsize R\scriptsize OBOSURFER}}\label{subsec:robosurfer}
\subsubsection{Preparation of the initial fitting set}\label{subsubsec:rs_init}
{\scshape Robosurfer} requires an initial fitting set to begin PES development, this was generated using a version of the geometry set generation code first described in Ref.~\citenum{Gyori2022JCP_manyHF}. As this program was originally intended for generating datasets of two approaching fragments, we have used a dummy atom during generation, which was stripped from the resulting geometries before the \textit{ab initio} computations. The \textit{ab initio} optimized minimum geometry was used as the reference geometry.
The center of mass distance between the dummy center with a mass of 9 amu and methanol was sampled with 0.01 \AA\ steps between $[0.998,9.608]$ \AA. At each distance, 20 000 random orientations were generated for the methanol and finally the Cartesian coordinates of each geometry were perturbed with random numbers uniformly distributed in $[-0.05,+0.05)$ \AA, resulting in a total of 17 240 000 samples before pruning. The pruning step was configured to discard geometries closer than 0.015 to each other in terms of the EW-RMSD ($a=2$ \AA) distance metric~\cite{Gyori2020JCTC}, yielding 18 346 initial geometries.
\par
After dummy atom removal and the computation of \textit{ab initio} energies for these geometries 530 of them were randomly chosen to be included in the first fitting set, while the initial spares set consisted of all 18346 geometries.

\subsubsection{Process of PES development I: initial phase}\label{subsubsec:robosurfer_usage_initial}
Over the course of PES development we have taken the approach of starting with a small fitting set, low degree polynomials, and a high fitting error target, then gradually increasing both the flexibility of the fitting function and the accuracy we demand from the PES, as the fitting set grew.
We have also followed a heuristic of not using fewer fitting points than 1.5 times the number of polynomial coefficients. We note that the PES is fitted to the total \textit{ab initio} energies, not the relative energies of the fitting set.
As a consequence, the RMS errors reported in this subsection are computed from the ``absolute" energies returned by the PES, and are therefore inclusive of the errors caused by the shift in the absolute energy of the minimum geometry. Except where otherwise indicated, PESs were fitted using the DGELS LLS solver.
Since geometries are always appended to the end of the fitting set, the ordering of them is stable, which lets us use their position in the file as an identifier. For example \#200 would be the 201\textsuperscript{st} geometry in the fitting set file, since we are counting from zero. This notation is used below to refer to fitting point that are outliers in the fit (fitting error $>5\times$ the RMS fitting error of points in its energy range).
\par
Using the initial geometry sets discussed above, PES development was started with vibrationally ground-state QCT, a primary PI-EW-RMSD geometry selection threshold of 0.013 and a targeted fitting error ($E_\text{targ}$) of 1 kcal/mol (349.8 \cm). Initial fitting parameters were $a=2.1$ bohr, a 4\textsuperscript{th} degree polynomial fit and no extra monomials, resulting in 317 polynomial coefficients. After 8 iterations, the fitting set grew to 952 geometries and {\scshape Robosurfer} was paused.
\par
To hasten the appropriate sampling of PES regions somewhat far from the equilibrium structure, more geometries were generated with the same method that was used for initial geometry generation, except with different parameters: only 10 000 orientations, larger coordinate perturbations in the range of $[-0.1,+0.1)$ \AA, and a larger EW-RMSD pruning threshold of 0.0275, which favored the generation of much more dissimilar geometries. This yielded 5 396 geometries, which were appended to the spares set.
\par
{\scshape Robosurfer} was then resumed and after 195 more iterations 3 201 fitting points were in use, with a RMS weighted error (RMSWE) of 0.98 kcal/mol, while the spares held 42 979 geometries. As the RMSWE of the fitting set was close to the $E_\text{targ}$ in use at the time (1 kcal/mol), we considered the fit saturated and increased the polynomial degree to 5, resulting in 1 053 coefficients and a RMSWE of 0.22 kcal/mol.
\par
{\scshape Robosurfer} was resumed for another 141 iterations, the fitting set grew to 3 358 geometries and the spares to 57 708, while the RMSWE rose to only 0.28 kcal/mol, giving room for reducing $E_\text{targ}$, which was first changed to 0.9 kcal/mol, then over the course of 332 iterations it was gradually lowered to 0.65 kcal/mol, where it was held for 811 iterations.
At this point, the fitting and spare sets contained 4 874 and 118 353 geometries, respectively, while the RMSWE of the fitting set had risen to 0.43 kcal/mol, approaching $E_\text{targ}$ again, prompting us to raise the polynomial degree to 6, yielding 3 250 coefficients.
\par
\begin{figure}[!htbp]
    \centering
\includegraphics[width=0.4\linewidth]{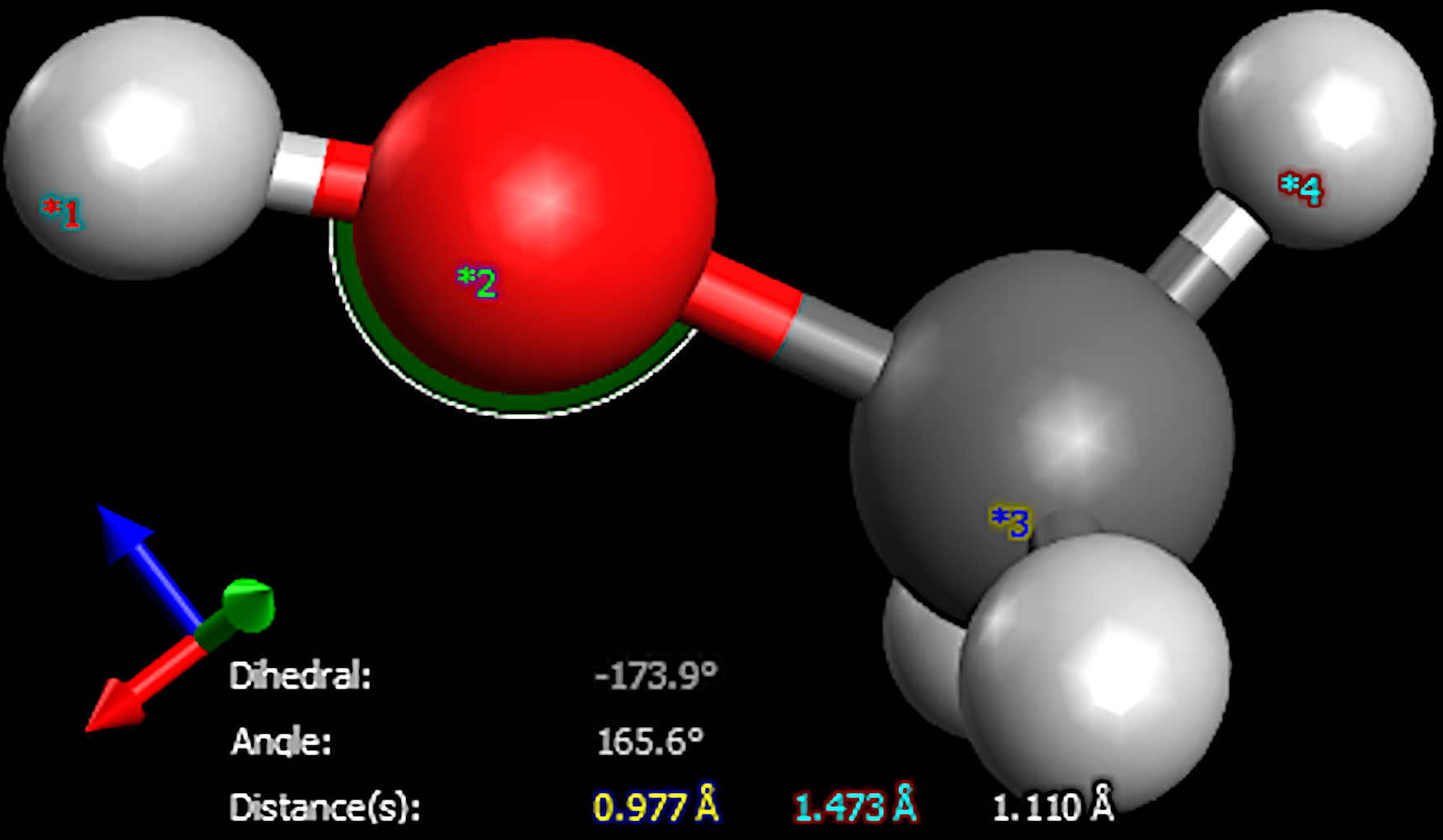}    
\includegraphics[width=0.37\linewidth]{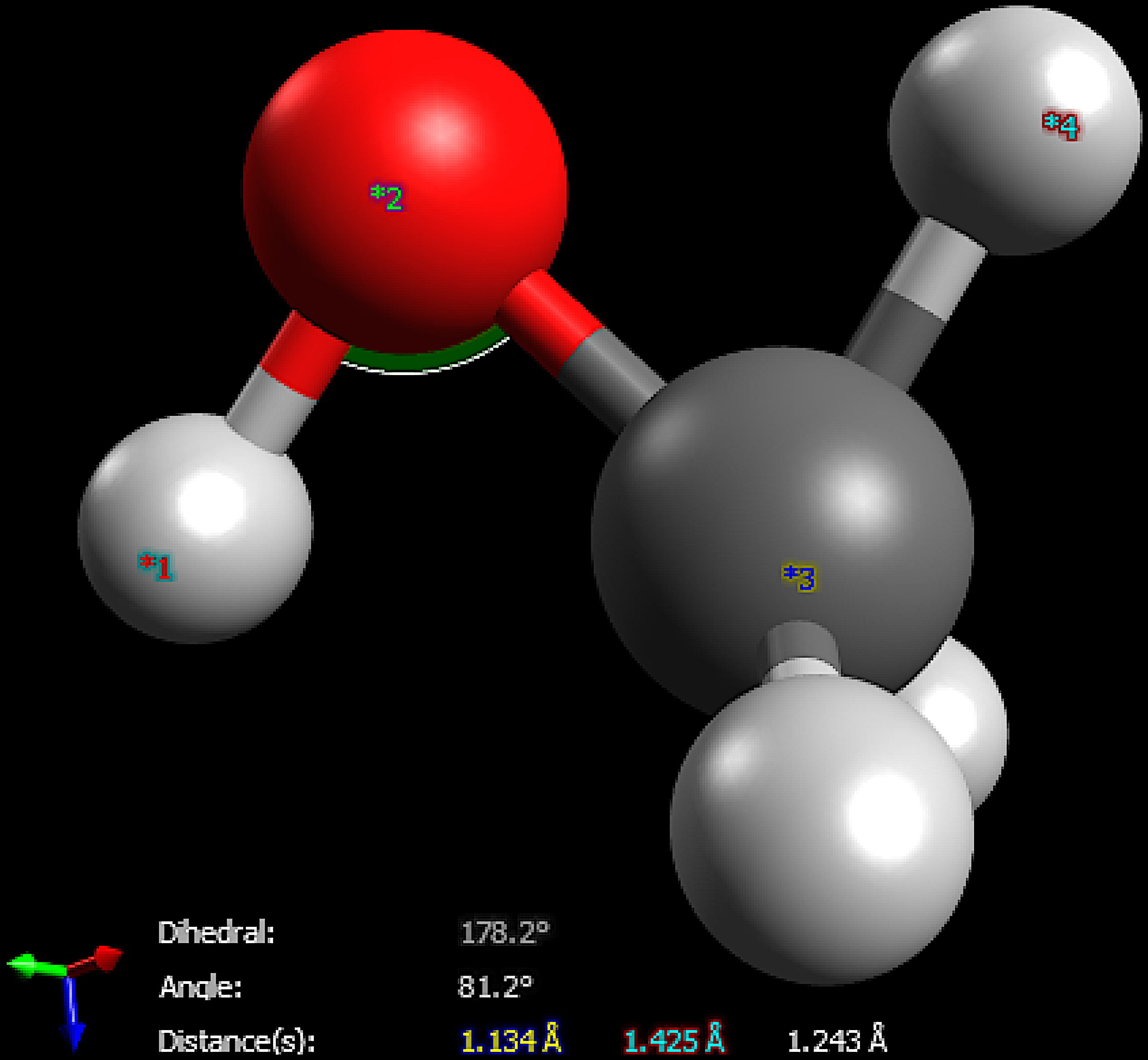}    
\includegraphics[width=0.4\linewidth]{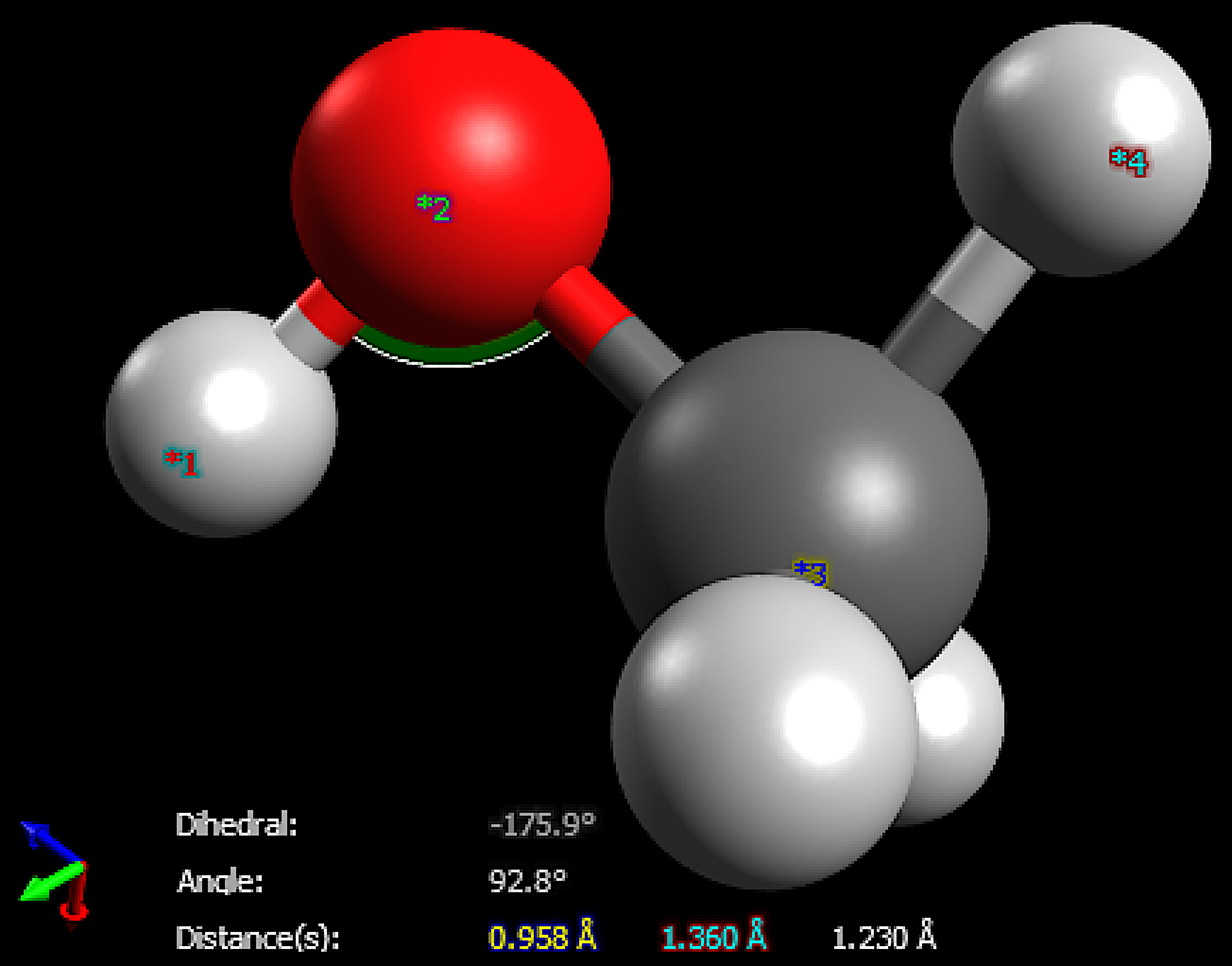}    
    \caption{
Geometries of fitting set members which were used for targeted sampling at the early stages of PES development due to their outlier status in the fit. Upper left panel: fitting point \#3787, upper right panel: fitting point \#8853, lower panel: fitting point \#1324.
    }
    \label{fig:outliers_1}
\end{figure}

After 122 more iterations, at fitting and spares set sizes of 5 342 and 120 899 geometries the RMSWE of the fitting set was still only 0.09 kcal/mol, allowing $E_\text{targ}$ to be lowered to first 0.6 kcal/mol, followed by 83 iterations, then to 0.5 kcal/mol followed by 123 iterations, to 0.4 kcal/mol followed by 123 iterations, to 0.3 kcal/mol followed by 236 iterations, to 0.25 kcal/mol followed by 130 iterations, and finally to 0.23 kcal/mol followed by 549 iterations.
At this point the spares set had 134 953 geometries, while the fitting set had 8 097 members and a RMSWE of 0.14 kcal/mol. Here, we observed that a notable fraction of the outliers in the fit were geometries (Fig.~\ref{fig:outliers_1}) with COH bond angles close to 180° and 90°.
\par
This prompted us to manually generate additional geometries around one of these outliers (Fig.~\ref{fig:outliers_1} upper left panel) using the random displacement tool described above, yielding \ild3 600 geometries which were appended to the spares.
Test fits with different values for $a$ had shown that increasing it to 3 bohr reduced the RMSWE of the fitting set to 0.08 kcal/mol (\ild28 \cm), therefore {\scshape Robosurfer} was continued with $a=3$ bohr for 133 iterations, after which $E_\text{targ}$ was lowered to 0.21 kcal/mol and 136 further iterations were run. Here, another batch of \ild5 300 geometries sampled around an outlier (Fig.~\ref{fig:outliers_1} lower panel) were appended to the spares. Further increasing $a$ to 3.5 bohr slightly lowered the fitting errors of the fitting set, and 112 iterations were run before the vicinity of a third outlier (Fig.~\ref{fig:outliers_1} upper right panel) was sampled, and \ild3 900 points were appended to the spares. By this time, the RMSWE of the fit had risen to 0.13 kcal/mol, and in an effort to keep the fitting error low $a$ was raised again, to 4 bohr, but this reduced the RMSWE only slightly.
\par
319 iterations later, when the spares set held 148 873 geometries and the fitting set consisted of 12 110 points with a RMSWE of 0.14 kcal/mol, $E_\text{targ}$ was lowered to 0.2 kcal/mol and 37 more iterations were ran, but by this time the number of new geometries selected in each iteration had fallen greatly, suggesting that the geometries commonly generated by running the geometry optimizer and vibrationally ground-state QCT have mostly been sampled to the density implied by the primary geometry selection PI-EW-RMSD threshold. Therefore, it was reduced from 0.013 to 0.012, allowing more dense sampling. After 29 iterations, $E_\text{targ}$ was reduced first to 0.19 kcal/mol, then after 53 iterations to 0.18 kcal/mol, followed by 57 further iterations which lead to 155 411 points in the spares and a fitting set with 13 489 points and a RMSWE of 0.15 kcal/mol.
\par
Next, 57 iterations were run with $E_\text{targ}=0.17$ kcal/mol, yielding spares and fitting sets of 157~207 and 14 200 points respectively, but by this time the fitting error of the fitting set had risen quite close to $E_\text{targ}$, signaling that the available flexibility of the fitting function in use was again saturated by the fitting set. We considered raising the polynomial degree to 7 (9 343 coefficients), but a test fit with the DGELSY LLS solver showed that the numerical rank of the resulting fitting matrix would have only been 3 493, i.e. 62\% of the monomials would have been ill-determined by the fitting set. While running {\scshape Robosurfer} with DGELSY was the straightforward choice to alleviate concerns about the numerical stability of the fit, we were reluctant to take this path due to the slowness of DGELSY and its poor scaling with the number of CPU cores in use.
\par
This prompted us to also check the rank deficiency of the fitting matrix for the 6\textsuperscript{th} degree fit that had been in use, which revealed that only 2 421 out of 3 250 coefficients were strongly determined, forcing us to consider not only the fitting error but also numerical stability while looking for alternative ways of adding flexibility to the fitting function.

\subsubsection{Process of PES development II: fitting error reduction}\label{subsubsec:robosurfer_usage_errorred}
\begin{figure}[!htbp]
    \centering
\includegraphics[width=0.38\linewidth]{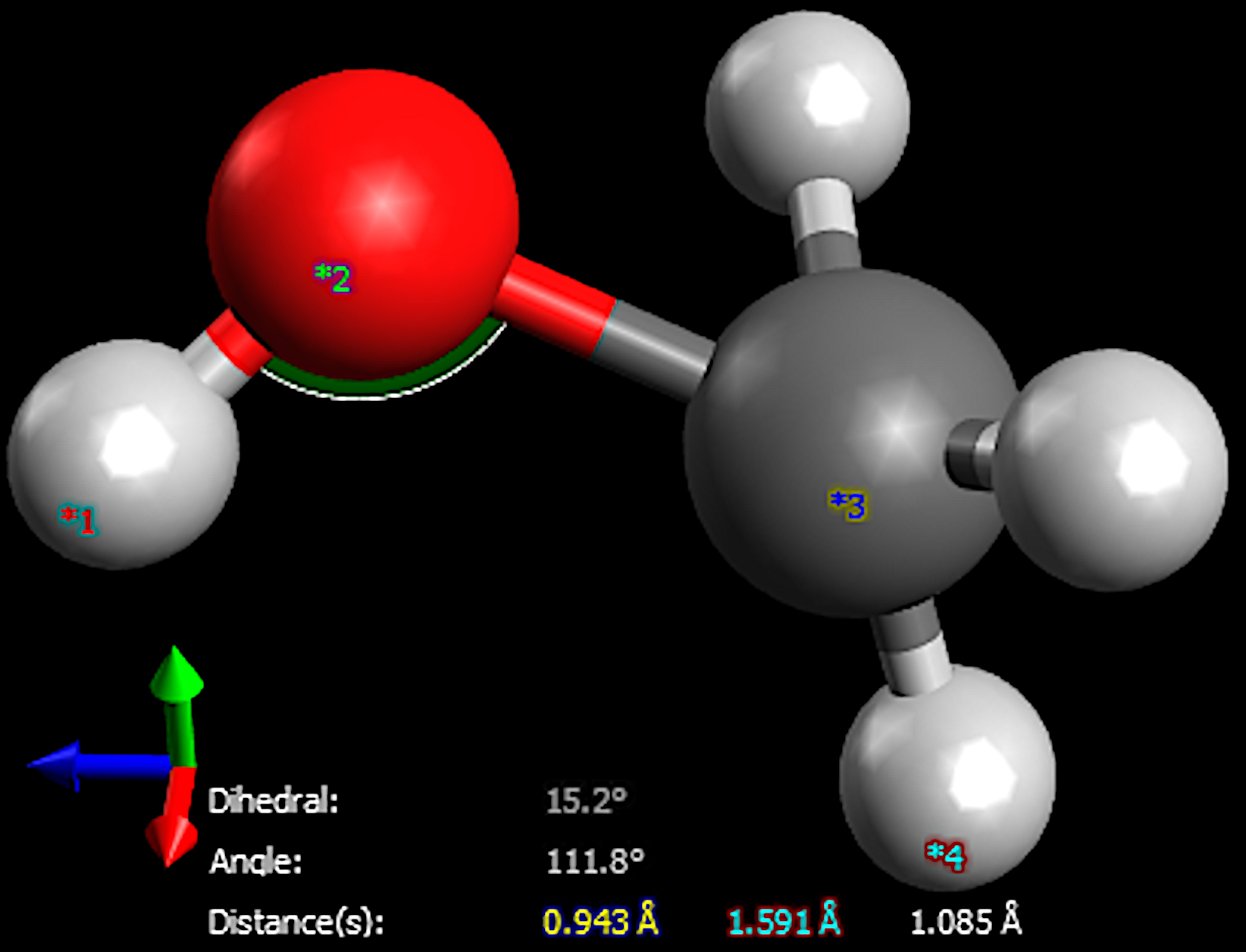}
\includegraphics[width=0.41\linewidth]{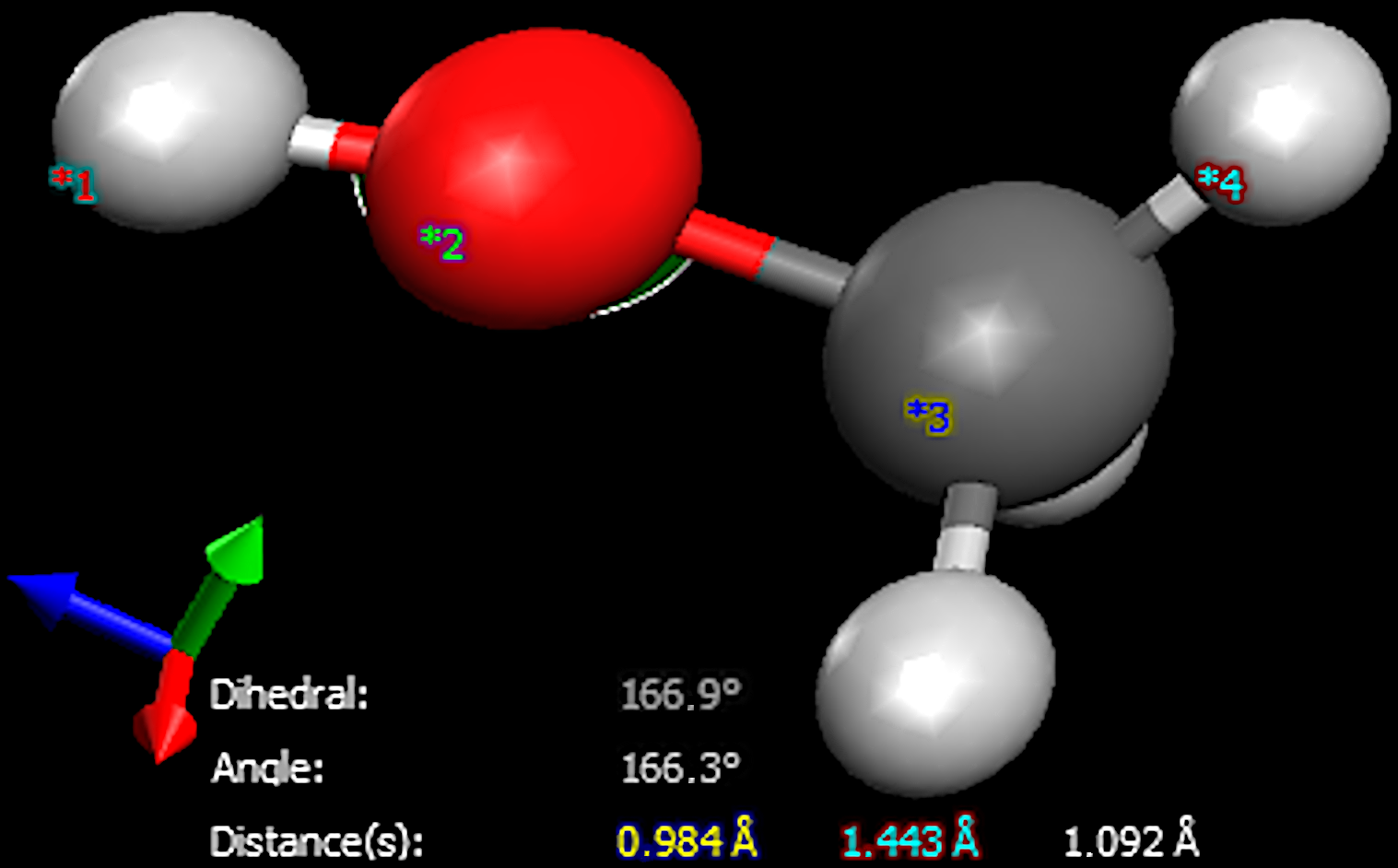}    
    \caption{
Geometries of the first pair of fitting set members which were used for targeted sampling at the middle stage of PES development due to their outlier status in the fit. Left panel: fitting point \#1736, right panel: fitting point \#2055.
    }
    \label{fig:outliers_2}
\end{figure}

First, we attempted to improve the numerical rank of the LLS matrix by continuing PES development with the DGELSY LLS solver, the hypothesis being that discarding the weakly determined monomials could result in higher fitting errors at geometries that would make those monomials more strongly determined, causing their addition to the fitting set. While fitting the 14~200 point fitting set with DGELSY did reveal an outlier (Fig.~\ref{fig:outliers_2} left panel), moving 100 points into the fitting set did not meaningfully improve the rank deficiency or change the fitting errors. Then, thinking that the last two series of {\scshape Robosurfer} iterations might have added suboptimally chosen geometries, the fitting set was truncated back to its 13 489 geometry state, and iterations with DGELSY were run until it again reached 14 300 geometries, but this has not yielded an improvement either and the fitting set was once again reset to 13 489 geometries.
\par
After generating \ild3 400 samples around the outlier in the left panel of Fig.~\ref{fig:outliers_2} and appending them to the spares set, {\scshape Robosurfer} iterations were continued with DGELSY and $E_\text{targ}$ raised back to 0.19~kcal/mol, until the fitting set grew to 14 401 points, at which point the spares set contained 159~522 geometries. Here the fit had a RMSWE of 0.16 kcal/mol, and we had to accept that approaching our planned fitting accuracy of 10 \cm\ would not be possible without a 7\textsuperscript{th} degree main polynomial. Possible explanations for the apparent difficulty of keeping the fitting error of the fitting set low are given in Section 2.2 of the main article.
\begin{figure}[!htbp]
    \centering
\includegraphics[width=0.8\linewidth]{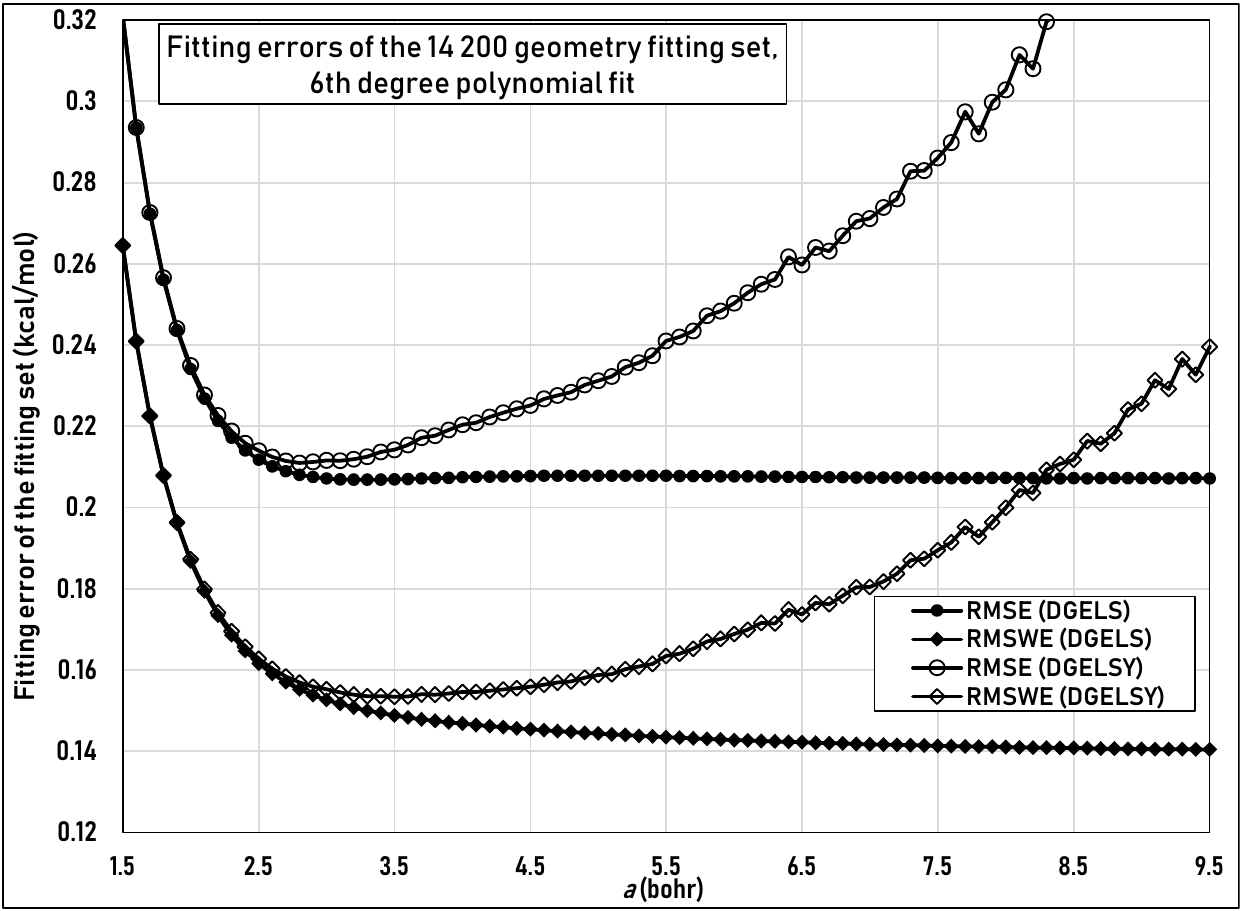}    
    \caption{
RMS fitting error and RMS weighted fitting error of a 14~200 geometry fitting set in function of the distance transformation parameter used in the 7\textsuperscript{th} degree fitting polynomial, computed with both the DGELS and DGELSY linear least squares solvers.
    }
    \label{fig:RMS_VS_a_14200}
\end{figure}
\par
Test fits done using the earlier 14 200 geometry fitting set had indicated (Fig.~\ref{fig:RMS_VS_a_14200}) that while the simpler DGELS LLS solver (which assumes full numerical rank) favors very large $a$ distance transformation parameters, the more sophisticated DGELSY solver (which only uses the numerically well-conditioned subset of the coefficients) reveals that raising $a$ (which corresponds to a more diffuse polynomial basis) beyond 2-3 bohr makes the fit increasingly ill-conditioned. For the above reasons, we chose to continue with a 7\textsuperscript{th} degree fit, $a=2.5$ bohr and the DGELSY solver.
\par
\begin{figure}[!htbp]
    \centering
\includegraphics[width=0.38\linewidth]{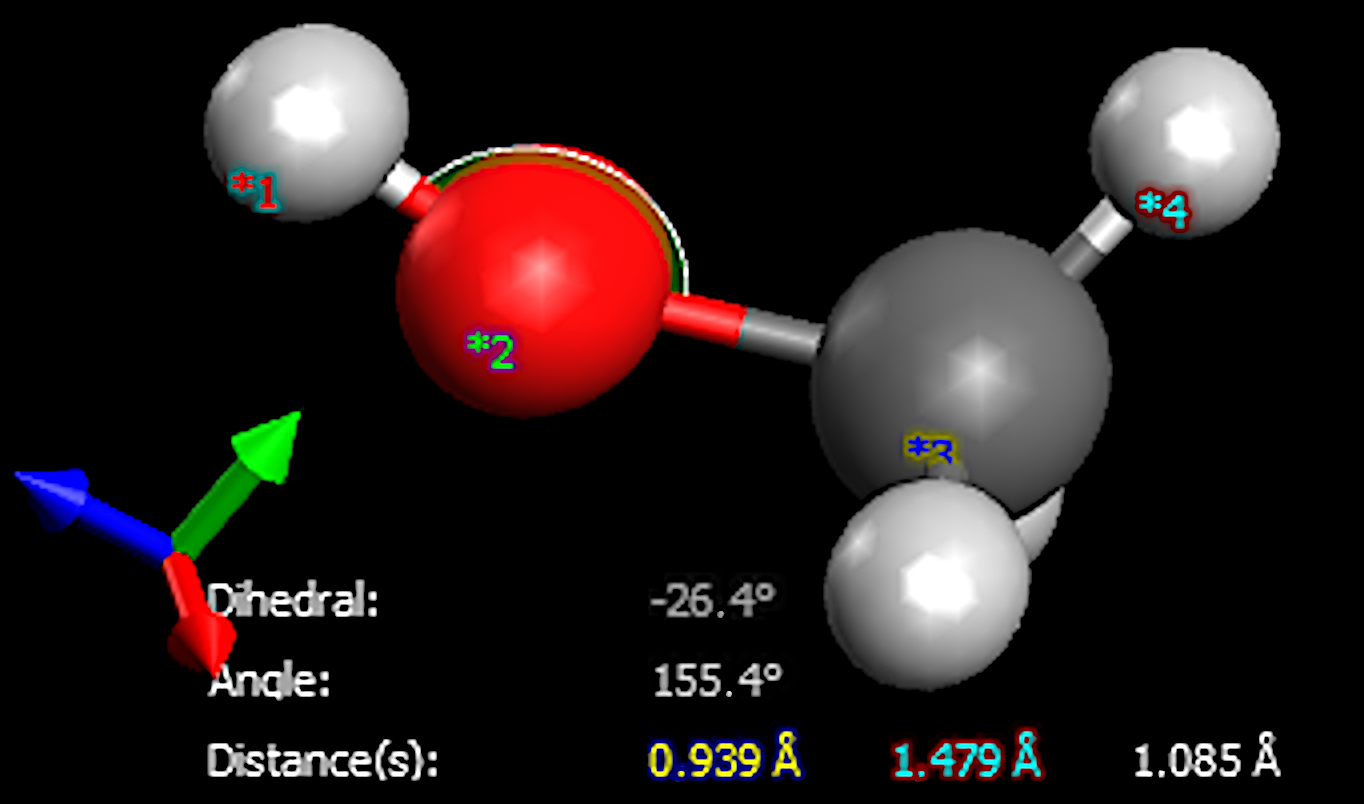}
\includegraphics[width=0.34\linewidth]{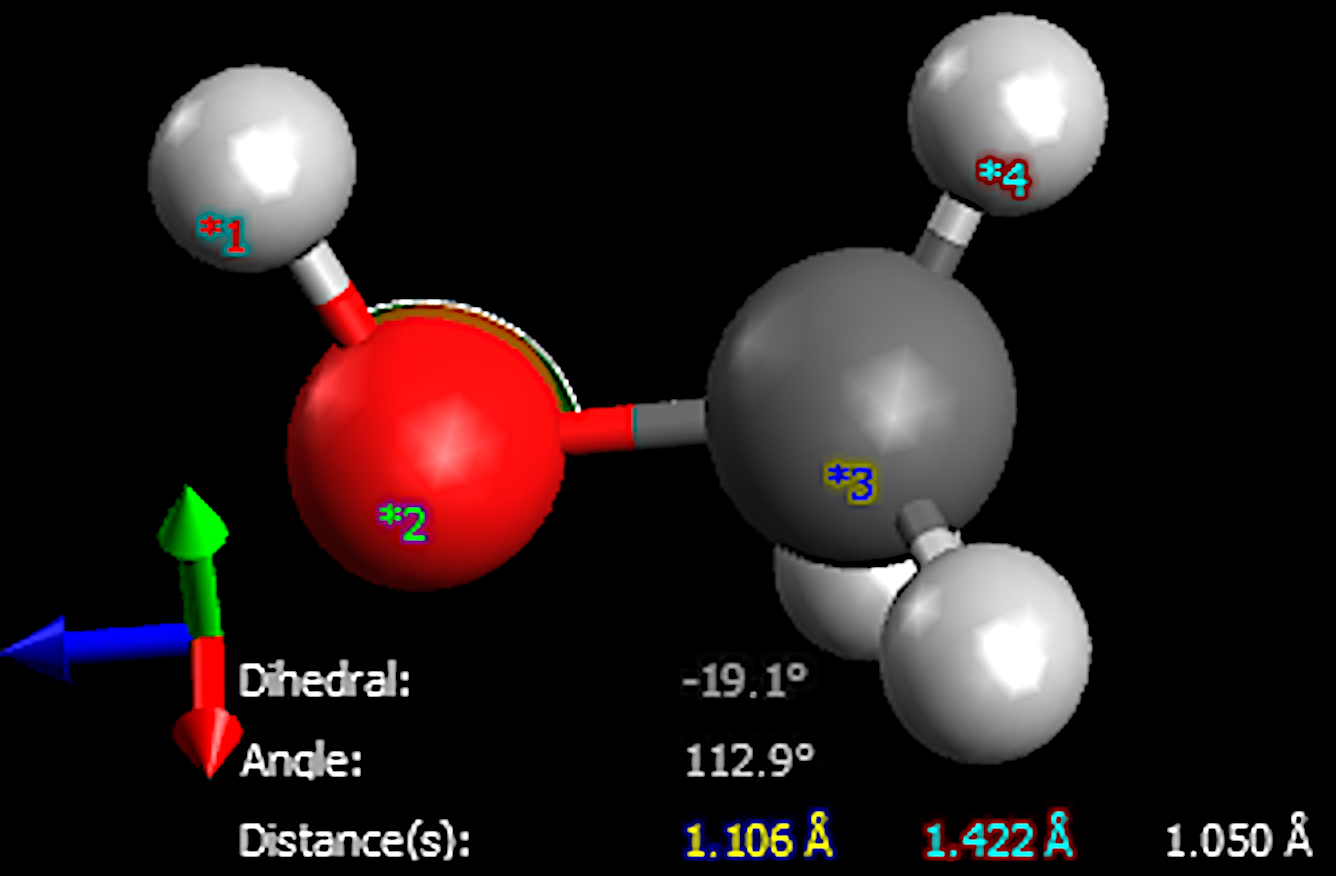}    
    \caption{
Geometries of the second pair of fitting set members which were used for targeted sampling at the middle stage of PES development due to their outlier status in the fit. Left panel: fitting point \#8197, right panel: fitting point \#3655.
    }
    \label{fig:outliers_3}
\end{figure}

This yielded a much lower RMSWE of 0.05 kcal/mol for the 14 401 geometry fitting set, with a numerical rank of 5 942 out of 9 343 coefficients, but several outliers were observed with relative energies $<37.2$ kcal/mol. We generated 2 888 additional samples around one of them  (Fig.~\ref{fig:outliers_2} right panel), appended these to the spares set and resumed running {\scshape Robosurfer} with the primary PI-EW-RMSD threshold reset back to 0.013, until we reached 16 436 fitting points. One of the original outliers from the 14 401 geometry fit that persisted up to this point (Fig.~\ref{fig:outliers_3} left panel) was then the subject of the usual targeted sampling, which added \ild3 112 points to the spares. {\scshape Robosurfer} was then continued until the fitting set grew to 17 401 geometries.
\par
\begin{figure}[!htbp]
    \centering
\includegraphics[width=0.85\linewidth]{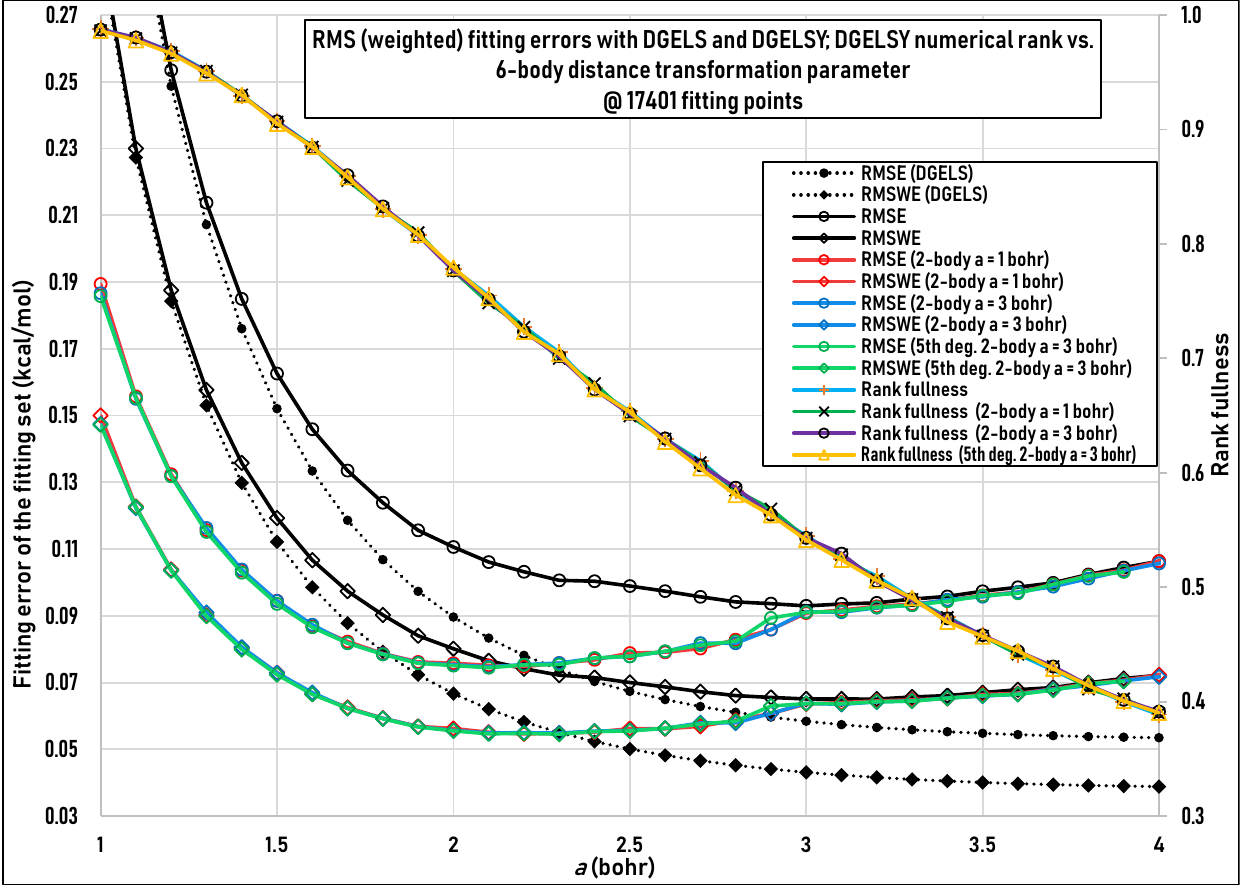}    
    \caption{
Left vertical axis: RMS fitting errors and RMS weighted fitting errors of the 17~401 geometry fitting set in function of the distance transformation parameter used in the 7\textsuperscript{th} degree main (6-body) fitting polynomial, computed with the DGELSY linear least squares (LLS) solver unless otherwise indicated. Right vertical axis: ratio between the LLS matrix numerical rank from DGELSY and the number of polynomial coefficients. Curves with no $a_{\text{2-body}}$ indicated belong to fits with no 2-body extra polynomial, curves where only $a_{\text{2-body}}$ is indicated belong to fits with a 2\textsuperscript{nd} degree 2-body extra polynomial.
    }
    \label{fig:RMS_VS_a_17401}
\end{figure}

Here, the spares set held 163 653 geometries and the RMSWE of the fitting set was 0.07 kcal/mol (\ild24 \cm), already much higher than our goal of \ild10 \cm. Since further increasing the main polynomial degree to 8 was not practically feasible, we began to increase the flexibility of the fitting function by introducing the extra monomials described in Section~\ref{subsubsec:robosurfer_config}, starting with 2-body extra monomials up to 2\textsuperscript{nd} degree. The $1/r_{ij}$ modifier for the 2-body extra monomials was enabled and left enabled until much later in PES development.
We performed test fits (Fig.~\ref{fig:RMS_VS_a_17401}) to find optimal values for $a_{\text{2-body}}$ and $a$ and found that the addition of 2-body extra monomials has a very notable effect on the relationship between fitting errors and $a$, despite the comparatively minuscule number of associated coefficients (12) at 2\textsuperscript{nd} degree.
The extra 2-body monomials significantly improve the achievable fitting errors for $a<3$ bohr and shift the optimal value from \ild3 bohr to 2.1--2.3 bohr, where the LLS matrix has a much more favorable numerical rank (\ild70\% rank completeness), as reported by DGELSY.
\par
At this point we experimented with increasing the degree of the 2-body extra polynomial to 5, but as seen on Fig.~\ref{fig:RMS_VS_a_17401} the changes in fitting errors and rank completeness are both negligible, and the choice of $a_{\text{2-body}}$ between 1 and 3 bohr is similarly inconsequential. Based on these tests, we continued developing the PES with $a_{\text{2-body}}=2.5$ bohr and $a=2.1$ bohr.
\par
This initially resulted in an RMSWE of 0.055~kcal/mol (19.1~\cm), but with multiple outliers. After 1000 geometries had been moved into the fitting set, the RMSWE rose to 21.6~\cm\ but only two outliers remained. One of these (Fig.~\ref{fig:outliers_3} right panel) was chosen for targeted sampling and extra geometries were generated around it as usual, while {\scshape Robosurfer} was still running. When the fitting set reached 18 462 geometries, the extra geometries were appended to the spares set of the 17~401 geometry fitting set and {\scshape Robosurfer} was resumed with this spares set until there were 19~401 geometries in the fitting set, 166~220 in the spares set, and the RMSWE was 23.6~\cm.
\par
\begin{figure}[!htbp]
    \centering
\includegraphics[width=0.35\linewidth]{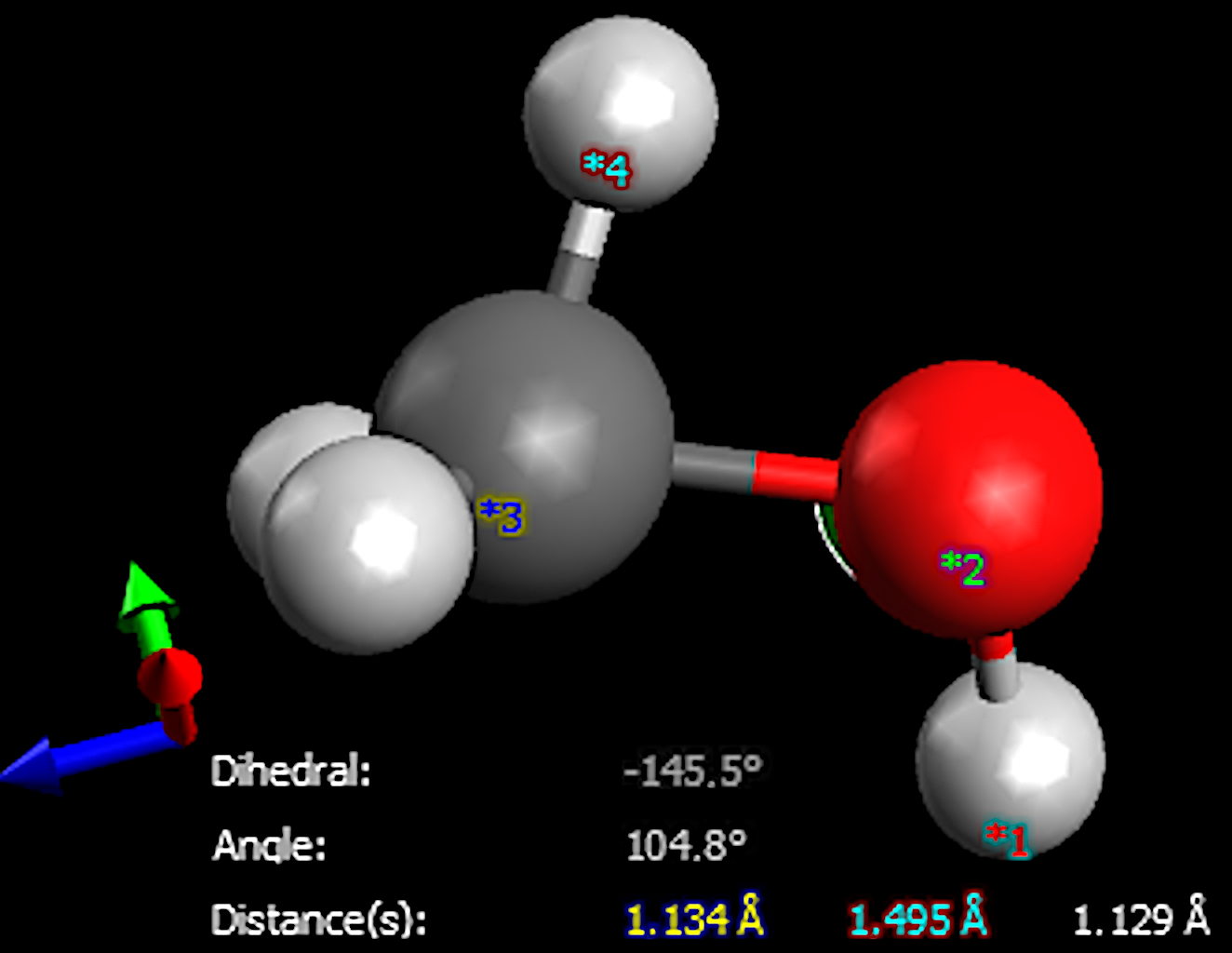}
\includegraphics[width=0.38\linewidth]{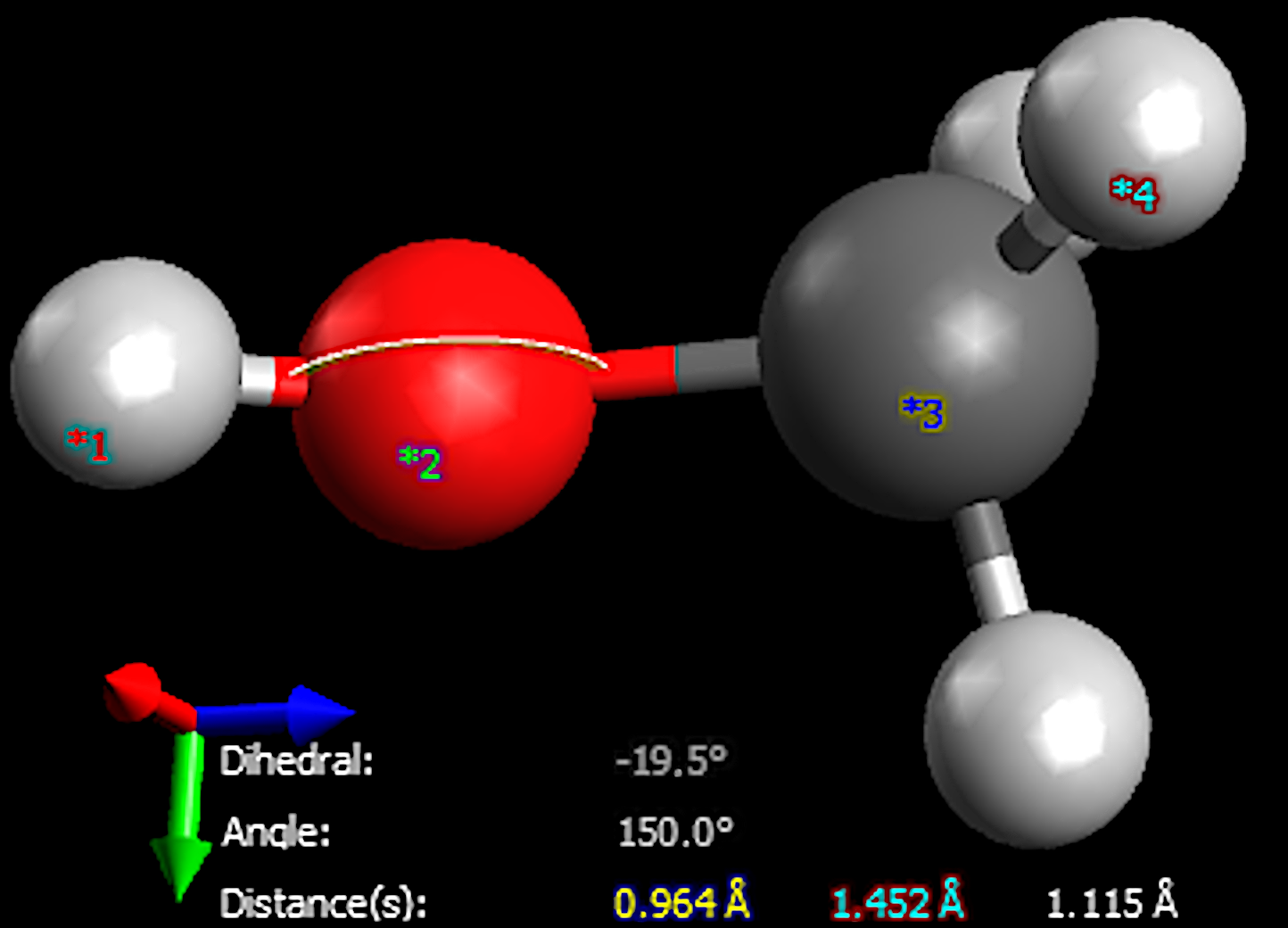}
    \caption{
Geometries of the third pair of fitting set members which were used for targeted sampling at the middle stage of PES development due to their outlier status in the fit. Left panel: fitting point \#18507, right panel: fitting point \#16520.
    }
    \label{fig:outliers_4}
\end{figure}

Here, extra geometries were generated around one of the two outliers (Fig.~\ref{fig:outliers_4} left panel) that were present in the 19~401 point fit, they were appended to the spares, and {\scshape Robosurfer} was resumed with $E_\text{targ}$ lowered from 0.19 to 0.13~kcal/mol. By the time the fitting set grew to 20 401 geometries, its RMSWE had grown to 25.2~\cm, prompting us to investigate options for increasing the flexibility of the fitting function again.

\par
\begin{figure}[!htbp]
    \centering
\includegraphics[width=0.85\linewidth]{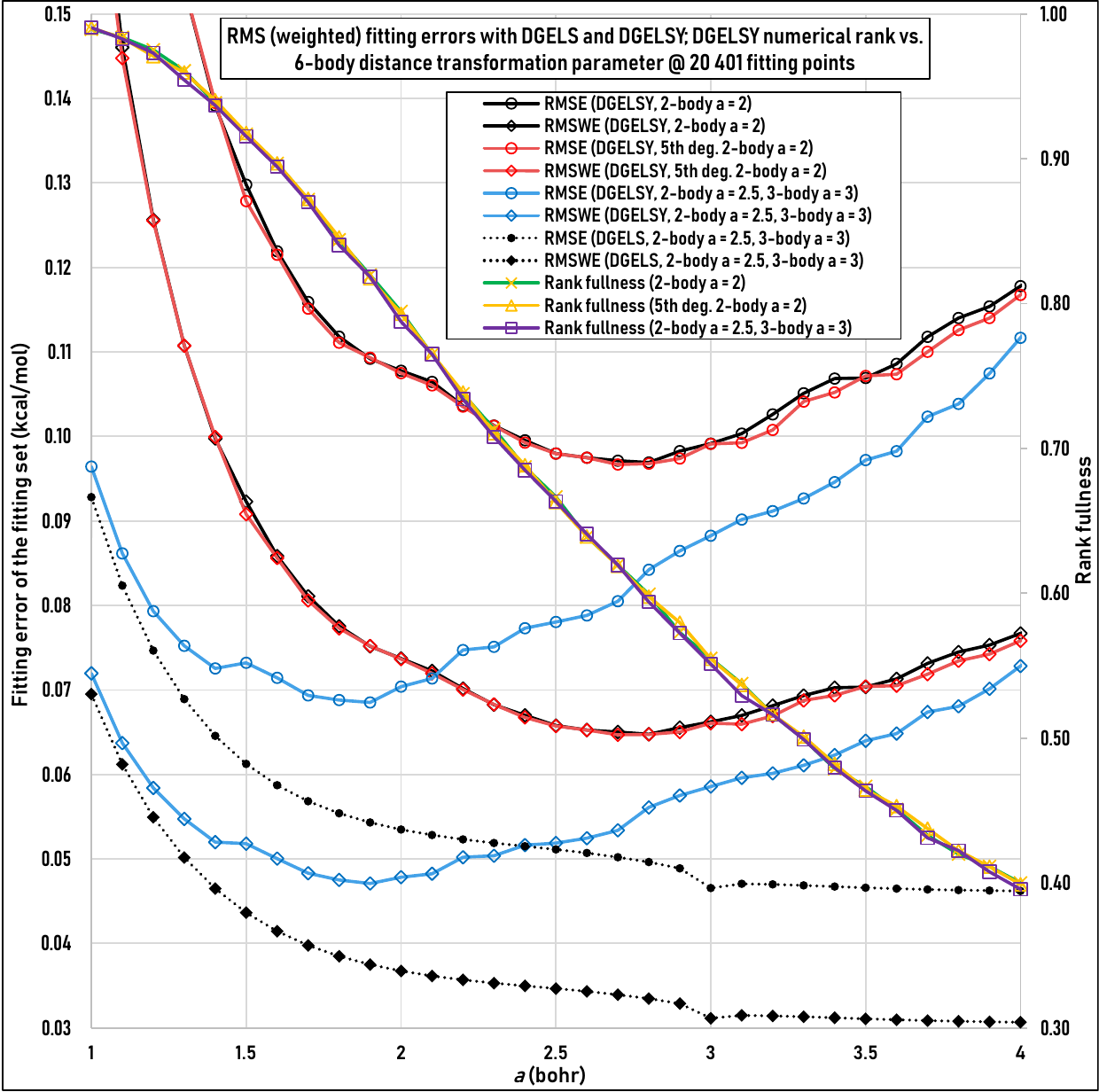}    
    \caption{
Left vertical axis: RMS fitting errors and RMS weighted fitting errors of the 20~401 geometry fitting set in function of the distance transformation parameter used in the 7\textsuperscript{th} degree main (6-body) fitting polynomial, computed with the DGELS and DGELSY linear least squares solvers. Right vertical axis: ratio between the LLS matrix numerical rank from DGELSY and the number of polynomial coefficients. Curves with no $a_{\text{3-body}}$ indicated belong to fits with no 3-body extra polynomial, curves where the degree of the 2-body extra polynomial is not indicated belong to fits with a 2\textsuperscript{nd} degree 2-body extra polynomial.
    }
    \label{fig:RMS_VS_a_20401}
\end{figure}

Test fits (Fig.~\ref{fig:RMS_VS_a_20401}) confirmed that raising the degree of the 2-body extra polynomial still offered negligible benefits, but adding the 3-body extra polynomial resulted in a substantial reduction in fitting errors (\ild8.8 \cm\ RMSWE), even at the 3\textsuperscript{rd} degree where it only contributes 53 coefficients. Just like when the 2-body extra polynomial was added, the shape of the RMSE vs. $a$ curve was altered by the 3-body extra polynomial. The optimal value of $a$ again shifted towards lower values: to 1.9 bohr from 2.1 bohr.
This was a welcome change, since reducing $a$ improves the numerical rank of the fitting matrix, furthermore we noted that with the 3-body extra polynomial present, the fitting errors of PESs obtained with the much faster DGELS LLS solver rise much slower at low values of $a$.
The acceptably high DGELSY rank completeness (\ild80\% at $a=1.9$ bohr) has allowed us to return to using the much faster DGELS for the rest of the {\scshape Robosurfer} iterations in this study, resulting in a RMSWE of 13.1 \cm\ for the 20~401 geometry fitting set, if $a_{\text{3-body}}$ is set to 3 bohr and $a_{\text{2-body}}$ is kept at 2.5 bohr.
\par
With these fitting parameters, {\scshape Robosurfer} was resumed for 55 iterations, yielding a fitting set of 21 619 geometries, but the rate at which new points were being added to the fitting set was low. Therefore, a fitting set size of 22 401 was reached by lowering $E_\text{targ}$ from 0.13 to 0.07~kcal/mol, and from 22 401 geometries, the primary PI-EW-RMSD threshold was lowered from 0.013 to 0.012. When the fitting set reached 24 401 geometries the RMSWE was still only 17.7 \cm\ but there were some outliers in the fit, one of which (Fig.~\ref{fig:outliers_4} right panel) had an \textit{ab initio} energy $<37.2$~kcal/mol, prompting us to generate 3 075 extra samples around it and append these to the spares set.
\par
Development was then continued with $E_\text{targ}=0.085$ kcal/mol and a PI-EW-RMSD threshold of 0.0115 until 26 401 points were in the fitting set, which was achieved in 34 iterations. Here, the RMSWE of the fitting set was 19.1 \cm, the spares set held 169 370 geometries, and we begun our first attempts at validating the accuracy of the PES using a variety of techniques such as the potential energy curves (PECs) of 1-dimensional slices (Fig.~\ref{fig:1dscan_bond} and ~\ref{fig:1dscan_angle}), harmonic vibrational frequency computations, and preliminary variational computations. These indicated that the PES was not satisfactory at 26 401 points, and we aimed to achieve a lower fitting error, leading us to consider our options for increasing the flexibility of the fitting function for the final time.
\begin{figure}[!htbp]
    \centering
\includegraphics[width=0.85\linewidth]{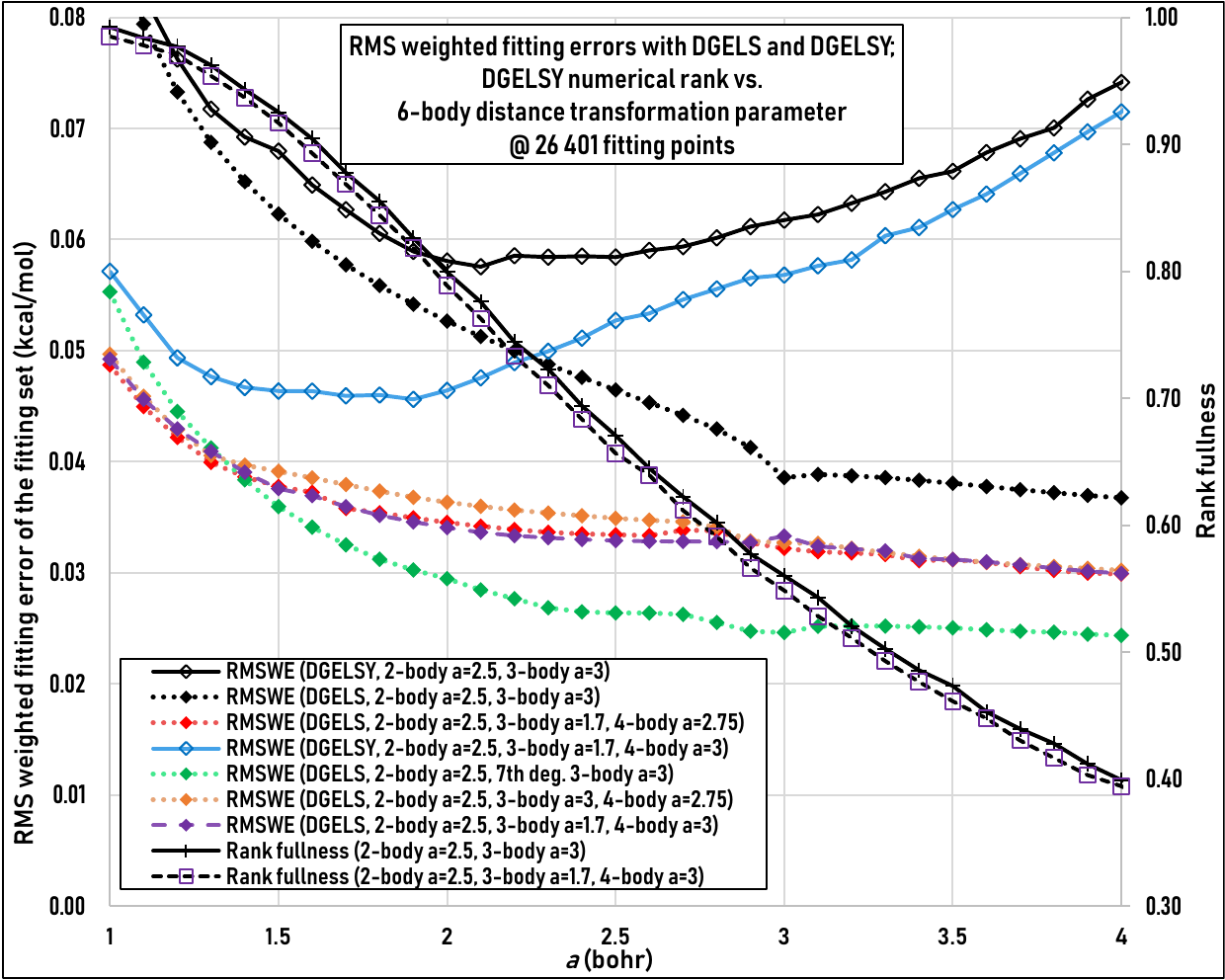}
    \caption{
Left vertical axis: RMS weighted fitting errors of the 26~401 geometry fitting set in function of the distance transformation parameter used in the 7\textsuperscript{th} degree main (6-body) fitting polynomial, computed with the DGELS and DGELSY linear least squares solvers. Right vertical axis: ratio between the LLS matrix numerical rank from DGELSY and the number of polynomial coefficients. Curves with no $a_{\text{4-body}}$ indicated belong to fits with no 4-body extra polynomial, curves where the degree of the 3-body extra polynomial is not indicated belong to fits with a 3\textsuperscript{rd} degree 3-body extra polynomial.
    }
    \label{fig:RMS_VS_a_26401}
\end{figure}
\par
After many test fits (a small subset of which are shown in Fig.~\ref{fig:RMS_VS_a_26401}), we found that while increasing the degree of the 2-body extra polynomial still only had a small effect, expanding the 3-body extra polynomial up to 7\textsuperscript{th} degree did offer a substantial improvement in fitting errors. Introducing the 4\textsuperscript{th} degree 4-body extra polynomial ($a_{\text{4-body}}=2.75$ bohr) instead of increasing the degree of the 3-body extra polynomial yielded a slightly less impressive, but still good improvement in PES fidelity.
Despite the lesser improvement, we choose to continue with the introduction of the 4-body extra polynomial, as it changed the RMSE vs. $a$ curve more towards having low fitting errors at the small $a$ values which result in a high numerical rank.
\par
Tuning the values of $a_{\text{3-body}}$ and $a_{\text{4-body}}$ yielded additional reductions in fitting errors, resulting in our final set of distance transformation parameters being $a_{\text{2-body}}=2.5$ bohr, $a_{\text{3-body}}=1.7$ bohr, $a_{\text{4-body}}=3$ bohr, and $a=1.9$ bohr. For the 26 401 geometry fitting set, these parameters achieve a RMSWE of 16.0 \cm\ and a numerical rank of 7 901 out of 9 652 coefficients (82\%) when used with the DGELSY LLS solver, and a RMSWE of 12.1 \cm\ with DGELS.

\subsubsection{Process of PES development III: finalization}\label{subsubsec:robosurfer_usage_final}
\begin{figure}[!htbp]
    \centering
    \includegraphics[width=0.35\linewidth]{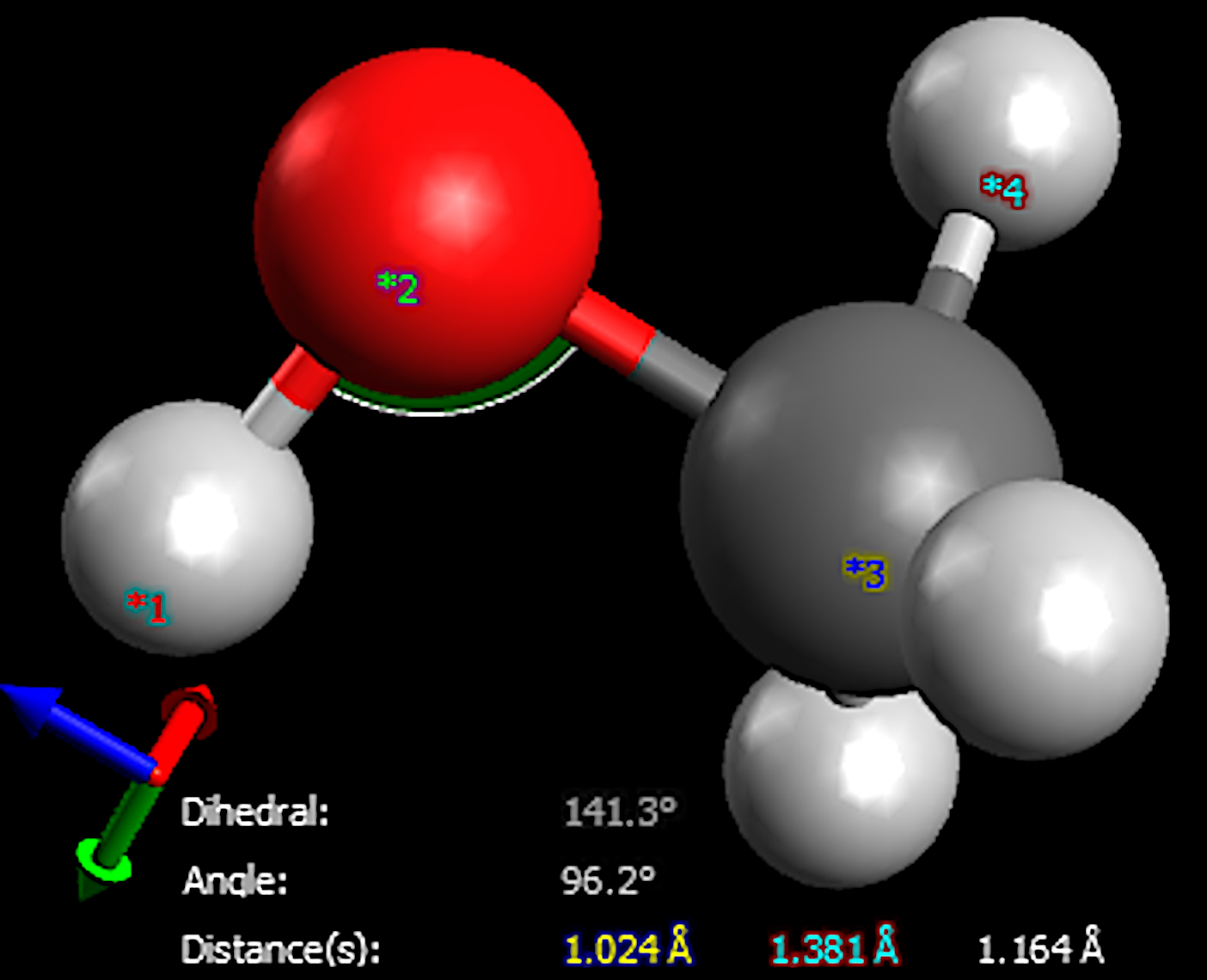}
    \caption{
Geometry \#2455 of the fitting set, which was used for targeted sampling at the final stage of PES development due to its outlier status in the fit.
    }
    \label{fig:outlier_fin}
\end{figure}

With the final fitting function and parameters established, {\scshape Robosurfer} iterations were continued from the 26~401 geometry fitting set with $E_\text{targ}=0.075$ kcal/mol (26.2 \cm) and in 17 iterations 2 000 points were added to the fitting set. Here, we noticed an outlier in the lower energy region (Fig.~\ref{fig:outlier_fin}), which was used to generate extra samples, and development was continued with $E_\text{targ}=0.07$ kcal/mol (24.5 \cm) until 29 401 fitting points, where $E_\text{targ}$ was again lowered to 0.065 kcal/mol (22.7 \cm).
\par
At a fitting set size of 30 401 geometries, where its RMSWE was 14.8 \cm\ and there were 171~501 geometries in the spares set, we stopped {\scshape Robosurfer} to decide if the quality of the PES was satisfactory and computed a wider set of tests, which included 1D PECs, harmonic frequencies, and a preliminary full-dimensional variational computation (Table~\ref{tbl:prelim_12Dvar}).
Unfortunately, our tests indicated that the vibrational energy levels of the PES deviated more from experimental data than the levels computed with PES13, especially for low energy (within 210 \cm\ from the ZPVE) modes, and the harmonic frequencies (Table~\ref{tbl:harmfreq-evolution}) were also unsatisfactory. This prompted us to continue running Robosurfer, but only 451 geometries were added over 25 iterations, therefore we started lowering $E_\text{targ}$ in steps of 0.005 kcal/mol to 0.055 kcal/mol (19.2 \cm), which resulted in the addition of 2 361 geometries to the fitting set in only 38 iterations. At this point, the RMSWE of the fitting set (15.1 \cm) was quite close to $E_\text{targ}$ and the RMSE had long exceeded it (24.2 \cm), therefore we only reduced $E_\text{targ}$ by 0.874 \cm\ before continuing PES development for 21 iterations.
\begin{table}[htbp!]
\caption{%
Vibrational energies (relative to ZPVE) of low energy torsional states obtained from preliminary 12-dimensional variational computations with the $b=4$ basis sets performed using PES fitted to two non-final fitting sets and our final fitting set, compared to energies computed with PES13 and experimental data. 
The PESs employed in this table were fitted using the DGELS LLS solver. 
MAE and MAX errors are the mean and maximum absolute differences from experiment. All values are in \cm.
}\label{tbl:prelim_12Dvar}
\begin{tabular}{r||r|ccc|r}
\hline\hline
& PES13$^a$ & 
\makecell{
30 401 geometry \\ fitting set
}
& 
\makecell{
36 416 geometry \\ fitting set
}
&
\makecell{
Final\\(39 401 geometry)\\fitting set
}
& Experiment\tabularnewline
\hline
 & 9.128 & 8.842 & 8.837 & 8.858 & 9.122\tabularnewline
 & 9.130 & 8.844 & 8.858 & 8.860 & 9.122\tabularnewline
 & 208.0 & 211.6 & 211.1 & 210.9 & 208.9\tabularnewline
 & 208.0 & 211.6 & 211.1 & 210.9 & 208.9\tabularnewline
 & 292.5 & 295.6 & 295.3 & 295.1 & 294.5\tabularnewline
 & 353.6 & 356.1 & 355.6 & 355.4 & 353.2\tabularnewline
 & 509.5 & 511.8 & 511.4 & 511.2 & 510.3\tabularnewline
 & 509.5 & 511.8 & 511.4 & 511.2 & 510.3\tabularnewline
 & 749.7 & 751.6 & 751.2 & 751.1 & 751.0\tabularnewline
 & 749.7 & 751.7 & 751.3 & 751.1 & 751.0\tabularnewline
 & 1044.7 & 1044.6 & 1044.5 & 1044.5 & 1046.7\tabularnewline
 & 1045.1 & 1046.4 & 1046.0 & 1045.8 & 1047.6\tabularnewline
 \hline
MAE &1.1 & 1.5 & 1.2 & 1.1 & 0\tabularnewline
MAX &2.5 & 2.9 & 2.4 & 2.2 & 0\tabularnewline
\hline\hline
\end{tabular}
\begin{flushleft}
$^a$The values are taken from Table 4 of Ref.~\citenum{Sunaga2024JCTC}.
\end{flushleft}
\end{table}
\par
Here, we noticed that some of the trajectories run by {\scshape Robosurfer} had been terminating before the number of timesteps we have specified, which was traced down to a quirk of the QCT MD program: since it was originally developed for bimolecular collisions, by default it terminated trajectories where the largest interatomic distance exceeded the largest distance of the initial geometry of the trajectory by $>1$ bohr. This makes perfect sense for terminating reactive trajectories after the products have separated, but since the initial geometry of a monomolecular trajectory only contains short interatomic distances, the vibrational motion can have a large enough amplitude to trigger this termination condition, hindering the effective sampling of large amplitude motions. After fixing this issue, PES development was continued from the 34 186 geometry fitting set with $E_\text{targ}=0.05$ kcal/mol (17.5 \cm) for 8 iterations, every iteration running 18 trajectories of 500~000 timesteps.
\par
After this, to further focus additional sampling efforts into the vibrationally relevant regions of the PES, the use of vibrationally excited trajectories was implemented in Robosurfer, and 12 of the 18 trajectories in each iteration were started from initial conditions where one of the 12 normal modes were excited by one quantum. This noticeably increased the rate at which new geometries were added to the fitting set, indicating that vibrationally excited trajectories were indeed discovering geometries that are important, but rarely sampled by ground-state QCT. With these settings, {\scshape Robosurfer} reached the final fitting set size of 39 401 geometries in 52 iterations.
\begin{table}[htbp!]
\caption{%
Harmonic frequencies of methanol (in \cm) at the equilibrium structure computed using the geometry optimization module of {\scshape Robosurfer} (optg) and PESs taken from the final phase of PES development, which were fitted to increasingly numerous geometry-energy pairs. The PESs used in this table were fitted with the DGELS LLS solver, the optimizations and numerical finite difference Hessian evaluations were done in quadruple precision arithmetic due to numerical noise. The reference values were computed with {\scshape Molpro} at the \emph{ab initio} level of theory used for PES development, CCSD(T)-F12b/cc-pVTZ-F12. The mean absolute errors (MAE) and maximum absolute errors (MAX) of the harmonic frequencies in \cm\ are also listed.
}\label{tbl:harmfreq-evolution}
\adjustbox{max width=\textwidth}{
\begin{tabular}{r||r|cccccccccccc}
\hline\hline
\makecell{Harm\\vib. \#} & \makecell{Ref.\\(Molpro)} & \makecell{30 401\\geoms.} & \makecell{31 534\\geoms.} & \makecell{33 193\\geoms.} & \makecell{34 180\\geoms.} & \makecell{35 030\\geoms.} & \makecell{35 194\\geoms.} & \makecell{35 401\\geoms.} & \makecell{36 146\\geoms.} &  \makecell{37 401\\geoms.} & \makecell{38 698\\geoms.} & \makecell{39 401\\geoms.}\tabularnewline
\hline
1 & 293.1  & 297.6  & 298.2  & 297.5  & 297.4  & 297.2  & 297.2  & 297.1  & 297.0  & 296.4  & 296.0  & 295.8\tabularnewline
2 & 1062.2 & 1061.6 & 1061.4 & 1060.8 & 1060.8 & 1061.0 & 1061.0 & 1060.9 & 1060.8 & 1060.7 & 1060.6 & 1060.5\tabularnewline
3 & 1090.1 & 1091.2 & 1091.2 & 1091.1 & 1090.9 & 1090.9 & 1090.8 & 1090.9 & 1090.8 & 1091.2 & 1090.9 & 1090.9\tabularnewline
4 & 1181.7 & 1178.0 & 1178.0 & 1178.4 & 1178.8 & 1178.5 & 1178.4 & 1178.2 & 1177.8 & 1177.6 & 1177.4 & 1177.5\tabularnewline
5 & 1383.7 & 1376.9 & 1377.2 & 1377.4 & 1377.4 & 1377.5 & 1377.6 & 1377.6 & 1377.8 & 1377.8 & 1378.1 & 1378.3\tabularnewline
6 & 1485.4 & 1486.5 & 1484.3 & 1485.2 & 1485.0 & 1484.5 & 1484.4 & 1485.0 & 1485.3 & 1485.4 & 1485.5 & 1485.7\tabularnewline
7 & 1511.1 & 1513.2 & 1512.6 & 1513.4 & 1513.0 & 1512.7 & 1512.3 & 1512.3 & 1512.5 & 1512.4 & 1511.9 & 1512.4\tabularnewline
8 & 1521.6 & 1524.4 & 1522.7 & 1522.6 & 1522.3 & 1522.1 & 1522.3 & 1522.3 & 1523.1 & 1522.7 & 1522.5 & 1522.2\tabularnewline
9 & 3015.9 & 3020.9 & 3019.5 & 3019.5 & 3019.5 & 3018.6 & 3018.6 & 3018.3 & 3018.8 & 3017.4 & 3017.4 & 3017.4\tabularnewline
10 & 3075.7 & 3073.7 & 3072.9 & 3072.6 & 3072.5 & 3072.9 & 3072.9 & 3072.7 & 3073.5 & 3073.5 & 3073.7 & 3073.8\tabularnewline
11 & 3136.7 & 3132.7 & 3132.4 & 3132.6 & 3132.2 & 3132.4 & 3132.3 & 3132.3 & 3132.3 & 3132.9 & 3133.3 & 3134.1\tabularnewline
12 & 3866.4 & 3861.9 & 3862.2 & 3862.6 & 3862.2 & 3862.8 & 3862.9 & 3862.9 & 3862.8 & 3863.1 & 3863.1 & 3863.1\tabularnewline
\hline
MAE & 0 & 3.17 & 2.98 & 2.87 & 2.84 & 2.66 & 2.63 & 2.60 & 2.65  & 2.41 & 2.25 & 2.19\tabularnewline
MAX & 0 & 6.82 & 6.47 & 6.24 & 6.24 & 6.23 & 6.08 & 6.12 & 5.88  & 5.90 & 5.62 & 5.41\tabularnewline
\hline\hline
\end{tabular}
}
\end{table}
\par
\begin{figure}[!htbp]
    \centering
\includegraphics[width=0.7\linewidth]{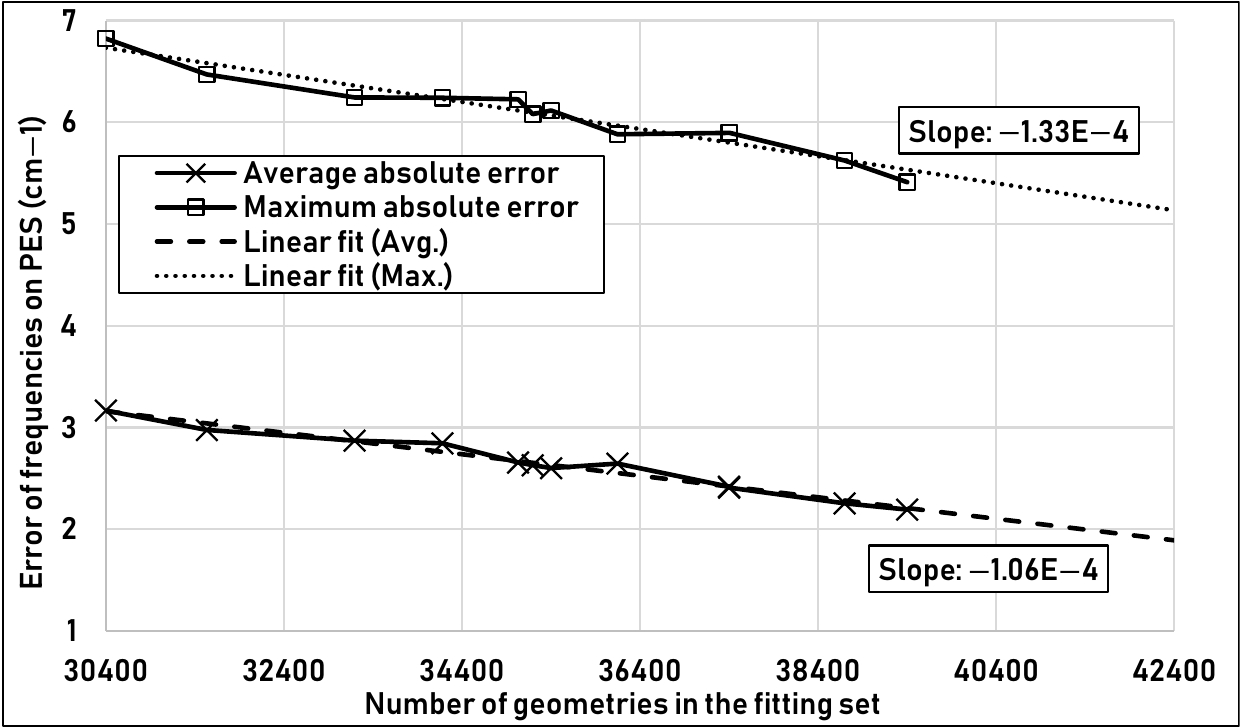}    
    \caption{
Evolution of the maximum and average absolute errors of the harmonic vibrational frequencies computed using PESs over the course of the final phase of PES development. These frequency errors belong to PESs fitted with DGELS, the geometry optimizations and numerical Hessian evaluations on the PESs were evaluated using quadruple-precision arithmetic. The reference \emph{ab initio} harmonic frequencies are shown in Table~\ref{tbl:harmfreq-evolution}.
    }
    \label{fig:vib_evolve}
\end{figure}

Over the course of these iterations, the errors of the harmonic frequencies computed using the PES were regularly evaluated (Table~\ref{tbl:harmfreq-evolution}), where a slow but steady improvement of the frequencies can be observed, at a rate of approx. $10^{-4}$ \cm\ per fitting set geometry (Fig.~\ref{fig:vib_evolve}), which would have likely continued with further {\scshape Robosurfer} iterations.

%
%
\section{ Comparison between linear least squares solvers and polynomial evaluation precisions}

For the PES fitting process, two LAPACK linear least squares (LLS) solvers are available in the {\scshape Robosurfer} code, DGELS and DGELSY. DGELS uses an ordinary QR decomposition assuming that the matrix is full rank. Actual matrices are not full rank, which may warrant the use of DGELSY. DGELSY can handle rank deficiency correctly by performing the QR factorization with column pivoting, but it is much more computationally costly.

Harmonic frequencies computed using different PES variants (DGELS vs. DGELSY, as well as different fitting set sizes and polynomial evaluation precisions of the PES) and the {\scshape Molpro} code are listed in Table \ref{tbl:harm_freq_dgels}. 
Attempting to run the geometry optimization with polynomial coefficients from DGELS and double precision polynomial evaluation results in poor convergence of the structure and large values for the rotational and translational frequencies, which should, in theory, become zero as shown by the values for quadruple precision polynomial evaluation. 
This implies that the PESs fitted using DGELS may include a lot of numerical noise. 
The coefficients obtained with DGELSY also yield slightly large rotational and translational frequencies in double precision, but to a much lesser extent. 
Both DGELSY PESs show larger fitting errors for the 1st mode (torsional motion) than the combination of DGELS and quadruple precision evaluation. However, this turned out not to be an issue for our purposes, as evidenced by the good agreement of our torsional computation (Table \ref{tbl:vib1}). The harmonic frequencies of other normal modes agree with the {\scshape Molpro} values within a few \cm.

%
%
\begin{table}[htbp!]
\caption{
Harmonic frequencies of methanol (in \cm) at the equilibrium structure computed using the geometry optimization module of {\scshape Robosurfer} (optg) and various PESs, and the {\scshape Molpro} code (CCSD(T)-F12b/cc-pVTZ-F12 and Hessian projection). DGELS and DGELSY are employed as LLS solvers, and double and quadruple precisions are used for evaluating the polynomials when the resulting PESs are used.
The labels a-f correspond to the frequencies of rotational and translational motion, while 1-12 correspond to the vibrational harmonic frequencies. 
The mean absolute errors (MAE) and maximum absolute errors (MAX) of the harmonic frequencies in \cm\ are also listed.
}\label{tbl:harm_freq_dgels}
\begin{tabular}{c| r rrrr| rrr}
\hline\hline
 &  & \multicolumn{4}{c|}{30401} & \multicolumn{3}{c}{39401}\tabularnewline
 &  & DGELS & DGELSY & DGELS & DGELSY &  DGELSY &DGELS & DGELSY\tabularnewline
\# & {\scshape Molpro} & doub & doub & quad &quad &  doub  &quad & quad\tabularnewline
 \hline
a & 0 & --102.60 & --5.24 & --0.04 & --0.04 & --15.89  &--0.04 & --0.03\tabularnewline
b & 0 & --51.27 & --2.78 & --0.02 & --0.02 & --9.99 & --0.02 & --0.02\tabularnewline
c & 0 & 10.54 & 4.92 & 0.01 & 0.01 & --6.14  &0.01 & 0.01\tabularnewline
d & 0 & 58.71 & 5.75 & 0.01 & 0.02 &  --4.35 &0.01 & 0.02\tabularnewline
e & 0 & 96.81 & 8.04 & 0.03 & 0.03 &  --2.98 &0.03 & 0.03\tabularnewline
f & 0 & 129.53 & 13.76 & 0.04 & 0.03 & 8.09  &0.04 & 0.03\tabularnewline
1 & 293.06 & 302.44 & 300.53 & 297.59 & 300.56 & 298.73  &295.80 & 298.59\tabularnewline
2 & 1062.17 & 1061.97 & 1064.05 & 1061.62 & 1064.03 &  1063.30& 1060.48 & 1063.31\tabularnewline
3 & 1090.10 & 1099.08 & 1091.14 & 1091.21 & 1091.11 &  1091.94 &1090.90 & 1091.97\tabularnewline
4 & 1181.67 & 1177.56 & 1178.30 & 1178.01 & 1178.26 &  1177.73 &1177.48 & 1177.76\tabularnewline
5 & 1383.68 & 1382.66 & 1377.33 & 1376.86 & 1377.32 & 1378.83  &1378.27 & 1378.94\tabularnewline
6 & 1485.40 & 1486.72 & 1487.18 & 1486.51 & 1487.13 & 1485.42 & 1485.75 & 1485.47\tabularnewline
7 & 1511.08 & 1514.78 & 1513.10 & 1513.20 & 1513.08 & 1512.45  &1512.40 & 1512.54\tabularnewline
8 & 1521.64 & 1526.28 & 1525.09 & 1524.37 & 1525.05 & 1524.95  &1522.20 & 1524.99\tabularnewline
9 & 3015.85 & 3022.09 & 3014.59 & 3020.86 & 3014.57 &  3014.64 &3017.37 & 3014.66\tabularnewline
10 & 3075.65 & 3076.12 & 3071.52 & 3073.74 & 3071.50 & 3073.12  &3073.78 & 3073.13\tabularnewline
11 & 3136.65 & 3139.23 & 3132.78 & 3132.72 & 3132.75 &  3134.45 &3134.11 & 3134.46\tabularnewline
12 & 3866.44 & 3862.71 & 3862.64 & 3861.91 & 3862.61 &   3863.14&3863.11 & 3863.11\tabularnewline
\hline
ZPVE & 11311.81 & 11468.46 & 11309.24 & 11309.41 & 11309.10 &11309.46   &11305.93 & 11309.57\tabularnewline
MAE &  & 3.86 & 3.37 & 3.17 & 3.37 &  2.61  &2.19 & 2.61\tabularnewline
MAX &  & 9.38 & 7.47 & 6.82 & 7.50 &   5.67 &5.41 & 5.53\tabularnewline
\hline\hline
\end{tabular}
\end{table}

In the GENIUSH computation, the DGELSY PES and double precision polynomial evaluation are employed. Since the center of mass is fixed at the origin of the body-fixed Cartesian coordinate when the PES routine is called, the translational and rotational harmonic frequency would not be a huge matter. However, using quadruple precision for the polynomial in the GENIUSH computation is also a future option because the bottleneck of the variational rovibrational computation is the Lanczos iteration, even if the polynomial is obtained at the quadruple precision.

In {\scshape Molpro}, the rotational and translational modes are projected out of the Hessian by default. We also computed the harmonic frequencies without this projection, but the differences between them were only a few $1.0\times10^{-2}$ \cm\ (see harm\_freq.tar.gz of Supplementary Information).

%
%
\section{One-dimensional PEC slices of the full-dimensional potential energy surfaces}
\begin{figure}[!htbp]
    \centering
\includegraphics[width=1.0\linewidth]{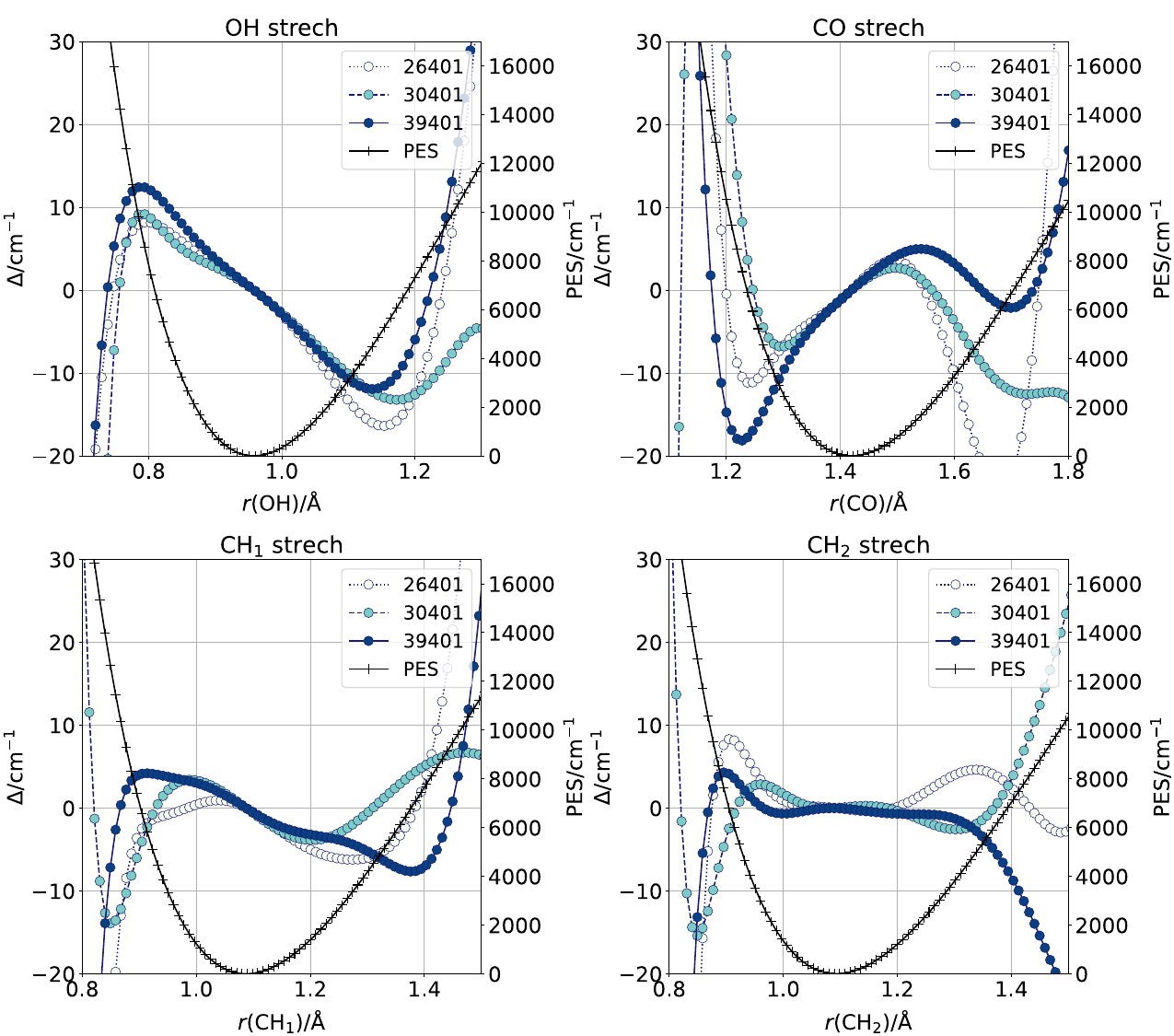}
    \caption{
Evolution of one-dimensional potential energy slices for the non-redundant bond lengths of methanol (cf. Fig.~1 in the main text). 
The other coordinates are fixed at the equilibrium geometry ($\tau=180^\circ$) obtained at each PES or \textit{ab initio} calculation. The PESs obtained with the {\scshape Robosurfer} (with 26 401, 30 401, and 39 401 geometries in the fitting set) and \textit{ab initio} calculations are shifted so that each minimum value can be zero. The left axis represents the difference between the \textit{ab initio} computations and PESs obtained by the {\scshape Robosurfer} (legends: 26401, 30401, and 39401). The right axis represents the \textit{ab initio} PES at the CCSD(T)-F12b/cc-pVTZ-F12 level (legend: PES).    
    }
    \label{fig:1dscan_bond}
\end{figure}

\begin{figure}[!htbp]
    \centering
\includegraphics[width=1.0\linewidth]{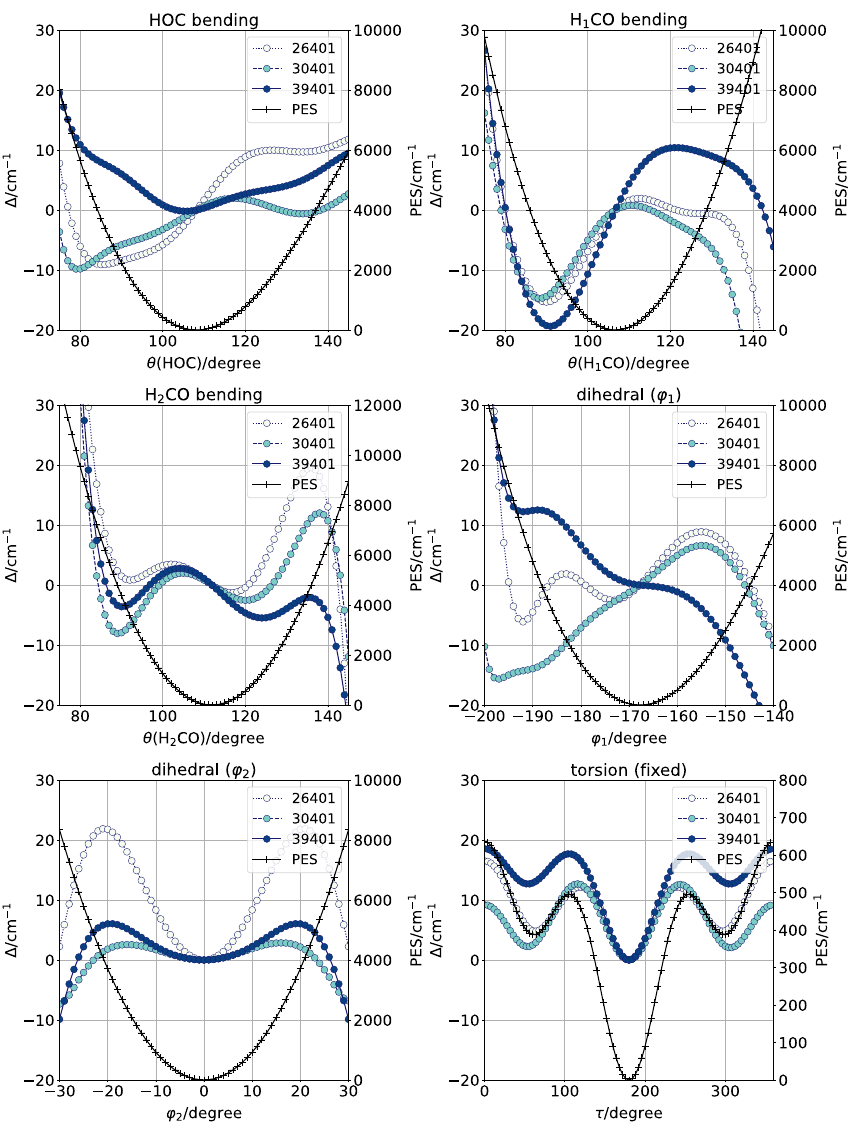}
    \caption{
Evolution of one-dimensional potential energy slices for the non-redundant bond angles of methanol (cf. Fig.~1 in the main text). 
The other coordinates are fixed at the equilibrium geometry ($\tau=180^\circ$) obtained at each PES or \textit{ab initio} calculation. The PESs obtained with the {\scshape Robosurfer} (with 26 401, 30 401, and 39 401 geometries in the fitting set) and \textit{ab initio} calculations are shifted so that each minimum value can be zero. The left axis represents the difference between the \textit{ab initio} computations and PESs obtained by the {\scshape Robosurfer} (legends: 26401, 30401, and 39401). The right axis represents the \textit{ab initio} PES at the CCSD(T)-F12b/cc-pVTZ-F12 level (legend: PES).    
    }
    \label{fig:1dscan_angle}
\end{figure}

\begin{figure}[!htbp]
    \centering
\includegraphics[width=0.5\linewidth]{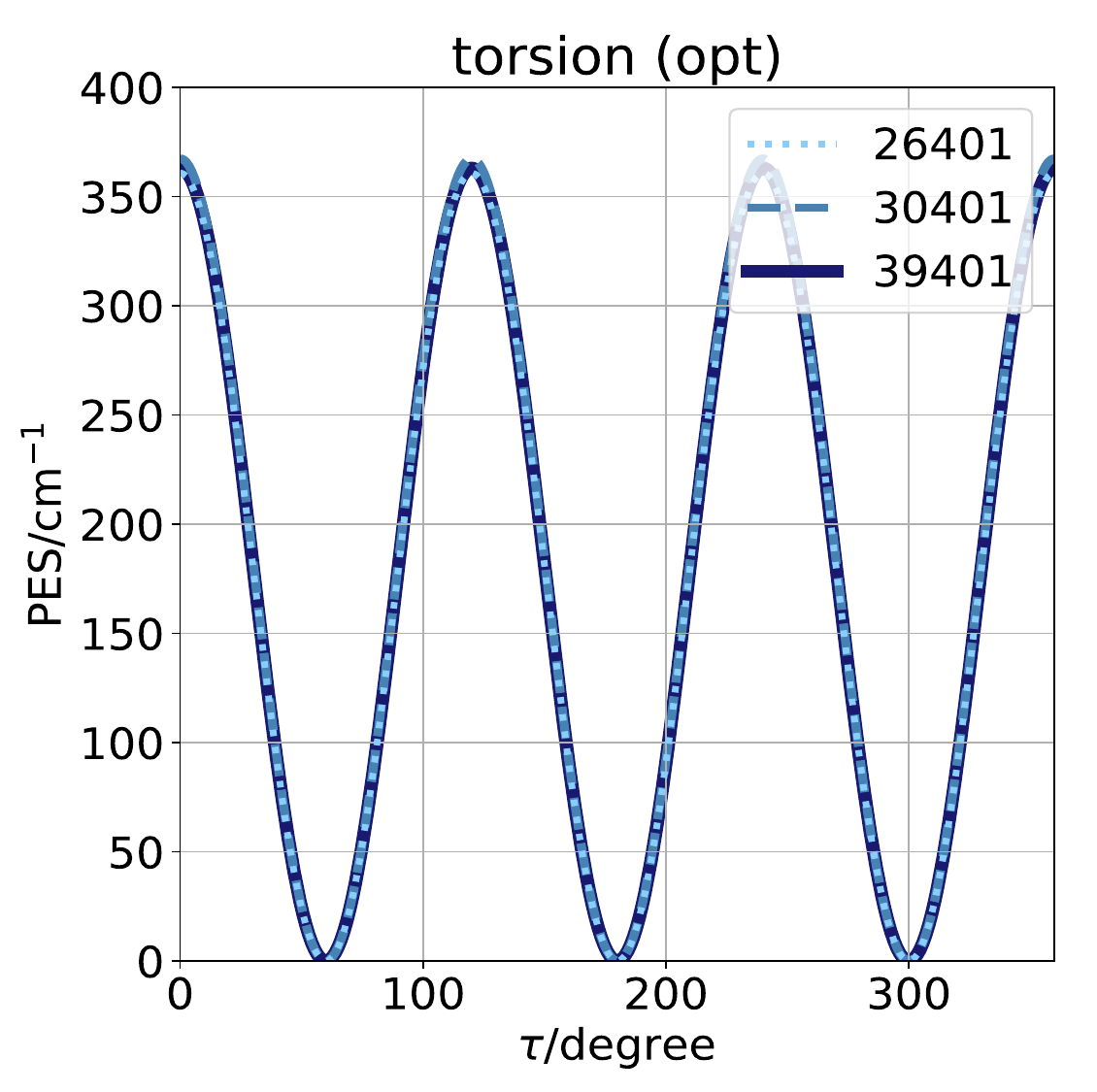}
    \caption{    
    One-dimensional potential energy slices for the torsional angle of methanol. 
    The other 11D coordinates are optimized at the given torsional angle, and the PES is shifted as the minimum value is zero. The torsional barriers of the PESs with 26401, 30401, 39401 geometry points, and the \textit{ab initio} calculation are 360.8, 367.4, 363.1, and 351.3 (in \cm) respectively. 
    }
    \label{fig:tor}
\end{figure}

Figures~\ref{fig:1dscan_bond} and~\ref{fig:1dscan_angle} visualize the potential energy curves (PECs) along the redundant internal coordinates, which are defined in the main text. In these 1D PECs, the other internal coordinates are fixed to the one in an equilibrium structure. The deviations from the \textit{ab initio} energies are also plotted. We can see that deviations from the \textit{ab initio} energy increase at coordinates far from the equilibrium structure, but the maximum deviation is only a few tens \cm\ in most of the regions shown in these figures. This indicates that the PES is very well sampled and that holes do not exist in this region at least. The systematic improvement with respect to the number of geometry points is not very obvious from these figures, which would imply that the PESs with fewer fitting set points are already well developed. However, we found that the errors of the harmonic frequencies obtained from the \emph{ab initio} values were gradually decreasing during the PES development.
Although the deviation for the torsional paths (10-20 \cm) is relatively large with respect to the PEC energy (less than 700~\cm), 
the torsional energy levels in the vibrational computation agree with the experiment very well (cf. Table~\ref{tbl:vib1}). 
Figure~\ref{fig:tor} is obtained by relaxing other 11D coordinates along the torsional angle. The periodic three minima are reproduced, which provides numerical verification that the permutational symmetry is correctly maintained by the PIP PES.

%
%

%
%

%
%

\clearpage

\section{ Employed masses and unit conversion peractor }
In the variational computations, we used atomic masses for the nuclei, $m_\mr{O}$ = 15.994 914 161 957 u, $m_\mr{C}$ = 12 u, $m_\mr{H}$ = 1.007 825 032 23 u. 1 $E_\mr{h}$ = 219474.63 \cm\ and 1 u = 1822.888 $m_\mr{e}$ was employed for the unit conversion, where $m_\mr{e}$ is the mass of an electron. 

%
%
\section{ Convergence of the vibrational energies with respect to the basis size }
The effect of the discrete variable representation (DVR) grid points on the torsional energy level is investigated (Table \ref{tbl:1D}). 
For grid numbers that are multiples of 3 (e.g., 27 and 33 grid points), the degeneracy is improved because the grid points are equally distributed over three equivalent minima. Although some splitting is found for 35 grid points, 33 grid points and functions would be enough to obtain the states up to around 2800 \cm\ within the error of 0.1 \cm. 

%
%
\begin{table}[htbp!]
\caption{
Vibrational energy levels of $\mathrm{CH}_3 \mathrm{OH}$, in $\mathrm{cm}^{-1}$, referenced to the ZPVE (\#1) using a 1D vibrational model 
for an increasing number of DVR grid points. 
The GENIUSH code and the developed PES using {\scshape Robosurfer} (39401 geometries) are used, fitted with DGELSY.
The $E$ states whose degeneracies are splitting by more than 0.01 \cm\ are highlighted in bold.
}\label{tbl:1D}
\begin{tabular}{c rrrrr}
\hline\hline
DVR & 27 & 29 & 31 & 33 & 35\tabularnewline
\hline
1 & 124.555 & 124.554 & 124.554 & 124.554 & 124.554 \tabularnewline
2 & 9.307 & 9.307 & 9.307 & 9.307 & 9.307\rule[0mm]{0mm}{5mm}\tabularnewline
3 & 9.307 & 9.308 & 9.308 & 9.307 & 9.307 \tabularnewline
4 & 204.769 & 204.768 & 204.768 & 204.768 & 204.768\rule[0mm]{0mm}{5mm}\tabularnewline
5 & 204.769 & 204.768 & 204.769 & 204.768 & 204.768 \tabularnewline
6 & 291.574 & 291.570 & 291.570 & 291.571 & 291.570\rule[0mm]{0mm}{5mm}\tabularnewline
7 & 347.470 & 347.473 & 347.473 & 347.473 & 347.473\rule[0mm]{0mm}{4mm}\tabularnewline
8 & 506.034 & \textbf{506.029} & 506.034 & 506.035 & 506.034\rule[0mm]{0mm}{5mm}\tabularnewline
9 & 506.034 & \textbf{506.041} & 506.035 & 506.035 & 506.036 \tabularnewline
10 & 748.255 & 748.255 & 748.246 & 748.256 & 748.256\rule[0mm]{0mm}{5mm}\tabularnewline
11 & 748.255 & 748.257 & 748.266 & 748.256 & 748.256 \tabularnewline
12 & 1047.569 & 1047.594 & 1047.594 & 1047.608 & 1047.594\rule[0mm]{0mm}{5mm}\tabularnewline
13 & 1047.884 & 1047.860 & 1047.860 & 1047.845 & 1047.860\rule[0mm]{0mm}{4mm}\tabularnewline
14 & 1403.021 & \textbf{1402.985} & 1403.021 & 1403.021 & \textbf{1403.001}\rule[0mm]{0mm}{5mm}\tabularnewline
15 & 1403.021 & \textbf{1403.059} & 1403.022 & 1403.021 & \textbf{1403.042} \tabularnewline
16 & 1813.635 & 1813.637 & \textbf{1813.588} & 1813.636 & 1813.637\rule[0mm]{0mm}{5mm}\tabularnewline
17 & 1813.635 & 1813.637 & \textbf{1813.686} & 1813.636 & 1813.637 \tabularnewline
18 & 2279.070 & 2279.346 & 2279.347 & 2279.317 & 2279.346\rule[0mm]{0mm}{5mm}\tabularnewline
19 & 2279.639 & 2279.380 & 2279.380 & 2279.408 & 2279.380\rule[0mm]{0mm}{4mm}\tabularnewline
20 & 2800.064 & \textbf{2799.861} & 2800.098 & 2800.098 & \textbf{2800.021}\rule[0mm]{0mm}{5mm}\tabularnewline
21 & 2800.064 & \textbf{2800.325} & 2800.099 & 2800.098 & \textbf{2800.176} \tabularnewline
22 & 3374.243 & \textbf{3375.751} & \textbf{3375.585 }& 3375.786 & 3375.787\rule[0mm]{0mm}{5mm}\tabularnewline
23 & 3374.243 &\textbf{ 3375.770} & \textbf{3375.980} & 3375.786 & 3375.787 \tabularnewline
24 & 3918.838 & 4004.995 & 4006.309 & 4006.284 & 4006.331\rule[0mm]{0mm}{5mm}\tabularnewline
25 & 4103.253 & 4005.132 & 4006.439 & 4006.497 & 4006.460\rule[0mm]{0mm}{4mm}\tabularnewline
\hline\hline
\end{tabular}
\end{table}

\section{Symmetrization in Hessian and coefficient vectors}
We developed an algorithm to enforce the numerical symmetry of the Hessian and the coefficient vectors (ave‑$L$). ave‑$L$ is obtained using the $\bar{\bos{G}}^\mr{s}$ [and $\bar{\bos{F}}^\mr{s}$] matrices, which are obtained by averaging $\bos{G}^{\mr{s}}(\tau)$ [and $\bos{F}^{\mr{s}}(\tau)$] over $\tau=60^\circ, 180^\circ, 300^\circ$. In this implementation, the symmetrically equivalent vibrational coordinates, degenerate modes, and the number of matrix elements with zero value are considered, which suppress the numerical noises occurring during the GF procedure. This is a more general and stable version to treat symmetry than that we implemented in Ref. \citenum{Sunaga2024JCTC}, which works only with \Ctv.

%
%
\section{Comparison between PES13, PES25, and experiment}
Table \ref{tbl:harm_freq_aveL} lists the harmonic frequencies associated with ave-$L$. The harmonic frequencies obtained using PES13 and PES25 qualitatively agree, but a large deviation is found, e.g., 20.3~\cm\ for $\nu_6$.

%
%
\begin{table}[!htbp]
\caption{%
Harmonic frequencies (\cm) associated with ave-$L$ and their numerical degeneracies. The results of PES25 ($\tilde{\nu}_{25}$) are compared with the values in Ref.~\citenum{Sunaga2024JCTC} using the PES13\cite{Qu2013MP_CH3+OH} ($\tilde{\nu}_{13}$). 
The corresponding state label and description, the dominant internal coordinates defined in the main text, and the symmetry label in the $C_{3\text{v}}(\text{M})$ MS group are also provided.
The $\nu_k,\ (k=1,2,\cdots,11)$ labeling follows the literature convention, corresponding to \Cs\ symmetry, to have comparable assignments for the vibrational states (Tables~\ref{tbl:vib1}-\ref{tbl:vib3}). The $\nu_k \:(k=1,2,\cdots,8)$ vibrational modes are symmetric, whereas $\nu_k \:(k=9,10,11)$ are anti-symmetric for reflection to the COH plane. 
}
\label{tbl:harm_freq_aveL}
\begin{tabular}{@{}clcccc@{}}
\hline\hline
state & description & coord & sym & $\tilde{\nu}_{13}$ \cite{Sunaga2024JCTC} & $\tilde{\nu}_{25}$\tabularnewline   %
 \hline
$\nu_1$& $\nu$(\ce{OH}) & $r_\mathrm{OH}$ & $a_1$ & 3869.820   &   3864.309 \tabularnewline %
$\nu_2$ & $\nu$(\ce{CH3})$_{\mathrm{asym}}$  & $r_\mathrm{CH}$ &$e$ & 3104.820   & 3099.061   \tabularnewline %
$\nu_9$  & $\nu$(\ce{CH3})$_{\mathrm{asym}}$ & $r_\mathrm{CH}$&$e$ & 3104.820   & 3099.060   \tabularnewline %
$\nu_3$& $\nu$(\ce{CH3})$_{\mathrm{sym}}$& $r_\mathrm{CH}$ &  $a_1$ & 3026.114   &  3024.560  \tabularnewline %
$\nu_4$&$\delta$(\ce{CH3})$_{\mathrm{asym}}$ & $\varphi_2$ &  $e$ & 1510.890   &   1517.638 \tabularnewline  %
$\nu_{10}$ & $\delta$(\ce{CH3})$_{\mathrm{asym}}$ & $\varphi_1$ &$e$ & 1510.887   &  1517.638  \tabularnewline %
$\nu_5$&$\delta$(\ce{CH3})$_{\mathrm{sym}}$ & $\theta_\mathrm{HCO}$ &  $a_1$ & 1475.296   & 1486.352   \tabularnewline  %
$\nu_6$&$\delta$(\ce{COH}) & $\theta_\mathrm{COH}$ &  $a_1$ & 1259.733   &  1280.059  \tabularnewline  %
$\nu_{11}$& $\rho$(\ce{CH3})$_{\mathrm{}}$& $\theta_\mathrm{HCO}$ &  $e$ & 1196.977   &  1183.725  \tabularnewline%
$\nu_7$  &$\rho$(\ce{CH3})$_{\mathrm{}}$ &$\theta_\mathrm{HCO}$ & $e$ & 1196.976   &  1183.725  \tabularnewline%
$\nu_8$& $\nu$(\ce{CO})$_{\mathrm{}}$&  $r_\mathrm{CO}$ & $a_1$ & 1066.481   &  1076.047  \tabularnewline %
\hline\hline
\end{tabular}
\end{table}

Tables \ref{tbl:vib1}-\ref{tbl:vib3} list all computed vibrational energies up to $2494 \mathrm{~cm}^{-1}$ beyond the ZPVE using the GENIUSH-Smolyak program, the PES13 and PES25 of $\mathrm{CH}_3 \mathrm{OH}$. The 1st-100th states on PES13 were provided in Table S7 of the Supporting Information of Ref.~\citenum{Sunaga2024JCTC}. The other states are newly computed in this study.
The detail of the assignment is explained in Section 4.3 in Ref.~\citenum{Sunaga2024JCTC}.

We reported in Ref.~\citenum{Sunaga2024JCTC} that 
 our computation for the states (no. 43, 46-47, 48) agreed better with the older experiment reported in $1974$~\cite{Serrallach1974JMS} than with the more recent experiment reported in 2003~\cite{Temsamani2003JMS}. 
In contrast, the results for these states of PES25 agree with the experiment in 2003 within a few \cm. 
Since the results of PES25 agree with the experiment in the other energy ranges as well, the agreement with the older experiment would be due to the imperfection of PES13. 
The experimental data were found for all the low-energy states, \#2-54, except for the state assigned as $7_1,\nu_{\tau}=2,\Gamma=E$ (no. 43-44). If the energy difference between the states \#43-44 and \#45 is sub-1 \cm\ as shown in our computation on PES25, it would be difficult to assign these two VBOs separately in the experiment.

The vibrational energies obtained in a matrix isolation environment \cite{Perchard2008CP_2400} are shown in parentheses, assuming they are $A_1$ or $A_2$ states, except for the doubly-degenerate \#150 state. The torsional splitting may be hindered in a matrix environment. 
The possible matrix effects would be comparable with the discrepancy of our computation from the gas phase experiment, a few \cm.

%
%
\begin{table}
\caption{
 Vibrational energies, $\tilde{\nu}$ in \cm, referenced to the zero-point vibrational energy (ZPVE, 11119.58~\cm\ for PES25, and 11108.05~\cm\ for PES13) of CH$_3$OH in 12D computed with the GENIUSH-Smolyak program with the $b=7$ basis sets on the PES25 and PES13\cite{Qu2013MP_CH3+OH}.
Comparisons with the experimental vibrational band origins are also shown.  $\delta_X=\tilde{\nu}_{\rm{exp}} -\tilde{\nu}_{\mathrm{PES}X}$, in \cm, where $X=13,25$. 
Assignments are obtained using the wavefunction of PES25's results. 
 SAMs: assignment of the curvilinear normal modes $1_n, 2_n, ..., 11_n$ $(n=0,1,\ldots)$ in Table~\ref{tbl:harm_freq_aveL},
 zero excitation ($n=0$) is noted as `0'.  `$[\left.\ldots\right]$' labels the largest contribution(s) from strongly mixed states. 
 $\Gamma$: $C_{3\text{v}}(\text{M})$ label according to Table~\ref{tbl:harm_freq_aveL} and the status of the torsional part. The details are discussed in Section~S1 of the Supporting Information of Ref.~\citenum{Sunaga2024JCTC}.
 All degeneracies for the $E$ states converged better than 0.01~\cm, except for the pairs listed in the footnotes. 
}\label{tbl:vib1}
\begin{tabular}{lc cc rr rr rc}
 \hline\hline
\# & $\nu_{\tau}$ & SAMs & $\Gamma$ & PES13 &  $\delta_{13}$  & PES25 &  $\delta_{25}$  & $\tilde{\nu}_{\rm{exp}}$ & \tabularnewline
 \hline
1 & 0 & 0 & $A_1$ & 0.0 &   & 0.0 &    &  & \tabularnewline
2-3 & 0 & 0 & $E$ & 9.117 & [0.0] & 8.856 & [0.3] & 9.122 & \citenum{Moruzzi1995_ch3oh}\tabularnewline
4-5 & 1 & 0 & $E$ & 208.0 & [0.9] & 210.9 & [--2.0] & 208.9 & \citenum{Moruzzi1995_ch3oh}\tabularnewline
6 & 1 & 0 & $A_2$ & 292.4 & [2.1] & 295.1 & [--0.6] & 294.5 & \citenum{Moruzzi1995_ch3oh}\tabularnewline
7 & 2 & 0 & $A_1$ & 353.6 & [--0.3] & 355.3 & [--2.1] & 353.2 & \citenum{Moruzzi1995_ch3oh}\tabularnewline
8-9 & 2 & 0 & $E$ & 509.3 & [1.0] & \textsuperscript{\emph{d}}511.0 & [--0.7] & 510.3 & \citenum{Moruzzi1995_ch3oh}\tabularnewline
10-11 & 3 & 0 & $E$ & 749.5 & [1.6] & 750.8 & [0.2] & 751.0 & \citenum{Moruzzi1995_ch3oh}\tabularnewline
12 & 0 & $8_1$ & $A_1$ & 1036.5 & [--2.2] & 1034.5 & [--0.2] & 1034.4 & \citenum{Lees2002PRA}\tabularnewline
13-14 & 0 & $8_1$ & $E$ & 1045.9 & [--3.2] & 1043.3 & [--0.7] & 1042.6 & \citenum{Lees2002PRA}\tabularnewline
15 & 3 & 0 & $A_2$ & 1044.2 & [2.4] & 1045.4 & [1.2] & 1046.7 & \citenum{Moruzzi1995_ch3oh}\tabularnewline
16 & 4 & 0 & $A_1$ & 1045.0 & [2.6] & 1046.5 & [1.1] & 1047.6 & \citenum{Moruzzi1995_ch3oh}\tabularnewline
17 & 0 & $6_1,7_1,11_1$ & $A_1$ & 1059.7 & [15.0] & 1074.7 & [0.0] & 1074.7 & \citenum{Lees2002PRA}\tabularnewline
18-19 & 0 & $6_1,7_1,11_1$ & $E$ & 1064.2 & [15.0] & 1079.1 & [0.2] & 1079.3 & \citenum{Lees2002PRA}\tabularnewline
20-21 & 0 & $7_1,11_1$ & $E$ & 1154.0 & [2.4] & 1155.0 & [1.5] & 1156.5 & \citenum{Lees2002PRA}\tabularnewline
22 & 0 & $7_1,11_1$ & $A_2$ & 1163.8 & [0.2] & 1162.1 & [1.9] & 1164.0 & \citenum{Lees2002PRA}\tabularnewline
23-24 & 1 & $8_1$ & $E$ & 1242.4 & [0.6] &  1245.8 & [--2.8] & 1243.0 & \citenum{Moruzzi1995_ch3oh}\tabularnewline
25-26 & 0 & $6_1$ & $E$ & 1285.8 & [11.7] &  1298.2 & [--0.7] & \textsuperscript{\emph{e}}1297.5  & \citenum{Lees2004JMS} \tabularnewline
27 & 0,1 & $6_1,7_1,11_1$ & $A_1$ & 1309.8 & [10.8] &  1319.9 & [0.7] &  \textsuperscript{\emph{f}}1320.6 & \citenum{Lees2004JMS} \tabularnewline
28 & 1 & $7_1,8_1,11_1$ & $A_2$ & 1322.7 & [--2.1] & 1320.4 & [0.2] & \textsuperscript{\emph{g}}1320.6 & \citenum{Lees2004JMS}\tabularnewline
29-30 & 1 & $6_1$ & $E$ & 1336.7 & [2.8] & 1343.5 & [--4.0] & \textsuperscript{\emph{h}}1339.5 & \citenum{Serrallach1974JMS}\tabularnewline
31 & 1 & $8_1$ & $A_2$ & 1328.9 & [16.1] & 1344.9 & [0.1] & \textsuperscript{\emph{i}}1345.0 & \citenum{Serrallach1974JMS}\tabularnewline
32 & 0,2 & $6_1,7_1,11_1$ & $A_1$ & 1372.8 & [--3.1] & 1369.3 & [0.4] & \textsuperscript{\emph{j}}1369.7 & \citenum{Lees2004JMS}\tabularnewline
33 & 2 & $8_1$ & $A_1$ & 1387.5 & [1.4] & 1392.1 & [--3.2] & 1388.9 & \citenum{Moruzzi1995_ch3oh}\tabularnewline
34-35 & 4 & 0 & $E$ & \textsuperscript{\emph{a}}1391.4 & [4.8] & 1392.2 & [4.0] & 1396.2 & \citenum{Fehrensen2003JCP}  \tabularnewline
36-37 & 1 & $7_1,11_1$ & $E$ & \textsuperscript{\emph{b}}1428.7 & [--1.6] & 1426.2 & [1.0] & \textsuperscript{\emph{k,l}}1427.1 & \citenum{Lees2004JMS} \tabularnewline
38 & 0 & $5_1$ & $A_1$ & 1444.7 & [8.6] & 1450.3 & [3.0] & \textsuperscript{\emph{l}}1453.3 & \citenum{Temsamani2003JMS}\tabularnewline
39-40 & 0 & $5_1$ & $E$ & 1455.1 & [7.0] & 1458.7 & [3.4] & 1462.1 & \citenum{Temsamani2003JMS}\tabularnewline
41-42 & 0 & $4_1,10_1$ & $E$ & 1462.7 & [11.2] & 1471.7 & [2.2] & \textsuperscript{\emph{l}}1473.9 & \citenum{Temsamani2003JMS}\tabularnewline
43-44 & 2 & $7_1,11_1$ & $E$ & \textsuperscript{\emph{c}}1471.8 &   &  1477.4 &   &   & \tabularnewline
45 & 0 & $4_1,10_1$ & $A_2$ & 1468.1 & [13.4] & 1477.8 & [3.7] & 1481.5 & \citenum{Temsamani2003JMS}\tabularnewline
46-47 & 0 & $4_1,10_1$ & $E$ & 1477.7 & [5.6] & 1481.8 & [1.5] & \textsuperscript{\emph{l}}1483.3 & \citenum{Temsamani2003JMS}\tabularnewline
48 & 0 & $4_1,10_1$ & $A_1$ & 1481.6 & [4.5] & 1485.4 & [0.7] & \textsuperscript{\emph{l}}1486.1 & \citenum{Temsamani2003JMS}\tabularnewline
49 & 1 & $6_1$ & $A_2$ & 1531.6 & [9.9] &  1541.7 & [--0.2] & \textsuperscript{\emph{m}}1541.5 & \citenum{Lees2004JMS}\tabularnewline
50 & 2 & $6_1$ & $A_1$ & 1540.7 & [6.5] &  1547.4 & [--0.1] &  \textsuperscript{\emph{n}}1547.2 & \citenum{Lees2004JMS}\tabularnewline
\hline\hline
\end{tabular}
\begin{flushleft}
$^a$~1391.400, 1391.415 
$^b$~1428.660, 1428.673 
$^c$~1471.807, 1471.820 
$^d$~511.029, 511.039 \par
$^e$~{Assignment of Ref. \citenum{Lees2004JMS} is $7_1,\nu_{\tau}=1,E$.} \par
$^f$~{Ref. \citenum{Lees2004JMS} note that subbands for which one or more of the initial lines that originate from a $v_t = 0$ lower level of low energy and low quantum number are observed in the line of $^g$} \par
$^g$~{Assignment of Ref. \citenum{Lees2004JMS} is $6_1,\nu_{\tau}=0,A$.} \par
$^h$~{Ref. \citenum{Lees2004JMS} also reports another experimental value: 1335.2.} \par
$^i$~{Assignment of Ref. \citenum{Serrallach1974JMS} is $7_1,\nu_{\tau}=1$.} \par
$^j$~{Assignment of Ref. \citenum{Lees2004JMS} is $\nu_{\tau}=0,A$, and the SAM is not assigned there.} \par
$^k$~{Assignment of Ref. \citenum{Lees2004JMS} is $\nu_{\tau}=1,E$, and the SAM is not assigned there.} \par
$^l$~{Ref. \citenum{Serrallach1974JMS} also reports experimental values: 1414.0, 1454.5, 1465(3), 1477.2, 1479.5 in \cm\ assignable to our no. 36-37, 38, 41-42, 46-47, and 48, energy levels, respectively.} \par
$^m$~{Assignment of Ref. \citenum{Lees2004JMS} is $\nu_{\tau}=1,A$, and the SAM is not assigned there.} \par
$^n$~{Assignment of Ref. \citenum{Lees2004JMS} is $\nu_{\tau}=2,A$, and the SAM is not assigned there.} \par
\end{flushleft}
\end{table}

\begin{table}
\caption{Vibrational states of $\mathrm{CH}_3 \mathrm{OH} \ldots$ [Table \ref{tbl:vib1} continued.]}\label{tbl:vib2}
\begin{tabular}{lc cc rr rr rc}
 \hline\hline
\# & $\nu_{\tau}$ & SAMs & $\Gamma$ & PES13 &   $\delta_{13}$ &   PES25 &  $\delta_{25}$  & $\tilde{\nu}_{\rm{exp}}$ & \tabularnewline
 \hline
51-52 & 2 & $8_1$ & $E$ & 1543.5 & [4.6] & 1549.9 & [--1.8] & 1548.1 & \citenum{Moruzzi1995_ch3oh}\tabularnewline
53-54 & 1 & $6_1$ & $E$ & 1577.7 & [--2.4] &  1574.5 & [0.9] & 1575.3 & \citenum{Lees2004JMS}\tabularnewline
55-56 & 1,2 & $5_1,7_1,11_1$ & $E$ & 1656.3 &   &  1661.1 &   &   & \tabularnewline
57-58 & 1 & $5_1$ & $E$ & 1662.8 &   &  1663.9 &   &   & \tabularnewline
59 & 1 & $4_1,10_1$ & $A_2$ & 1668.7 &   &  1672.2 &   &   & \tabularnewline
60 & 1 & $4_1,10_1$ & $A_1$ & 1668.0 &   &  1682.5 &   &   & \tabularnewline
61-62 & 1 & $4_1,10_1$ & $E$ & 1676.6 &   &  1685.0 &   &   & \tabularnewline
63 & 1,2 & $6_1,7_1,11_1$ & $A_2$ & 1712.2 &   &  1714.7 &   &   & \tabularnewline
64 & 2 & $6_1,7_1,11_1$ & $A_1$ & 1745.3 &   &  1742.2 &   &   & \tabularnewline
65-66 & 2 & $6_1$ & $E$ & \textsuperscript{\emph{a}}1731.9 & [10.8] &  \textsuperscript{\emph{b}}1742.4 & [0.3] & \textsuperscript{\emph{d}}1742.6 & \citenum{Lees2004JMS} \tabularnewline
67 & 1 & $5_1$ & $A_2$ & 1748.1 &   &  1746.6 &   &   & \tabularnewline
68-69 & 1 & $4_1,10_1$ & $E$ & 1756.0 &   &  \textsuperscript{\emph{c}}1765.1 &   &   & \tabularnewline
70-71 & 3 & $8_1$ & $E$ & 1782.3 &   &  1778.5 &   &   & \tabularnewline
72-73 & 5 & 0 & $E$ & 1794.7 &   &  1803.0 &   &   & \tabularnewline
74 & 2 & $5_1$ & $A_1$ & 1809.7 &   &  1807.7 &   &   & \tabularnewline
75-76 & 2 & $4_1,10_1$ & $E$ & 1819.2 &   &  1827.1 &   &   & \tabularnewline
77 & 3 & $7_1,11_1$ & $A_1$ & 1907.6 &  &  1909.7 &   &   & \tabularnewline
78 & 3 & $7_1,11_1$ & $A_2$ & 1907.4 &   &  1909.8 &   &   & \tabularnewline
79-80 & 2,3 & $6_1,7_1,11_1$ & $E$ & 1942.1 &   &  1950.9 &   &   & \tabularnewline
81-82 & 2 & $5_1$ & $E$ & 1974.8 &   &  1963.1 &   &   & \tabularnewline
83-84 & 2 & $4_1,10_1$ & $E$ & 1958.9 &   &  1969.1 &   &   & \tabularnewline
85 & 2 & $4_1,10_1$ & $A_2$ & 1983.7 &   &  1991.2 &   &   & \tabularnewline
86 & 2 & $4_1,10_1$ & $A_1$ & 1985.8 &   &  1993.8 &   &   & \tabularnewline
87-88 & 3 & $6_1$ & $E$ & 1986.6 &   &  1997.5 &   &   & \tabularnewline
89 & 0 & $8_2$ & $A_1$ & 2060.8 & [--5.8] & 2058.1 & [--3.1] & 2055.0 & \citenum{Lees2002PRA}\tabularnewline
90-91 & 0 & $8_2$ & $E$ & 2070.3 & [--6.7] & 2066.8 & [--3.2] & 2063.6 & \citenum{Lees2002PRA}\tabularnewline
92 & 3 & $8_1$ & $A_2$ & 2077.1 &   &  2080.9 &   &   & \tabularnewline
93 & 4 & $8_1$ & $A_1$ & 2077.9 &   &  2081.9 &   &   & \tabularnewline
94 & 0 & {[}$6_1+8_1,8_2${]} & $A_1$ & 2091.9 & [8.4] & 2102.2 & [--2.0] & \textsuperscript{\emph{e}}2100.2 & \citenum{Lees2002PRA}\tabularnewline
95-96 & 0 & {[}$7_1+8_1,8_1+11_1${]} & $E$ & 2096.5 & [8.4] & 2106.4 & [--1.6] & 2104.9 & \citenum{Lees2002PRA}\tabularnewline
97 & 0 & {[}$6_1+7_1,6_1+11_1${]} & $A_1$ & 2123.9 & [21.2] & 2144.8 & [0.3] & \textsuperscript{\emph{f}}2145.1 & \citenum{Lees2002PRA}\tabularnewline
98-99 & 0 & {[}$6_1+7_1,6_1+11_1${]} & $E$ & 2126.1 & [21.3] & 2146.9 & [0.4] & 2147.4 & \citenum{Lees2002PRA}\tabularnewline
100-101 & 0 & {[}$7_1+8_1, 8_1+11_1${]} & $E$ & 2183.2 & [--0.2] & 2182.5 & [0.5] & \textsuperscript{\emph{g}}2182.9 & \citenum{Lees2002PRA}\tabularnewline
102 & 0 & {[}$7_1+8_1, 8_1+11_1${]} & $A_2$ & 2192.8 & [--2.2] & 2190.1 & [0.5] & \textsuperscript{\emph{h}}2190.6 & \citenum{Lees2002PRA}\tabularnewline
\hline\hline
\end{tabular}
\begin{flushleft}
$^a$~1731.850, 1731.861
$^b$~1742.346, 1742.378
$^c$~1765.133, 1765.145 \par
$^d$~{Assignment of Ref. \citenum{Lees2004JMS} is $\nu_{\tau}=2,E$, and the SAM is not assigned there.} \par
$^e$~{Assignment of Ref. \citenum{Lees2002PRA} is $7_1+8_1,\nu_{\tau}=0,A$.} \par
$^f$~{Assignment of Ref. \citenum{Lees2002PRA} is $7_2,\nu_{\tau}=0,A$.} \par
$^g$~{Assignment of Ref. \citenum{Lees2002PRA} is $7_1+8_1,\nu_{\tau}=0,E$.} \par
$^h$~{Assignment of Ref. \citenum{Lees2002PRA} is $7_2,\nu_{\tau}=0,A$.} \par
\end{flushleft}
\end{table}

\begin{table}
\caption{
Vibrational states of $\mathrm{CH}_3 \mathrm{OH} \ldots$ [Table \ref{tbl:vib2} continued.]
The subscripts of Ne and N$_2$ refer to the experimental values obtained in the matrix-isolation spectroscopy in solid Ne and N$_2$, respectively.
}\label{tbl:vib3}
\begin{tabular}{lc cc rr rr rc}
 \hline\hline
\# & $\nu_{\tau}$ & SAMs & $\Gamma$ & PES13 &  $\delta_{13}$ &  PES25 &  $\delta_{25}$  & $\tilde{\nu}_{\rm{exp}}$ & \tabularnewline
 \hline
103-104 & 3 & $5_1$ & $E$ & 2192.8 &   &  2203.0 &   &   & \tabularnewline
105 & 3 & $4_1,10_1$ & $A_2$ & 2196.8 &   &  2205.8 &   &   & \tabularnewline
106 & 3 & $4_1,10_1$ & $A_1$ & 2196.8 &   &  2206.2 &   &   & \tabularnewline
107-108 & 3,4 & $7_1,11_1$ & $E$ & 2205.4 &   &  2210.4 &   &   & \tabularnewline
109-110 & 0 & $7_2,11_2$ & $E$ & 2210.9 &   &  2223.8 &   &   & \tabularnewline
111 & 0,1 & {[}$7_1+11_1,6_1,11_1${]} & $A_2$ & 2220.0 &   &  2233.7 &   &   & \tabularnewline
112-113 & 3,4 & $7_1,11_1$ & $E$ & 2247.6 &   &  2234.6 &   &   & \tabularnewline
114-115 & 3 & $4_1,10_1$ & $E$ & 2227.9 &   &  2235.1 &   &   & \tabularnewline
116 & 5 & 0 & $A_2$ & 2241.3 &   &  2241.4 &   &   & \tabularnewline
117 & 6 & 0 & $A_1$ & 2241.2 &   &  2241.5 &   &   & \tabularnewline
118-119 & 1 & $8_2$ & $E$ & 2264.4 &   &  2267.5 &   &   & \tabularnewline
120 & 0 & {[}$7_1+11_1,7_2,11_2${]} & $A_1$ & 2298.1 &   &  2301.6 &   &   & \tabularnewline
121 & 4 & $6_1$ & $A_1$ & 2293.0 &   &  2306.2 &   &   & \tabularnewline
122 & 3 & $6_1$ & $A_2$ & 2294.0 &   &  2306.2 &   &   & \tabularnewline
123-124 & 0 & {[}$7_2,11_2${]} & $E$ & 2305.6 &  & 2306.3 &  & (2305.0)\textsubscript{Ne} & \citenum{Perchard2008CP_2400}\tabularnewline
125-126 & 0 & {[}$6_1+8_1${]} & $E$ & 2314.8 &   &  2326.3 &   &   & \tabularnewline
127 & 1 & $8_2$ & $A_2$ & 2347.5 &   &  2342.2 &   &   & \tabularnewline
128 & 0 & {[}$6_1+8_1${]} & $A_1$ & 2336.8 &  & 2342.3 &  & (2360.2)\textsubscript{Ne} & \citenum{Perchard2008CP_2400}\tabularnewline
129-130 & 1 & {[}$6_1+8_1${]} & $E$ & 2364.5 &   &  2368.8 &   &   & \tabularnewline
131 & 1 & $8_2$ & $A_2$ & 2356.1 &   &  2372.0 &   &   & \tabularnewline
132 & 0 & $6_2$ & $A_1$ & 2366.6 &  & 2383.6 &  & \textsuperscript{\emph{a}}(2386.4)\textsubscript{Ne} & \citenum{Perchard2008CP_2400}\tabularnewline
133-134 & 0,1 & $6_2,7_2,11_2$ & $E$ & 2377.9 &   &  2392.3 &   &   & \tabularnewline
135 & 0 & {[}$7_1+8_1, 8_1+11_1${]} & $A_1$ & 2401.0 &  & 2397.9 &  & \textsuperscript{\emph{b}}(2393.6)\textsubscript{Ne} & \citenum{Perchard2008CP_2400}\tabularnewline
136 & 0 & {[}$6_1+7_1,6_1+11_1${]} & $A_2$ & 2383.5 &   &  2403.9 &   &   & \tabularnewline
137-138 & 0,1 & $6_2,7_2,11_2$ & $E$ & 2395.3 &   &  2410.4 &   &   & \tabularnewline
139 & 2 & $8_2$ & $A_1$ & 2409.3 &   &  2414.9 &   &   & \tabularnewline
140-141 & 4 & $8_1$ & $E$ & 2422.6 &   &  2426.6 &   &   & \tabularnewline
142-143 & 1 & {[}$7_1+8_1, 8_1+11_1${]} & $E$ & 2456.8 &   &  2452.6 &   &   & \tabularnewline
144 & 1 & {[}$7_1+11_1${]} & $A_1$ & 2446.0 &   &  2453.4 &   &   & \tabularnewline
145-146 & 0 & {[}$6_1+7_1,7_1+11_1${]} & $E$ & 2447.1 &   &  2461.5 &   &   & \tabularnewline
147 & 0 & $5_1+8_1$ & $A_1$ & 2473.0 &  & 2478.9 &  & (2475.2)\textsubscript{Ne} & \citenum{Perchard2008CP_2400}\tabularnewline
148 & 0,1 & {[}$6_1+11_1,7_1+11_1${]} & $A_2$ & 2472.0 &  & 2481.0 &  & (2489.9)\textsubscript{Ne} & \citenum{Perchard2008CP_2400}\tabularnewline
149 & 0 & $5_1+8_1$ & $E$ & 2483.6 &   &  2487.2 &   &   & \tabularnewline
150 & 0,2 & {[}$6_1+7_1,7_1+11_1${]} & $E$ & 2486.5 &   &  2494.3 &   & \textsuperscript{\emph{c}}(2495.9)\textsubscript{N$_2$}  & \citenum{Perchard2008CP_2400}\tabularnewline
\hline\hline
\end{tabular}
\begin{flushleft}
$^a$~{Assignment of Ref. \citenum{Perchard2008CP_2400} is $6_1+7_1$.}\par
$^b$~{Ref. \citenum{Perchard2008CP_2400} does not show the assignment.}\par
$^c$~{Assignment of Ref. \citenum{Perchard2008CP_2400} is $6_1+11_1$ and $8_1+10_1$ (doubly degenerate).}
\end{flushleft}
\end{table}

\end{document}